\documentstyle[12pt,graphicx]{article}
\begin{document}
\baselineskip 18pt
\def\today{\ifcase\month\or
 January\or February\or March\or April\or May\or June\or
 July\or August\or September\or October\or November\or December\fi
 \space\number\day, \number\year}

%
\def\thebibliography#1{\section*{References\markboth
 {References}{References}}\list
 {[\arabic{enumi}]}{\settowidth\labelwidth{[#1]}
 \leftmargin\labelwidth
 \advance\leftmargin\labelsep
 \usecounter{enumi}}
 \def\newblock{\hskip .11em plus .33em minus .07em}
 \sloppy
 \sfcode`\.=1000\relax}
\let\endthebibliography=\endlist
\def\beq{\begin{equation}}
\def\eeq{\end{equation}}
\def\beqn{\begin{eqnarray}}
\def\eeqn{\end{eqnarray}}
\def\rmuu{\gamma^{\mu}}
\def\rmud{\gamma_{\mu}}
\def\PL{{1-\gamma_5\over 2}}
\def\PR{{1+\gamma_5\over 2}}
\def\sinW2{\sin^2\theta_W}
\def\AEM{\alpha_{EM}}
\def\mul{M_{\tilde{u} L}^2}
\def\mur{M_{\tilde{u} R}^2}
\def\mdl{M_{\tilde{d} L}^2}
\def\mdr{M_{\tilde{d} R}^2}
\def\mz2{M_{z}^2}
\def\c2b{\cos 2\beta}
\def\au{A_u}
\def\ad{A_d}
\def\cob{\cot \beta}
\def\v#1{v_#1}
\def\tb{\tan\beta}
\def\epem{$e^+e^-$}
\def\KK{$K^0$-$\overline{K^0}$}
\def\wi{\omega_i}
\def\xj{\chi_j}
\def\Wmu{W_\mu}
\def\Wnu{W_\nu}
\def\m#1{{\tilde m}_#1}
\def\mH{m_H}
\def\mw#1{{\tilde m}_{\omega #1}}
\def\mx#1{{\tilde m}_{\chi^{0}_#1}}
\def\mc#1{{\tilde m}_{\chi^{+}_#1}}
\def\mwi{{\tilde m}_{\omega i}}
\def\mxi{{\tilde m}_{\chi^{0}_i}}
\def\mci{{\tilde m}_{\chi^{+}_i}}
\def\mz{M_z}
\def\sw{\sin\theta_W}
\def\cw{\cos\theta_W}
\def\cb{\cos\beta}
\def\sb{\sin\beta}
\def\rwi{r_{\omega i}}
\def\rxj{r_{\chi j}}
\def\rfp{r_f'}
\def\Kik{K_{ik}}
\def\Fq2{F_{2}(q^2)}
\def\f{\({\cal F}\)}
\def\d1{{\f(\tilde c;\tilde s;\tilde W)+ \f(\tilde c;\tilde \mu;\tilde W)}}
\def\tw{\tan\theta_W}
\def\sec2w{sec^2\theta_W}

\begin{titlepage}

\begin{center}
{\Large {\bf  {Coupling the Supersymmetric 210 Vector Multiplet to Matter
in SO(10)}}}\\
\vskip 0.5 true cm
\vspace{2cm}
\renewcommand{\thefootnote}
{\fnsymbol{footnote}}
 Pran Nath and Raza M. Syed
\vskip 0.5 true cm
\end{center}

\noindent
{ Department of Physics, Northeastern University,
Boston, MA 02115-5000, USA} \\
\vskip 1.0 true cm
\centerline{\bf Abstract}
\medskip
\noindent
An analysis of the couplings of the 210 dimensional SO(10) vector
multiplet to matter is given.
 Specifically we give an $SU(5)\times U(1)$ decomposition of
 the  vector couplings $\overline{16}_{\pm}-16_{\pm}-210$,
  where $16_{\pm}$
 is the semispinor of $SO(10)$ chirality ${\pm}$, using a recently
 derived basic theorem. The analysis is
 carried out using the Wess-Zumino gauge. However, we also
 consider the more general situation where all components of the
 vector multiplet enter in the couplings with the chiral fields.
 Here  elimination of the  auxiliary fields leads to a sigma
 model type nonlinear Lagrangian.
 Interactions of  the type analysed here may find
 applications in effective theories with the 210 vector arising
 as a condensate. The analysis presented  here completes the
 explicit computation of all lowest order couplings involving
the $16_{\pm}$ of spinors with Higgs and vectors multiplets using
the basic theorem.
\end{titlepage}

\section{Introduction}
  In the usual couplings of vector bosons, the  vector bosons belong to the adjoint representation
  of the gauge group (${\widehat V}_a$, a=1,..,N) and thus have one to
  one correspondence with the number of generators of the gauge
  group ($ T^a$, a=1,..,N). This allows one to form the Lie valued quantity
   ${\widehat V}={\widehat V}_aT^a$ which enters prominently in the construction of Yang-Mills
  gauge interactions which describe the self
   interactions of the gauge bosons. Further, one also utilizes
   the Lie valued quantities to couple the vector bosons to matter.
 In supersymmetric theories one essentially uses the same strategy
 in that one also uses Lie valued quantities and further one uses
 the Wess-Zumino gauge\cite{wz} in which the vector multiplet is reduced
 to just three components, ${\cal V}^{\mu}, \lambda, D$, where
 ${\cal V}^{\mu}$ is the spin 1 vector field, $\lambda$ is a spin $\frac{1}{2}$
  Majorana fields and $D$ is an auxiliary field.
  The question arises how one may construct the couplings of
  a vector multiplet which does not belong to the adjoint representation.
  We focus here on the group SO(10) which is one of the
  groups under considerable scrutiny as it is a possible grand
  unification group for the unification of the electroweak and of the
  strong interactions\cite{georgi}.
   Thus, for example, in SO(10) one has a $16$ dimensional spinor
  representation  which can accomodate a full one generation of quarks
  and leptons and its vector couplings  have the following decomposition
 \beq
 \overline{16}\times 16 =1+ 45 +210
 \eeq
  Thus while it is straightforward to couple 1 and 45 plet of vectors
  with the $ \overline{16}\times 16$ using the usual Yang-Mills construction,
  and for the case of supersymmetry using the supersymmetric Yang-Mills
  construction, the same procedure does not apply to the coupling of the
  vector 210 multiplet. Recently, we have given a complete computation
  of the couplings in the superpotential  which involve the 16 plet
  of matter\cite{ns1,ns2}. In Ref.\cite{ns1,ns2} a "basic theorem"
  using oscillator method\cite{sakita,wilczek} was developed which allowed
  one to carry out explicit analytic computations of the SO(10) couplings.
  Since $16\times 16=10+ 120+ \overline{126}$ we have
  given a complete determination of the couplings of matter -matter -Higgs
  couplings of the type $16-16-10$, $16-16-120$ and $16-16-\overline{126}$
  \cite{anderson,largereps}. The present analysis is motivated by similar considerations where
   we wish to give a complete analysis of the vector couplings of
   $\overline{16}\times 16$. While the vector couplings $\overline{16}-16-1$ and
   $\overline{16}-16-45$ are straightforwardly given by the standard
   analysis, this is not the case for the $\overline{16}-16-210$
   vector coupling. Here we need a new technique to  address this question.
   The purpose of this paper is to  do  just that. In this paper we
   consider the couplings of the supersymmetric vector 210 multiplet in SO(10).
   We focus on this construction both for the theoretical
   challenge of constructing such couplings as well as for the
   possibility that such interactions may surface in some future effective
   theories to describe fully all the degrees of freedom at some
   relevant energy scale.

The outline of the rest of the paper is as  follows: In Sec.2 we
follow the conventional approach and give the coupling of the
$210$ multiplet with $16$ plet of matter, i.e., we compute the
couplings $\overline{16}_{\pm}-16_{\pm}-210$ in the Wess-Zumino
gauge and we carry out a full $SU(5)\times U(1)$ decomposition of
it. Elimination of the auxiliary fields is carried out in Appendix
C. At the very outset we discard the constraint of gauge
invariance since the imposition of such a constraint is untenable
for the 210 multiplet. In Sec.3 we consider the more general
couplings of the $210$ multiplet retaining all the components of
the vector multiplet, i.e, we do not impose the Wess-Zumino gauge
constraint\cite{wz}. In the construction we use the superfield
   formalism\cite{Salam:1974yz} to guarantee
   that we have explicit supersymmetry at all stages in the theory.
In this case elimination of the auxiliary fields leads to a
non-linear Lagrangian with infinite order of nonlinearities in it.
The general technique underlying this procedure is illustrated in
Appendix G for the U(1) case. This analysis has some
   resemblance to the analysis of Ref.\cite{fayet} which also used a
   unconstrained vector multiplet, i.e., it did not impose the constraint
   of the
   Wess-Zumino gauge\cite{wz}. However, the analysis of Ref.\cite{fayet}
   did not include an explicit  mass term for the vector multiplet,
   nor the self interactions of the vector fields and it did not
   integrate the auxiliarly fields. In fact the motivation of the work of
   Ref.\cite{fayet} was very different in that the analysis of
   Ref.\cite{fayet} was geared to study
   spontaneous symmetry breaking and generation of vector boson
   masses in that context.
Returning to the $210$ multiplet we note that since the $210$ vector
multiplet interaction cannot be gauge invariant, one must view its
interactions only as effective interactions and thus the
appearance of sigma model type nonlinearities here are quite
acceptable.  In Sec.4 we give the conclusions. Appendix A is
devoted to notation and definition of the components of the vector
and chiral superfields. Normalization of the dynamical modes are
given in Appendix B. An elimination of the auxiliary fields appearing
in Sec.2 is given Appendix C and an
illustration of the SO(10) couplings of
Sec.3 is given in Appendix D. For completeness an $SU(5)\times U(1)$
decomposition of the singlet vector couplings and of the $45$ vector
couplings are given in Appendices E and F.

\section{Coupling of $\bf{210}$ vector multiplet to $\bf{16_{\pm}}$
plet of matter}
In usual formulations of particle interactions the vectors belong
to either singlets or the adjoint representations of the gauge
group of the theory under consideration. To couple the vectors to
 matter one forms a Lie valued quantity ${\widehat V}^aT_a$
where $T_a$ are the generators of the gauge group satisfying the
algebra $[T_a,T_b]=if_{abc} T_c$.
Then one couples the Lie valued quantity to the matter fields in
the form $\Phi^{\dagger}e^{g{\widehat V}}\Phi$ which can be shown
to be a gauge invariant combination. For the representations $1$
and $45$ on the right hand side of Eq.(1) one can carry out this
construction straightforwardly (see Appendices E and F).
However, this construction does not work for the $210$ vector
multiplet as  one cannot write a gauge invariant
Yang-Mills theory  for  it.
 Further, for the same reason one cannot write a
gauge invariant coupling of the $210$ vector  with
matter. To construct the 210 vector couplings, the technique
we adopt is to carry out a direct expansion in
powers of the vector supersuperfield. Thus we have

\begin{eqnarray}
{\mathsf L}_{V+\Phi}^{^{(210~
Interaction)}}=h_{ab}^{^{(210+)}}[<\widehat{\Phi}_{(+)a}|
\widehat{\Phi}_{(+)b}>+\frac{{\mathsf
g}^{^{(210)}}}{4!}<\widehat{\Phi}_{(+)a}|\widehat{\mathsf
V}_{\mu\nu\rho\lambda}\Gamma_{[\mu}\Gamma_{\nu}\Gamma_{\rho}
\Gamma_{\lambda]}|\widehat{\Phi}_{(+)b}>\nonumber\\
+\frac{1}{2}\left(\frac{{\mathsf
g}^{^{(210)}}}{4!}\right)^2<\widehat{\Phi}_{(+)a}|\widehat{\mathsf
V}_{\mu\nu\rho\lambda}\Gamma_{[\mu}\Gamma_{\nu}\Gamma_{\rho}
\Gamma_{\lambda]}\widehat{\mathsf
V}_{\alpha\beta\gamma\delta}\Gamma_{[\alpha}\Gamma_{\beta}\Gamma_{\gamma}
\Gamma_{\delta]}|\widehat{\Phi}_{(+)b}>]|_{\theta^2\bar
{\theta}^2}\nonumber\\
 +h_{ab}^{^{(210-)}}[<\widehat{\Phi}_{(-)a}|
\widehat{\Phi}_{(-)b}>+\frac{{\mathsf
g}^{^{(210)}}}{4!}<\widehat{\Phi}_{(-)a}|\widehat{\mathsf
V}_{\mu\nu\rho\lambda}\Gamma_{[\mu}\Gamma_{\nu}\Gamma_{\rho}
\Gamma_{\lambda]}|\widehat{\Phi}_{(-)b}>
\nonumber\\
+\frac{1}{2}\left(\frac{{\mathsf
g}^{^{(210)}}}{4!}\right)^2<\widehat{\Phi}_{(-)a}|\widehat{\mathsf
V}_{\mu\nu\rho\lambda}\Gamma_{[\mu}\Gamma_{\nu}\Gamma_{\rho}
\Gamma_{\lambda]}\widehat{\mathsf
V}_{\alpha\beta\gamma\delta}\Gamma_{[\alpha}\Gamma_{\beta}\Gamma_{\gamma}
\Gamma_{\delta]}|\widehat{\Phi}_{(-)b}>]|_{\theta^2\bar
{\theta}^2} +..\nonumber\\
\Gamma_{[\mu}\Gamma_{\nu}\Gamma_{\rho}
\Gamma_{\lambda]}=\frac{1}{4!}\sum_P(-1)^{\delta_P}
\Gamma_{\mu_{P(1)}}\Gamma_{\nu_{P(2)}}\Gamma_{\rho_{P(3)}}
\Gamma_{\lambda_{P(4)}}
\end{eqnarray}
where $\sum_P$ denoting the sum over all permutations and
$\delta_P$ takes on the value $0$ ($1$) for even (odd)
permutations. $\widehat{\mathsf V}_{\mu\nu\rho\lambda}$ ($\mu$,
$\nu$, $\lambda$, $\rho$=1,2,..,10) is the vector superfield,
$\widehat{\Phi}_{(+)a}$ is the $16_+$ chiral superfield (a is the
generation index) and $\Gamma_{\mu}$ satisfies a rank-10 Clifford
algebra, $[\Gamma_{\mu},\Gamma_{\nu}]$= $2\delta_{\mu\nu}$.
In the following we give a full exhibition of the couplings to
only linear order in the vector superfield in terms of its
$SU(5)\times U(1)$ decompostion but a similar analysis can be
done for couplings involving higher powers of the superfield.
 Thus using the analysis of Ref.\cite{ns1,ns2} we find that
 the $\overline{16}_+16_+$ couplings can be decomposed as follows

\begin{eqnarray}
h_{ab}^{^{(210+)}}<\widehat{\Phi}_{(+)a}|\widehat{\Phi}_{(+)b}>|_{\theta^2\bar
{\theta}^2}=h_{ab}^{^{(210+)}}[-\partial_A{\bf
A}^{\dagger}_{(+)a}\partial^A{\bf A}_{(+)b}
-\partial_A{\bf A}^{i\dagger}_{(+)a}\partial^A{\bf A}_{(+)bi}\nonumber\\
-\partial_A{\bf A}^{\dagger}_{(+)aij}\partial^A{\bf
A}_{(+)b}^{ij}-i\overline{\bf
{\Psi}}_{(+)aL}\gamma^A\partial_A{\bf
{\Psi}}_{(+)bL}\nonumber\\
-i\overline{\bf {\Psi}}_{(+)aL}^i\gamma^A\partial_A{\bf
{\Psi}}_{(+)biL}-i\overline{\bf
{\Psi}}_{(+)aijL}\gamma^A\partial_A{\bf {\Psi}}_{(+)bL}^{ij}]
+{\mathsf L}^{(210)}_{(1)auxiliary}\nonumber\\
{\mathsf
L}^{(210)}_{(1)auxiliary}=h_{ab}^{^{(210+)}}<F_{(+)a}|F_{(+)b}>
\end{eqnarray}
\begin{eqnarray}
h_{ab}^{^{(210-)}}<\widehat{\Phi}_{(-)a}|
\widehat{\Phi}_{(-)b}>|_{\theta^2\bar
{\theta}^2}=h_{ab}^{^{(210-)}}[-\partial_A{\bf
A}^{\dagger}_{(-)a}\partial^A{\bf A}_{(-)b}
-\partial_A{\bf A}^{\dagger}_{(-)ai}\partial^A{\bf A}_{(-)b}^i\nonumber\\
-\partial_A{\bf A}^{ij\dagger}_{(-)a}\partial^A{\bf
A}_{(-)bij}-i\overline{\bf {\Psi}}_{(-)aL}\gamma^A\partial_A{\bf
{\Psi}}_{(-)bL}\nonumber\\
-i\overline{\bf {\Psi}}_{(-)aiL}\gamma^A\partial_A{\bf
{\Psi}}_{(-)bL}^i-i\overline{\bf
{\Psi}}_{(-)aL}^{ij}\gamma^A\partial_A{\bf {\Psi}}_{(-)bijL}]
+{\mathsf L}^{(210)}_{(2)auxiliary}\nonumber\\
{\mathsf
L}^{(210)}_{(2)auxiliary}=h_{ab}^{^{(210-)}}<F_{(-)a}|F_{(-)b}>
\end{eqnarray}
\begin{equation}
{\bf {\Psi}}_{(\pm)a}=\left(\matrix{{\bf
\psi}_{(\pm)a\tilde{\alpha}}\cr
\overline{{\bf\psi}}^{\dot{\tilde\alpha}}_{(\pm)a}}\right)
\end{equation}
In the above the upper case Latin letters ($A$, $B$, $C$, $D$)
are the
Lorentz indices while the Greek letters with
tilde's($\tilde{\alpha}$, $\dot{\tilde {\beta}}$, ...)are Weyl
indices. Similarly we find
\begin{eqnarray}
\frac{h_{ab}^{^{(210+)}}{\mathsf
g}^{^{(210)}}}{4!}<\widehat{\Phi}_{(+)a}|\widehat{\mathsf
V}_{\mu\nu\rho\lambda}\Gamma_{[\mu}\Gamma_{\nu}\Gamma_{\rho}
\Gamma_{\lambda]}|\widehat{\Phi}_{(+)b}>|_{\theta^2\bar
{\theta}^2}~~~~~\nonumber\\
=h_{ab}^{^{(210+)}}{\mathsf
g}^{^{(210)}}\{[\frac{1}{2}\sqrt{\frac{5}{6}}\left(i{\bf
A}^{\dagger}_{(+)a}\stackrel{\leftrightarrow}{\partial}_A {\bf
A}_{(+)b}-\overline{\bf {\Psi}}_{(+)aL}\gamma_A{\bf
{\Psi}}_{(+)bL}\right)\nonumber\\
-\frac{1}{4\sqrt {30}}\left(i{\bf
A}^{\dagger}_{(+)aij}\stackrel{\leftrightarrow}{\partial}_A{\bf
A}_{(+)b}^{ij}-\overline{\bf {\Psi}}_{(+)aijL}\gamma_A{\bf
{\Psi}}_{(+)bL}^{ij}\right)\nonumber\\
+\frac{1}{2\sqrt {30}}\left(i{\bf
A}^{i\dagger}_{(+)a}\stackrel{\leftrightarrow}{\partial}_A{\bf
A}_{(+)bi}-\overline{\bf {\Psi}}_{(+)aL}^i\gamma_A{\bf
{\Psi}}_{(+)biL}\right)]{\cal V}^{'A}\nonumber\\
+[\frac{1}{\sqrt {6}}\left(i{\bf
A}^{i\dagger}_{(+)a}\stackrel{\leftrightarrow}{\partial}_A{\bf
A}_{(+)b}-\overline{\bf {\Psi}}_{(+)aL}^i\gamma_A{\bf
{\Psi}}_{(+)bL}\right)]{\cal V}^{'A}_i\nonumber\\
+[\frac{1}{\sqrt {6}}\left(i{\bf
A}^{\dagger}_{(+)a}\stackrel{\leftrightarrow}{\partial}_A{\bf
A}_{(+)bi}-\overline{\bf {\Psi}}_{(+)aL}\gamma_A{\bf
{\Psi}}_{(+)biL}\right)]{\cal V}^{'Ai}\nonumber\\
+[-\frac{1}{4\sqrt 2}\left(i{\bf
A}^{\dagger}_{(+)alm}\stackrel{\leftrightarrow}{\partial}_A{\bf
A}_{(+)b}-\overline{\bf {\Psi}}_{(+)almL}\gamma_A{\bf
{\Psi}}_{(+)bL}\right)\nonumber\\
+\frac{1}{24\sqrt 2}\epsilon_{ijklm}\left(i{\bf
A}^{i\dagger}_{(+)a}\stackrel{\leftrightarrow}{\partial}_A{\bf
A}_{(+)b}^{jk}-\overline{\bf {\Psi}}_{(+)aL}^i\gamma_A{\bf
{\Psi}}_{(+)bL}^{jk}\right)]{\cal V}^{'Alm}\nonumber\\
+[-\frac{1}{4\sqrt 2}\left(i{\bf
A}^{\dagger}_{(+)a}\stackrel{\leftrightarrow}{\partial}_A{\bf
A}_{(+)b}^{lm}-\overline{\bf {\Psi}}_{(+)aL}\gamma_A{\bf
{\Psi}}_{(+)bL}^{lm}\right)\nonumber\\
+\frac{1}{24\sqrt 2}\epsilon^{ijklm}\left(i{\bf
A}^{\dagger}_{(+)aij}\stackrel{\leftrightarrow}{\partial}_A{\bf
A}_{(+)bk}-\overline{\bf {\Psi}}_{(+)aijL}\gamma_A{\bf
{\Psi}}_{(+)bkL}\right)]{\cal V}^{'A}_{lm}\nonumber\\
 +[\frac{1}{6\sqrt
2}\left(i{\bf
A}^{\dagger}_{(+)aik}\stackrel{\leftrightarrow}{\partial}_A{\bf
A}_{(+)b}^{kj}-\overline{\bf {\Psi}}_{(+)aikL}\gamma_A{\bf
{\Psi}}_{(+)bL}^{kj}\right)\nonumber\\
-\frac{1}{2\sqrt 2}\left({\bf
A}^{j\dagger}_{(+)a}\stackrel{\leftrightarrow}{\partial}_A{\bf
A}_{(+)bi}+i\overline{\bf {\Psi}}_{(+)aL}^j\gamma_A{\bf
{\Psi}}_{(+)biL}\right)]{\cal V}^{'iA}_j\nonumber\\
+[\frac{1}{6\sqrt {6}}\epsilon_{ijklm}\left(i{\bf
A}^{i\dagger}_{(+)a}\stackrel{\leftrightarrow}{\partial}_A{\bf
A}_{(+)b}^{jn}-\overline{\bf {\Psi}}_{(+)aL}^i\gamma_A{\bf
{\Psi}}_{(+)bL}^{jn}\right)]{\cal V}^{'Aklm}_n\nonumber\\
+[-\frac{1}{6\sqrt {6}}\epsilon^{ijklm}\left(i{\bf
A}^{\dagger}_{(+)ain}\stackrel{\leftrightarrow}{\partial}_A{\bf
A}_{(+)bj}-\overline{\bf {\Psi}}_{(+)ainL}\gamma_A{\bf
{\Psi}}_{(+)bjL}\right)]{\cal V}^{'An}_{klm}\nonumber\\
+[-\frac{1}{4\sqrt {6}}\left(i{\bf
A}^{\dagger}_{(+)aij}\stackrel{\leftrightarrow}{\partial}_A{\bf
A}_{(+)b}^{kl}-\overline{\bf {\Psi}}_{(+)aijL}\gamma_A{\bf
{\Psi}}_{(+)bL}^{kl}\right)]{\cal V}^{'Aij}_{kl}\nonumber\\
-\frac{i}{2}\sqrt{\frac{5}{6}}\left[-{\bf
A}^{\dagger}_{(+)a}\overline{\bf {\Psi}}_{(+)bR}+\frac{1}{10}{\bf
A}^{\dagger}_{(+)aij}\overline{\bf
{\Psi}}_{(+)bR}^{ij}-\frac{1}{5}{\bf
A}^{i\dagger}_{(+)a}\overline{\bf
{\Psi}}_{(+)biR}\right]{\Lambda}^{'}_{L}\nonumber\\
+\frac{i}{\sqrt {3}}\left[ {\bf A}^{i\dagger}_{(+)a}\overline{\bf
{\Psi}}_{(+)bR}\right]{\Lambda}^{'}_{iL}+\frac{i}{\sqrt {3}}\left[
{\bf A}^{\dagger}_{(+)a}\overline{\bf
{\Psi}}_{(+)biR}\right]{\Lambda}^{'i}_{L}\nonumber\\
-\frac{i}{4}\left[{\bf A}^{\dagger}_{(+)alm}\overline{\bf
{\Psi}}_{(+)bR}-\frac{1}{6}\epsilon_{ijklm}{\bf
A}^{i\dagger}_{(+)a}\overline{\bf
{\Psi}}_{(+)bR}^{jk}\right]{\Lambda}^{'lm}_{L}\nonumber\\
-\frac{i}{4}\left[{\bf A}^{\dagger}_{(+)a}\overline{\bf
{\Psi}}_{(+)bR}^{lm}-\frac{1}{6}\epsilon^{ijklm}{\bf
A}^{\dagger}_{(+)aij}\overline{\bf
{\Psi}}_{(+)bkR}\right]{\Lambda}^{'}_{lmL}\nonumber\\
-\frac{i}{2}\left[-\frac{1}{3}{\bf
A}^{\dagger}_{(+)aik}\overline{\bf {\Psi}}_{(+)bR}^{kj}+{\bf
A}^{j\dagger}_{(+)a}\overline{\bf
{\Psi}}_{(+)biR}\right]{\Lambda}^{'i}_{jL}\nonumber\\
+\frac{i}{6\sqrt {3}}\epsilon_{ijklm}\left[ {\bf
A}^{i\dagger}_{(+)a}\overline{\bf
{\Psi}}_{(+)bR}^{jn}\right]{\Lambda}^{'klm}_{nL}-\frac{i}{6\sqrt
{3}}\epsilon^{ijklm}\left[ {\bf A}^{\dagger}_{(+)ain}\overline{\bf
{\Psi}}_{(+)bjR}\right]{\Lambda}^{'n}_{klmL}\nonumber\\
-\frac{i}{4\sqrt {3}}\left[ {\bf
A}^{\dagger}_{(+)aij}\overline{\bf
{\Psi}}_{(+)bR}^{kl}\right]{\Lambda}^{'ij}_{klL}\nonumber\\
 -\frac{i}{2}\sqrt{\frac{
5}{6}}\left[-\overline{\bf {\Psi}}_{(+)aL}{\bf
A}_{(+)b}+\frac{1}{10}\overline{\bf {\Psi}}_{(+)aijL}{\bf
A}^{ij}_{(+)b}-\frac{1}{5}\overline{\bf {\Psi}}_{(+)aL}^i{\bf
A}_{(+)bi}\right]{\Lambda}^{'}_{R}\nonumber\\
-\frac{i}{\sqrt {3}}\left[ \overline{\bf {\Psi}}_{(+)aL}^{i}{\bf
A}_{(+)b}\right]{\Lambda}^{'}_{iR}-\frac{i}{\sqrt {3}}\left[
{\overline{\bf
{\Psi}}_{(+)aL}\bf A}_{(+)bi}\right]{\Lambda}^{'i}_{R}\nonumber\\
+\frac{i}{4}\left[\overline{\bf {\Psi}}_{(+)almL}{\bf
A}_{(+)b}-\frac{1}{6}\epsilon_{ijklm}\overline{\bf
{\Psi}}_{(+)aL}^i{\bf A}^{jk}_{(+)b}\right]{\Lambda}^{'lm}_{R}\nonumber\\
+\frac{i}{4}\left[\overline{\bf {\Psi}}_{(+)aL}{\bf
A}_{(+)b}^{lm}-\frac{1}{6}\epsilon^{ijklm}\overline{\bf
{\Psi}}_{(+)aijL}{\bf A}_{(+)bk}\right]{\Lambda}^{'}_{lmR}\nonumber\\
+\frac{i}{2}\left[-\frac{1}{3}\overline{\bf {\Psi}}_{(+)aikL}{\bf
A}_{(+)b}^{kj}+\overline{\bf {\Psi}}_{(+)aL}^j{\bf
A}_{(+)bi}\right]{\Lambda}^{'i}_{jR}\}\nonumber\\
-\frac{i}{6\sqrt {3}}\epsilon_{ijklm}\left[ \overline{\bf
{\Psi}}_{(+)aL}^{i}{\bf
A}^{jn}_{(+)b}\right]{\Lambda}^{'klm}_{nR}+\frac{i}{6\sqrt
{3}}\epsilon^{ijklm}\left[ {\overline{\bf
{\Psi}}_{(+)ainL}\bf A}_{(+)bj}\right]{\Lambda}^{'n}_{klmR}\nonumber\\
+\frac{i}{4\sqrt {3}}\left[ \overline{\bf {\Psi}}_{(+)aijL}{\bf
A}^{kl}_{(+)b}\right]{\Lambda}^{'ij}_{klR}\}+{\mathsf L}^{(210)}_{(3)auxiliary}\nonumber\\
{\mathsf
L}^{(210)}_{(3)auxiliary}=\frac{h_{ab}^{^{(210+)}}{\mathsf
g}^{^{(210)}}}{4!2}<A_{(+)a}|\Gamma_{[\mu}\Gamma_{\nu}\Gamma_{\rho}
\Gamma_{\lambda]}|A_{(+)b}>D_{\mu\nu\rho\lambda}
\end{eqnarray}
where
\begin{eqnarray}
{\bf
A}^{i\dagger}_{(+)a}\stackrel{\leftrightarrow}{\partial}_A{\bf
A}_{(+)bi}\stackrel{def}={\bf A}^{i\dagger}_{(+)a}{\partial}_A{\bf
A}_{(+)bi}-\left({\partial}_A{\bf A}^{i\dagger}_{(+)a}\right) {\bf
A}_{(+)bi}\nonumber\\
\Lambda_{R,L}=\frac{1\pm \gamma_5}{2}\Lambda\nonumber\\
\Lambda_{jkl}^i=\left(\matrix{\lambda_{\tilde{\alpha}jkl}^i\cr
\overline{\lambda}^{\dot{\tilde\alpha}i}_{jkl}}\right)
\end{eqnarray}
and so on. Similarly for the $\overline{16}_-16_-$ couplings
we find
\begin{eqnarray}
\frac{h_{ab}^{^{(210-)}}{\mathsf
g}^{^{(210)}}}{4!}<\widehat{\Phi}_{(-)a}|\widehat{\mathsf
V}_{\mu\nu\rho\lambda}\Gamma_{[\mu}\Gamma_{\nu}\Gamma_{\rho}
\Gamma_{\lambda]}|\widehat{\Phi}_{(-)b}>|_{\theta^2\bar
{\theta}^2}~~~~~\nonumber\\
 =h_{ab}^{^{(210-)}}{\mathsf
g}^{^{(210)}}\{[\frac{1}{2}\sqrt{\frac{5}{6}}\left(i{\bf
A}^{\dagger}_{(-)a}\stackrel{\leftrightarrow}{\partial}_A {\bf
A}_{(-)b}-\overline{\bf {\Psi}}_{(-)aL}\gamma_A{\bf
{\Psi}}_{(-)bL}\right)\nonumber\\
-\frac{1}{4\sqrt {30}}\left(i{\bf
A}^{ij\dagger}_{(-)a}\stackrel{\leftrightarrow}{\partial}_A{\bf
A}_{(-)bij}-\overline{\bf {\Psi}}_{(-)aL}^{ij}\gamma_A{\bf
{\Psi}}_{(-)bijL}\right)\nonumber\\
+\frac{1}{2\sqrt {30}}\left(i{\bf
A}^{\dagger}_{(-)ai}\stackrel{\leftrightarrow}{\partial}_A{\bf
A}_{(-)b}^{i}+i\overline{\bf {\Psi}}_{(-)aiL}\gamma_A{\bf
{\Psi}}_{(-)bL}^i\right)]{\cal V}^{'A}\nonumber\\
+[\frac{1}{\sqrt {6}}\left(i{\bf
A}^{\dagger}_{(-)a}\stackrel{\leftrightarrow}{\partial}_A{\bf
A}_{(-)b}^{i}-\overline{\bf {\Psi}}_{(-)aL}\gamma_A{\bf
{\Psi}}_{(-)bL}^i\right)]{\cal V}^{'A}_i\nonumber\\
+[\frac{1}{\sqrt {6}}\left(i{\bf
A}^{\dagger}_{(-)ai}\stackrel{\leftrightarrow}{\partial}_A{\bf
A}_{(-)b}-\overline{\bf {\Psi}}_{(-)aiL}\gamma_A{\bf
{\Psi}}_{(-)bL}\right)]{\cal V}^{'Ai}\nonumber\\
+[\frac{1}{4\sqrt 2}\left({\bf
A}^{\dagger}_{(-)a}\stackrel{\leftrightarrow}{\partial}_A{\bf
A}_{(-)blm}+i\overline{\bf {\Psi}}_{(-)aL}\gamma_A{\bf
{\Psi}}_{(-)blmL}\right)\nonumber\\
-\frac{1}{24\sqrt 2}\epsilon_{ijklm}\left(i{\bf
A}^{ij\dagger}_{(-)a}\stackrel{\leftrightarrow}{\partial}_A{\bf
A}_{(-)b}^{k}-\overline{\bf {\Psi}}_{(-)aL}^{ij}\gamma_A{\bf
{\Psi}}_{(-)bL}^{k}\right)]{\cal V}^{'Alm}\nonumber\\
+[\frac{1}{4\sqrt 2}\left(i{\bf
A}^{lm\dagger}_{(-)a}\stackrel{\leftrightarrow}{\partial}_A{\bf
A}_{(-)b}-\overline{\bf {\Psi}}_{(-)aL}^{lm}\gamma_A{\bf
{\Psi}}_{(-)bL}\right)\nonumber\\
-\frac{1}{24\sqrt 2}\epsilon^{ijklm}\left(i{\bf
A}^{\dagger}_{(-)ai}\stackrel{\leftrightarrow}{\partial}_A{\bf
A}_{(-)bjk}-\overline{\bf {\Psi}}_{(-)aiL}\gamma_A{\bf
{\Psi}}_{(-)bjkL}\right)]{\cal V}^{'A}_{lm}\nonumber\\
+[\frac{1}{6\sqrt 2}\left(i{\bf
A}^{jk\dagger}_{(-)a}\stackrel{\leftrightarrow}{\partial}_A{\bf
A}_{(-)bki}-\overline{\bf {\Psi}}_{(-)aL}^{jk}\gamma_A{\bf
{\Psi}}_{(-)bkiL}\right)\nonumber\\
-\frac{1}{2\sqrt 2}\left(i{\bf
A}^{\dagger}_{(-)ai}\stackrel{\leftrightarrow}{\partial}_A{\bf
A}_{(-)b}^{j}-\overline{\bf {\Psi}}_{(-)aiL}\gamma_A{\bf
{\Psi}}_{(-)bL}^j\right)]{\cal V}^{'iA}_j\nonumber\\
+[\frac{1}{12\sqrt 6}\epsilon_{ijklm}\left(i{\bf
A}^{ij\dagger}_{(-)a}\stackrel{\leftrightarrow}{\partial}_A{\bf
A}_{(-)b}^n-\overline{\bf {\Psi}}_{(-)aL}^{ij}\gamma_A{\bf
{\Psi}}_{(-)bL}^n\right)]{\cal V}^{'Aklm}_{n}\nonumber\\
+[\frac{1}{12\sqrt 6}\epsilon^{ijklm}\left(i{\bf
A}^{\dagger}_{(-)an}\stackrel{\leftrightarrow}{\partial}_A{\bf
A}_{(-)bij}-\overline{\bf {\Psi}}_{(-)anL}\gamma_A{\bf
+{\Psi}}_{(-)bijL}\right)]{\cal V}^{'An}_{klm}\nonumber\\
+[-\frac{1}{4\sqrt 6}\epsilon_{ijklm}\left(i{\bf
A}^{kl\dagger}_{(-)a}\stackrel{\leftrightarrow}{\partial}_A{\bf
A}_{(-)bij}-\overline{\bf {\Psi}}_{(-)aL}^{kl}\gamma_A{\bf
{\Psi}}_{(-)bijL}\right)]{\cal V}^{'Aij}_{kl}\nonumber\\
-\frac{i}{2}\sqrt{\frac{5}{6}}\left[-{\bf
A}^{\dagger}_{(-)a}\overline{\bf {\Psi}}_{(-)bR}+\frac{1}{10}{\bf
A}^{ij\dagger}_{(-)a}\overline{\bf
{\Psi}}_{(-)bijR}-\frac{1}{5}{\bf
A}^{\dagger}_{(-)ai}\overline{\bf
{\Psi}}_{(-)bR}^i\right]{\Lambda}^{'}_{L}\nonumber\\
+\frac{i}{\sqrt 3}\left[{\bf A}^{\dagger}_{(-)a}\overline{\bf
{\Psi}}_{(-)bR}^i\right]{\Lambda}^{'}_{iL}+\frac{i}{\sqrt
3}\left[{\bf A}^{\dagger}_{(-)ai}\overline{\bf
{\Psi}}_{(-)bR}\right]{\Lambda}^{'i}_{L}\nonumber\\
+\frac{i}{4}\left[{\bf A}^{\dagger}_{(-)a}\overline{\bf
{\Psi}}_{(-)blmR}-\frac{1}{6}\epsilon_{ijklm}{\bf
A}^{ij\dagger}_{(-)a}\overline{\bf
{\Psi}}_{(-)bR}^{k}\right]{\Lambda}^{'lm}_{L}\nonumber\\
+\frac{i}{4}\left[{\bf A}^{lm\dagger}_{(-)a}\overline{\bf
{\Psi}}_{(-)bR}-\frac{1}{6}\epsilon^{ijklm}{\bf
A}^{\dagger}_{(-)ai}\overline{\bf
{\Psi}}_{(-)bjkR}\right]{\Lambda}^{'}_{lmL}\nonumber\\
-\frac{i}{2}\left[-\frac{1}{3}{\bf
A}^{jk\dagger}_{(-)a}\overline{\bf {\Psi}}_{(-)bkiR}+{\bf
A}^{\dagger}_{(-)ai}\overline{\bf
{\Psi}}_{(-)bR}^j\right]{\Lambda}^{'i}_{jL}\nonumber\\
+\frac{i}{6\sqrt 3}\epsilon_{ijklm}\left[{\bf
A}^{ij\dagger}_{(-)a}\overline{\bf
{\Psi}}_{(-)bR}^{n}\right]{\Lambda}^{'klm}_{nL} +\frac{i}{6\sqrt
3}\epsilon^{ijklm}\left[{\bf A}^{\dagger}_{(-)an}\overline{\bf
{\Psi}}_{(-)bijR}\right]{\Lambda}^{'n}_{klmL}\nonumber\\
-\frac{i}{4\sqrt 3}\left[{\bf A}^{kl\dagger}_{(-)a}\overline{\bf
{\Psi}}_{(-)bijR}\right]{\Lambda}^{'ij}_{klL}\nonumber\\
+\frac{i}{2}\sqrt{\frac{5}{6}}\left[-\overline{\bf
{\Psi}}_{(-)aL}{\bf A}_{(-)b}+\frac{1}{10}\overline{\bf
{\Psi}}_{(-)aL}^{ij}{\bf A}_{(-)bij}-\frac{1}{5}\overline{\bf
{\Psi}}_{(-)aiL}{\bf
A}_{(-)b}^i\right]{\Lambda}^{'}_{R}\nonumber\\
-\frac{i}{\sqrt 3}\left[\overline{\bf {\Psi}}_{(-)aL}{\bf
A}_{(-)b}^i\right]{\Lambda}^{'}_{iR} -\frac{i}{\sqrt
3}\left[\overline{\bf {\Psi}}_{(-)aiL}{\bf
A}_{(-)b}\right]{\Lambda}^{'i}_{R}\nonumber\\
-\frac{i}{4}\left[\overline{\bf {\Psi}}_{(-)aL}{\bf
A}_{(-)blm}-\frac{1}{6}\epsilon_{ijklm}\overline{\bf
{\Psi}}_{(-)aL}^{ij}{\bf A}^{k}_{(-)b}\right]{\Lambda}^{'lm}_{R}\nonumber\\
-\frac{i}{4}\left[\overline{\bf {\Psi}}_{(-)aL}^{lm}{\bf
A}_{(-)b}-\frac{1}{6}\epsilon^{ijklm}\overline{\bf
{\Psi}}_{(-)aiL}{\bf A}_{(-)bjk}\right]{\Lambda}^{'}_{lmR}\nonumber\\
+\frac{i}{2}\left[-\frac{1}{3}\overline{\bf
{\Psi}}_{(-)aL}^{jk}{\bf A}_{(-)bki}+\overline{\bf
{\Psi}}_{(-)aiL}{\bf A}_{(-)b}^j\right]{\Lambda}^{'i}_{jR}\nonumber\\
-\frac{i}{6\sqrt{3}}\epsilon_{ijklm}\left[\overline{\bf
{\Psi}}_{(-)aL}^{ij}{\bf A}^{n}_{(-)b}\right]{\Lambda}^{'klm}_{nR}
-\frac{i}{6\sqrt {3}}\epsilon^{ijklm}\left[\overline{\bf
{\Psi}}_{(-)anL}{\bf A}_{(-)bij}\right]{\Lambda}^{'n}_{klmR}\nonumber\\
+\frac{i}{4\sqrt {3}}\left[\overline{\bf{\Psi}}_{(-)aL}^{kl}{\bf
A}_{(-)bij}\right]{\Lambda}^{'ij}_{klR}\} +{\mathsf
L}^{(210)}_{(4)auxiliary}\nonumber\\
{\mathsf
L}^{(210)}_{(4)auxiliary}=\frac{h_{ab}^{^{(210-)}}{\mathsf
g}^{^{(210)}}}{4!2
}<A_{(-)a}|\Gamma_{[\mu}\Gamma_{\nu}\Gamma_{\rho}
\Gamma_{\lambda]}|A_{(-)b}>D_{\mu\nu\rho\lambda}
\end{eqnarray}
Further, the kinetic energy for the vector multiplet is given by
\begin{eqnarray}
 {\mathsf
L}_{V}^{^{(210~K.E.)}}=\frac{1}{64}\left[\widehat{\cal
W}^{\tilde{\alpha}}_{\mu\nu\rho\lambda} \widehat{\cal
W}_{\tilde{\alpha}\mu\nu\rho\lambda}|_{\theta^2}+\widehat{\overline{\cal
W}}_{\dot{\tilde{\alpha}}\mu\nu\rho\lambda}\widehat{\overline{\cal
W}}^{\dot{\tilde{\alpha}}}_{\mu\nu\rho\lambda}|_{\bar
{\theta}^2}\right]\nonumber\\
\widehat{\cal W}^{\tilde{\alpha}}_{\mu\nu\rho\lambda} =\overline
{\mathsf D}^2\mathsf D_{\tilde{\alpha}}\widehat{\mathsf
V}_{\mu\nu\rho\lambda},~~~ \widehat{\overline{\cal
W}}_{\dot{\tilde{\alpha}}\mu\nu\rho\lambda}={\mathsf D}^2\overline
{\mathsf D}_{\dot{\tilde{\alpha}}}\widehat{\mathsf
V}_{\mu\nu\rho\lambda}\nonumber\\
 {\mathsf
D}_{\tilde{\alpha}}=\frac{\partial}{\partial\theta^{\tilde{\alpha}}}+
i\sigma_{\tilde{\alpha}{\dot{\tilde {\alpha}}}}^A\overline
{\theta}^{\dot{\tilde {\alpha}}}\partial_A;~~~~ \overline {\mathsf
D}_{\dot{\tilde{\alpha}}}=-\frac{\partial}{\partial\overline
{\theta}^{\dot{\tilde {\alpha}}}}-i\theta ^{\tilde{\alpha}}
\sigma_{\tilde{\alpha}\dot{\tilde{\alpha}}}
\end{eqnarray}
Explicit evaluation of Eq.(1) gives
\begin{eqnarray}
{\mathsf L}_{V}^{^{(210~K.E.)}}=-\frac{1}{4}{\cal
V}_{AB\mu\nu\rho\sigma}{\cal
V}^{AB}_{\mu\nu\rho\sigma}-\frac{i}{2}\overline
{\Lambda}_{\mu\nu\rho\sigma}\gamma^A\partial_A\Lambda_{\mu\nu\rho\sigma}+{\mathsf L}^{(210)}_{(5)auxiliary}\nonumber\\
{\cal V}^{AB}_{\mu\nu\rho\sigma}=\partial^A{\cal
V}^B_{\mu\nu\rho\sigma}-\partial^B{\cal
V}^A_{\mu\nu\rho\sigma}\nonumber\\
{\mathsf L}^{(210)}_{(5)auxiliary}=\frac{1}{2}D_{\mu\nu\rho\sigma}D_{\mu\nu\rho\sigma}\nonumber\\
\Lambda_{\mu\nu\rho\sigma}=\left(\matrix{\lambda_{\tilde{\alpha}_{\mu\nu\rho\sigma}}\cr
\overline{\lambda}^{\dot{\tilde\alpha}}_{\mu\nu\rho\sigma}}\right)
\end{eqnarray}
Finally, the superpotential of the theory is taken to be
\begin{eqnarray}
{\mathsf L}^{(210)}_{\mathsf W}={\mathsf
W}(\widehat{\Phi}_{(+)},\widehat{\Phi}_{(-)})|_{\theta^2}+
h.c.\nonumber\\
{\mathsf W}(\widehat{\Phi}_{(+)},\widehat{\Phi}_{(-)})=
\mu_{ab}<\widehat{\Phi}_{(-)a}^*|B|\widehat{\Phi}_{(+)b}>\nonumber\\
{\mathsf W}({\bf A}_{(+)},{\bf A}_{(-)}) =i\mu_{ab}\left({\bf
A}_{(-)a}^{\bf{T}}{\bf A}_{(+)b}-\frac{1}{2}{\bf
A}_{(-)aij}^{\bf{T}}{\bf A}_{(+)b}^{ij}+{\bf
A}_{(-)a}^{i\bf{T}}{\bf A}_{(+)bi}\right)
\end{eqnarray}
where $\mu_{ab}$ is taken to be a symmetric tensor and $B$ is
the usual SO(10)charge conjugation operator . Thus we have
\begin{eqnarray}
{\mathsf L}_{\mathsf W}=-i\mu_{ab}\left( \overline{\bf
{\Psi}}_{(-)aR}{\bf {\Psi}}_{(+)bL}+\overline{\bf
{\Psi}}_{(-)aR}^i{\bf {\Psi}}_{(+)biL}-\frac{1}{2}\overline{\bf
{\Psi}}_{(-)aijR}{\bf
{\Psi}}_{(+)bL}^{ij}\right)\nonumber\\
+i\mu_{ab}^{*}\left( \overline{\bf {\Psi}}_{(-)aL}{\bf
{\Psi}}_{(+)bR}+\overline{\bf {\Psi}}_{(-)aiL}{\bf
{\Psi}}_{(+)bR}^i-\frac{1}{2}\overline{\bf
{\Psi}}_{(-)aL}^{ij}{\bf {\Psi}}_{(+)bijR}\right)
+{\mathsf L}^{(210)}_{(6)auxiliary}\nonumber\\
{\mathsf L}^{(210)}_{(6)auxiliary}=i\mu_{ab}[{\bf F}_{(-)a}{\bf
A}_{(+)b}+{\bf A}_{(-)a}^{\bf{T}}{\bf F}_{(+)b}-\frac{1}{2}{\bf
F}_{(-)aij}{\bf A}_{(+)b}^{ij}\nonumber\\
-\frac{1}{2}{\bf A}_{(-)aij}^{\bf{T}}{\bf F}_{(+)b}^{ij} + {\bf
F}_{(-)a}^i{\bf A}_{(+)bi}+{\bf A}_{(-)a}^{i\bf{T}}{\bf
F}_{(+)bi}] +h.c.
\end{eqnarray}
Elimination of the auxiliary fields is carried out in
appendix C.
\section{A more general analysis of  $\bf{16_{+}-\overline{16}_{+}-210}$
vector couplings} In this section we consider the couplings of the
unconstrained $210$ vector multiplet with matter, i.e., in the
analysis we use the full vector multiplet rather than the
truncated one under the constraint of the Wess-Zumino gauge.  An
illustration of this procedure is given in Appendix G for the
$U(1)$ case. The Lagrangian that governs the interactions of the
210 multiplet consists of the kinetic energy term for the 210
plet, self interactions, and interactions of the 210 plet with
$16$ and $\overline{16}$ of matter. For generality we also include
 a mass term for the 210 vector multiplet. As in Appendix G we will not
 impose the Wess-Zumino gauge but keep the full multiplet. Thus we take
 the Lagrangian governing the 210 vector multiplet to be
\begin{eqnarray}
{\mathsf L}^{(210)}={\mathsf L}_{V}^{'^{(210~K.E.)}}+{\mathsf
L}_{V}^{'^{(210~Mass)}}+ {\mathsf
L}_V^{'^{(210~Self-Interaction)}}
+{\mathsf L}_{V+\Phi}^{'^{(210~Interaction)}}+{\mathsf L}^{'^{(210~Self-Interaction)}}_{\Phi}\nonumber\\
{\mathsf L}_{V}^{'^{(210~K.E.)}}=\frac{1}{64}\left[\widehat{\cal
W}^{\tilde{\alpha}}_{\mu\nu\rho\lambda} \widehat{\cal
W}_{\tilde{\alpha}\mu\nu\rho\lambda}|_{\theta^2}+\widehat{\overline{\cal
W}}_{\dot{\tilde{\alpha}}\mu\nu\rho\lambda}\widehat{\overline{\cal
W}}^{\dot{\tilde{\alpha}}}_{\mu\nu\rho\lambda}|_{\bar
{\theta}^2}\right]\nonumber\\
{\mathsf L}_{V}^{'^{(210~Mass)}}=m^2\widehat{\mathsf
V}_{\mu\nu\rho\lambda}\widehat{\mathsf
V}_{\mu\nu\rho\lambda}|_{\theta^2\bar {\theta}^2}\nonumber\\
{\mathsf L}_V^{'^{(210~Self-Interaction)}}={\alpha}_1
\widehat{\mathsf V}_{\mu\nu\rho\lambda}\widehat{\mathsf
V}_{\rho\lambda\alpha\beta} \widehat{\mathsf
V}_{\alpha\beta\mu\nu} |_{\theta^2\bar{\theta}^2}
+{\alpha}_2\widehat{\mathsf V}_{\mu\nu\rho\lambda}
\widehat{\mathsf V}_{\rho\lambda\alpha\beta} \widehat{\mathsf
V}_{\alpha\beta\eta\tau}\widehat{\mathsf V}_{\eta\tau\mu\nu}
|_{\theta^2\bar{\theta}^2}\nonumber\\
{\mathsf
L}_{V+\Phi}^{'^{(210~Interaction)}}=\frac{h_{ab}^{^{(210)}}}{4!}\widehat{\Phi}_{a}^{\dagger}
\widehat{\mathsf
V}_{\mu\nu\rho\lambda}\Gamma_{[\mu}\Gamma_{\nu}\Gamma_{\rho}
\Gamma_{\lambda]}\widehat{\Phi}_{b}|_{\theta^2\bar {\theta}^2}\nonumber\\
\widehat{\mathsf
L}_{\Phi}^{'^{(210~Self-Interaction)}}=\widehat{\Phi}^{\dagger}_a\widehat{\Phi}_a|_{\theta^2\bar
{\theta}^2}~~~~~~~~
\end{eqnarray}
In the above the Greek subscripts($\alpha$, $\beta$, ...) are the SO(10)
indices. One
could, of course, add more interactions, for example, in
${\mathsf L}_V^{'^{(210~Self-Interaction)}}$ such as $V^5$
etc which are allowed once one gives up the Wess-Zumino gauge.
Similarly in ${\mathsf L}_{V+\Phi}^{'^{(210~Interaction)}}$ one may
add additional terms as well.
However, the line of construction
remains unchanged and the inclusion of additional terms only brings
in more complexity. Thus to  keep the analysis simple we omit such
terms. Evaluating Eq.(13) we get
\begin{eqnarray}
{\mathsf L}^{(210)}= -\frac{1}{4}{\cal V}_{AB\mathtt{XY}}{\cal
V}^{AB}_{\mathtt{XY}}-\frac{1}{2}m^2{\cal V}_{A\mathtt{XY}}{\cal
V}^{A}_{\mathtt{XY}}-\frac{1}{2}\partial^AB_{\mathtt{XY}}\partial_AB_{\mathtt{XY}}
-i\overline
{\Lambda}_{\mathtt{XY}}\gamma^A\partial_A\Lambda_{\mathtt{XY}}-m\overline
{\Lambda}_{\mathtt{XY}}\Lambda_{\mathtt{XY}}\nonumber\\
-\partial^AA^{\dagger}_a\partial_AA_a-i\overline
{\Psi}_{aL}\gamma^A\partial_A\Psi_{aL}
 -\frac{h_{ab}^{^{(210)}}}{24m}B_{\mathtt{XY}}(\partial^AA^{\dagger}_a)\widetilde{\Gamma}_{\mathtt{XY}}\partial_AA_b
\nonumber\\
-\frac{h_{ab}^{^{(210)}}}{96m}\partial^A\left(A_a^{\dagger}\widetilde{\Gamma}_{\mathtt{XY}}A_b\right)
\partial_AB_{\mathtt{XY}}+\frac{ih_{ab}^{^{(210)}}}{48}\left[A^{\dagger}_a\widetilde{\Gamma}_{\mathtt{XY}}
\partial^AA_b-(\partial^A
A_a^{\dagger})\widetilde{\Gamma}_{\mathtt{XY}}A_b\right]
{\cal V}_{A\mathtt{XY}}\nonumber\\
+\frac{h_{ab}^{^{(210)}}}{48m\sqrt{2}}\left[\left(\overline
{\Psi}_{aL}\gamma^A\widetilde{\Gamma}_{\mathtt{XY}}\Lambda_{L\mathtt{XY}}\right)\partial_AA_b+\partial_AA_a^{\dagger}
\left(\overline
{\Lambda}_{L\mathtt{XY}}\gamma^A\widetilde{\Gamma}_{\mathtt{XY}}\Psi_{bL}\right)\right]\nonumber\\
+\frac{h_{ab}^{^{(210)}}}{24m\sqrt{2}}\left[\left(\overline
{\Psi}_{aL}\gamma^A\widetilde{\Gamma}_{\mathtt{XY}}\partial_A\Lambda_{L\mathtt{XY}}\right)A_b-A_a^{\dagger}\left(\overline
{\Lambda}_{L\mathtt{XY}}\gamma^A\widetilde{\Gamma}_{\mathtt{XY}}\partial_A\Psi_{bL}\right)\right]\nonumber\\
-\frac{ih_{ab}^{^{(210)}}}{24\sqrt{2}}\left[\left(\overline
{\Psi}_{aL}\widetilde{\Gamma}_{\mathtt{XY}}\Lambda_{R\mathtt{XY}}\right)A_b-A_a^{\dagger}\left(\overline
{\Lambda}_{R\mathtt{XY}}\widetilde{\Gamma}_{\mathtt{XY}}\Psi_{bL}\right)\right]\nonumber\\
-\frac{h_{ab}^{^{(210)}}}{48}\overline
{\Psi}_{aL}\gamma^A\widetilde{\Gamma}_{\mathtt{XY}}\Psi_{bL}{\cal
V}_{A\mathtt{XY}}-\frac{ih_{ab}^{^{(210)}}}{24m}\overline
{\Psi}_{aL}\gamma^A\widetilde{\Gamma}_{\mathtt{XY}}\partial_A(\Psi_{bL})B_{\mathtt{XY}}\nonumber\\
-\frac{1}{m^3}\left(\frac{3\alpha_1}{4}B_{\mathtt{WY}}+\frac{\alpha_2}{m}B_{\mathtt{WX}}B_{\mathtt{XY}}\right)
\partial_AB_{\mathtt{YZ}}\partial^AB_{\mathtt{ZW}}-
\frac{3}{m}\left(\frac{\alpha_1}{2}B_{\mathtt{WY}}+\frac{\alpha_2}{m}B_{\mathtt{WX}}B_{\mathtt{XY}}\right)
{\cal V}_{
A\mathtt{YZ}}{\cal V}^A_{\mathtt{ZW}}\nonumber\\
+\frac{3}{m^3}\left(\alpha_1B_{\mathtt{WY}}+\frac{2\alpha_2}{m}B_{\mathtt{WX}}B_{\mathtt{XY}}\right)
\left(i\overline
{\Lambda}_{L\mathtt{YZ}}\gamma^A\partial_A\Lambda_{L\mathtt{ZW}}
-m\overline{\Lambda}_{\mathtt{YZ}}\Lambda_{\mathtt{ZW}}\right)\nonumber\\
+\frac{3}{m^2}\left(\frac{\alpha_1}{2}\delta_{\mathtt{WX}}+\frac{2\alpha_2}{m}B_{\mathtt{WX}}\right)
\left(\overline{\Lambda}_{L\mathtt{XY}}\gamma^A\Lambda_{L\mathtt{YZ}}\right){\cal
V}_{A\mathtt{ZW}}\nonumber\\
 + \frac{3\alpha_2}{2m^4}
\left(\overline{\Lambda}^c_{R\mathtt{WX}}\Lambda_{L\mathtt{XY}}\right)
\left(\overline{\Lambda}_{L\mathtt{YZ}}\Lambda_{R\mathtt{ZW}}^c\right)
+{\mathsf L}_{auxiliary}^{'~210}
\end{eqnarray}
Where we have defined for brevity
\begin{eqnarray}
\widetilde{\Gamma}_{\mathtt{XY}}=\Gamma_{[\mu}\Gamma_{\nu}\Gamma_{\rho}
\Gamma_{\lambda]};~~~B_{\mathtt{XY}}=B_{\mu\nu\rho\sigma};~~~{\cal
V}_{A\mathtt{XY}}{\cal V}^A_{\mathtt{YZ}}={\cal
V}_{A\mu\nu\rho\sigma}{\cal V}^A_{\rho\sigma\lambda\tau}
\end{eqnarray}
and so on. Further
\begin{eqnarray}
\Lambda=\left(\matrix{m\chi_{\tilde{\alpha}}\cr
\overline{\lambda}^{\dot{\tilde\alpha}}}\right),~~~
\Psi_a=\left(\matrix{\psi_{a\tilde{\alpha}}\cr
\overline{\psi}^{\dot{\tilde\alpha}}_a}\right),~~~ B=mC,\nonumber\\
 \Lambda^c={\cal C}\overline{\Lambda}^T,~~~ {\cal
C}=\left(\matrix{i\sigma^2&0\cr 0&i\overline{\sigma}^2}\right),~~~
 \overline{\Lambda}=\Lambda^{\dagger}\gamma^0
\end{eqnarray}
Note that each of the chiral fields, $S$ ($\equiv A, \Psi, F$) can
be expanded in terms of its SU(5) components as $|S_a>=|0>{\bf
S}_a+\frac{1}{2}b_i^{\dagger}b_j^{\dagger}|0>{\bf
S}_a^{ij}+\frac{1}{24}\epsilon^{ijklm}b_j^{\dagger}b_k^{\dagger}b_l^{\dagger}b_m^{\dagger}|0>{\bf
S}_{ai}$. We will expand some of the terms appearing in Eq.(14) in
appendix D.

 The Lagrangian containing the auxiliary fields
is given by
\begin{eqnarray}
{\mathsf
L}_{auxilliary}^{'~210}=\left(mB_{\mathtt{ZW}}+\frac{3\alpha_1}{2m^2}B_{\mathtt{WX}}B_{\mathtt{XZ}}
+\frac{2\alpha_2}{m^3}B_{\mathtt{WX}}B_{\mathtt{XY}}B_{\mathtt{YZ}}+\frac{h_{ab}^{^{(210)}}}{48}A^{\dagger}_a
\widetilde{\Gamma}_{\mathtt{ZW}}A_b \right)D_{\mathtt{ZW}}
+\frac{1}{2}D_{\mathtt{XY}}D_{\mathtt{XY}}\nonumber\\
+\left(\frac{1}{2}m^2\delta_{\mathtt{YW}}+\frac{3\alpha_1}{2m}B_{\mathtt{YW}}+\frac{3\alpha_2}
{m^2}B_{\mathtt{WX}}B_{\mathtt{XY}}\right)
\left(M_{\mathtt{YZ}}M_{\mathtt{ZW}}+N_{\mathtt{YZ}}N_{\mathtt{ZW}}\right)\nonumber\\
+i\left[\frac{h_{ab}^{^{(210)}}}{48}F^{\dagger}_a\widetilde{\Gamma}_{\mathtt{ZW}}A_b+\frac{3}{m^2}\left(
\frac{\alpha_1}{4}\delta_{\mathtt{WX}}+
\frac{\alpha_2}{m}B_{\mathtt{WX}}\right)\left(\overline{\Lambda}_{L\mathtt{XY}}\Lambda_{R\mathtt{YZ}}^c\right)
\right]\left(M_{\mathtt{ZW}}+iN_{\mathtt{ZW}}\right)\nonumber\\
-i\left[\frac{h_{ab}^{^{(210)}}}{48}A^{\dagger}_a\widetilde{\Gamma}_{\mathtt{ZW}}F_b+\frac{3}{m^2}\left(
\frac{\alpha_1}{4}\delta_{\mathtt{WX}}
+\frac{\alpha_2}{m}B_{\mathtt{WX}}\right)\left(\overline{\Lambda}^c_{R\mathtt{XY}}\Lambda_{L\mathtt{YZ}}\right)
\right]\left(M_{\mathtt{ZW}}-iN_{\mathtt{ZW}}\right)\nonumber\\
 +\frac{ih_{ab}^{^{(210)}}}{24m\sqrt 2}\left(\overline{\Lambda}_{L\mathtt{XY}}\Psi_{aR}\widetilde{\Gamma}_{\mathtt{XY}}
 \right)F_b
 -\frac{ih_{ba}^{^{(210)}}}{24m\sqrt 2}
\left(\widetilde{\Gamma}_{\mathtt{XY}}\overline{\Psi}_{aR}\Lambda_{L\mathtt{XY}}\right)F_b^{\dagger}\nonumber\\
+\frac{h_{ab}^{^{(210)}}}{24m}B_{\mathtt{XY}}F^{\dagger}_a\widetilde{\Gamma}_{\mathtt{XY}}F_b+F^{\dagger}_aF_a~~~~~
\end{eqnarray}
Finally, eliminating the auxiliary fields we obtain
\begin{eqnarray}
{\mathsf L}_{auxilliary}^{'}=
-\frac{1}{2}m^2B_{\mathtt{XY}}B_{\mathtt{YX}}-\frac{3\alpha_1}{2m}B_{\mathtt{XY}}B_{\mathtt{YZ}}B_{\mathtt{ZX}}
-\left(\frac{9{\alpha_1}^2}{8m^4}+\frac{2\alpha_2}{m^2}\right)B_{\mathtt{WX}}B_{\mathtt{XY}}B_{\mathtt{YZ}}B_{\mathtt{ZW}}
\nonumber\\
-\frac{3\alpha_1\alpha_2}{m^5}B_{\mathtt{VW}}B_{\mathtt{WX}}B_{\mathtt{XY}}B_{\mathtt{YZ}}B_{\mathtt{ZV}}
-\frac{2{\alpha_2}^2}{m^6}B_{\mathtt{UV}}B_{\mathtt{VW}}B_{\mathtt{WX}}B_{\mathtt{XY}}B_{\mathtt{YZ}}B_{\mathtt{ZU}}
\nonumber\\
-\frac{h_{ab}^{^{(210)}}h_{cd}^{^{(210)}}}{4608}\left(A^{\dagger}_a\widetilde{\Gamma}_{\mathtt{XY}}A_b\right)
\left(A^{\dagger}_c
\widetilde{\Gamma}_{\mathtt{XY}} A_d\right)
-\frac{mh_{ab}^{^{(210)}}}{48}\left(A^{\dagger}_a\widetilde{\Gamma}_{\mathtt{XY}}A_b\right)B_{\mathtt{XY}}\nonumber\\
-\frac{h_{ab}^{^{(210)}}\alpha_1}{32m^2}\left(A^{\dagger}_a\widetilde{\Gamma}_{\mathtt{XY}}A_b\right)B_{\mathtt{XZ}}
B_{\mathtt{ZY}}
-\frac{h_{ab}^{^{(210)}}\alpha_2}{24m^3}\left(A^{\dagger}_a\widetilde{\Gamma}_{\mathtt{XY}}A_b\right)
B_{\mathtt{XW}}B_{\mathtt{WZ}}B_{\mathtt{ZY}}\nonumber\\
-\frac{1}{2}m^2\left[{\mathbf K}_{\mathtt{WX}}({\mathbf
P}^{-1})_{\mathtt{WXYZ}}{\mathbf K}_{\mathtt{YZ}}+{\mathbf
J}_{\mathtt{WX}}({\mathbf P}^{-1})_{\mathtt{WXYZ}}{\mathbf
J}_{\mathtt{YZ}}\right]
-\frac{ih_{ab}^{^{(210)}}}{24m\sqrt 2}({\mathbf
Q}^{-1})_{ac}{\mathbf R}_{c}
\widetilde{\Gamma}_{\mathtt{XY}}\overline{\Psi}_{bR}\Lambda_{L\mathtt{XY}}
\nonumber\\
+\frac{h_{ab}^{^{(210)}}}{8m^4}A^{\dagger}_a\widetilde{\Gamma}_{\mathtt{UV}}({\mathbf
S}^{-1})_{bc}{\mathbf T}_{c}({\mathbf
P}^{-1})_{\mathtt{UVXY}}\left(
\frac{\alpha_1}{4}\delta_{\mathtt{YW}}+
\frac{\alpha_2}{m}B_{\mathtt{YW}}\right)\overline{\Lambda}_{L\mathtt{WZ}}\Lambda_{R\mathtt{ZX}}^c
\end{eqnarray}
where
\begin{eqnarray}
{\mathbf
P}_{\mathtt{UVXY}}=\delta_{\mathtt{UX}}\delta_{\mathtt{VY}}+\frac{3\alpha_1}{2m^3}\left(\delta_{\mathtt{UY}}
B_{\mathtt{VX}}+B_{\mathtt{UX}}\delta_{\mathtt{VY}}\right)+\frac{3\alpha_2}{m^4}\left(\delta_{\mathtt{UY}}
B_{\mathtt{VW}}B_{\mathtt{WX}}+B_{\mathtt{UW}}B_{\mathtt{WY}}
\delta_{\mathtt{XV}}\right)\nonumber\\
{\mathbf
Q}_{bc}=\delta_{bc}+\frac{h_{bc}^{^{(210)}}}{24m}\widetilde{\Gamma}_{\mathtt{XY}}B_{\mathtt{XY}}
-\frac{h_{ac}^{^{(210)}}h_{bd}^{^{(210)}}}{1152m^2}A^{\dagger}_a\widetilde{\Gamma}_{\mathtt{UV}}({\mathbf
P}^{-1})_{\mathtt{UVXY}}\widetilde{\Gamma}_{\mathtt{XY}}A_d\nonumber\\
{\mathbf
S}_{bc}=\delta_{bc}+\frac{h_{cb}^{^{(210)}}}{24m}\widetilde{\Gamma}_{\mathtt{XY}}B_{\mathtt{XY}}
-\frac{h_{cd}^{^{(210)}}h_{ab}^{^{(210)}}}{1152m^2}\widetilde{\Gamma}_{\mathtt{UV}}A_d({\mathbf
P}^{-1})_{\mathtt{UVXY}}A^{\dagger}_a\widetilde{\Gamma}_{\mathtt{XY}}\nonumber\\
{\mathbf R}_{c}=-\frac{ih_{ac}^{^{(210)}}}{24m\sqrt
2}\overline{\Lambda}_{L\mathtt{XY}}\Psi_{aR}\widetilde{\Gamma}_{\mathtt{XY}}+\frac{h_{ac}^{^{(210)}}}{8m^4}
A^{\dagger}_a\widetilde{\Gamma}_{\mathtt{UV}} ({\mathbf
P}^{-1})_{\mathtt{UVXY}}\left(
\frac{\alpha_1}{4}\delta_{\mathtt{YW}}+
\frac{\alpha_2}{m}B_{\mathtt{YW}}\right)\overline{\Lambda}_{L\mathtt{WZ}}\Lambda_{R\mathtt{ZX}}^c\nonumber\\
{\mathbf T}_{c}=\frac{ih_{ca}^{^{(210)}}}{24m\sqrt
2}\widetilde{\Gamma}_{\mathtt{XY}}\overline{\Psi}_{aR}\Lambda_{L\mathtt{XY}}+\frac{h_{ca}^{^{(210)}}}{8m^4}
\widetilde{\Gamma}_{\mathtt{UV}}A_a ({\mathbf
P}^{-1})_{\mathtt{UVXY}}\left(
\frac{\alpha_1}{4}\delta_{\mathtt{YW}}+
\frac{\alpha_2}{m}B_{\mathtt{YW}}\right)\overline{\Lambda}_{R\mathtt{WZ}}^c\Lambda_{L\mathtt{ZX}}\nonumber\\
{\mathbf
K}_{\mathtt{XY}}=-\frac{ih_{ab}^{^{(210)}}}{48m^2}\left[({\mathbf
Q}^{-1})_{ac}{\mathbf
R}_{c}\widetilde{\Gamma}_{\mathtt{XY}}A_b-A^{\dagger}_a\widetilde{\Gamma}_{\mathtt{XY}}({\mathbf
S}^{-1})_{bc}{\mathbf T}_{c} \right]\nonumber\\
-\frac{3i}{m^4}\left( \frac{\alpha_1}{4}\delta_{\mathtt{XU}}+
\frac{\alpha_2}{m}B_{\mathtt{XU}}\right)\left(\overline{\Lambda}_{L\mathtt{UV}}\Lambda_{R\mathtt{VY}}^c
-\overline{\Lambda}_{R\mathtt{UV}}^c\Lambda_{L\mathtt{VY}}\right)\nonumber\\
{\mathbf
J}_{\mathtt{XY}}=\frac{h_{ab}^{^{(210)}}}{48m^2}\left[({\mathbf
Q}^{-1})_{ac}{\mathbf
R}_{c}\widetilde{\Gamma}_{\mathtt{XY}}A_b+A^{\dagger}_a\widetilde{\Gamma}_{\mathtt{XY}}({\mathbf
S}^{-1})_{bc}{\mathbf T}_{c} \right]\nonumber\\
+\frac{3}{m^4}\left( \frac{\alpha_1}{4}\delta_{\mathtt{XU}}+
\frac{\alpha_2}{m}B_{\mathtt{XU}}\right)\left(\overline{\Lambda}_{L\mathtt{UV}}\Lambda_{R\mathtt{VY}}^c
+\overline{\Lambda}_{R\mathtt{UV}}^c\Lambda_{L\mathtt{VY}}\right)
\end{eqnarray}
\section{Conclusion}
In this paper we have given an explicit computation of the couplings  of
 the 210 dimensional SO(10) vector mutliplet.
 Specifically, we have computed the vector couplings
$\overline{16}_{\pm}-16_{\pm}-210$ in terms of its $SU(5)\times U(1)$
decomposition. We approached this coupling  from two view points.
First, we use the conventional approach of using the Wess-Zumino gauge.
However, since the $210$ couplings are not expected to be  gauge invariant
and hence such interactions are not expected to be renormalizable,
we  also consider a nonlinear sigma model type couplings of $210$ with
matter. Such couplings arise when we consider the  full $210$ multiplet
without using the Wess-Zumino gauge.  Here elimination of the
auxiliary fields leads to interactions of the vector multiplet with
the chiral fields with nonlinearities of infinite order as in a
nonlinear sigma model.
 Although couplings of the type discussed do not thus far
 appear in theories of fundamental interactions a 210 vector multiplet
 may arise as a condensate in effective theories. The analysis  we have
 presented here concludes our effort to give a complete analytic
 computation of all lowest order couplings involving the $16_{\pm}$ of
 matter with Higgs and vector multiplets. Although the analysis
 given here is specific to the case of the vector couplings
 $\overline{16}_{\pm}-16_{\pm}-210$ the techniques developed here are
 general and can be applied to other cases where the dimensionality of the
  vector mulitiplet does not equal the dimensionality of the
  adjoint representation of the group.
\begin{center}
{\bf ACKNOWLEDGEMENTS}
\end{center}
This research was supported in part by NSF grant  PHY-0139967.
Part of this work was done when one of the authors (PN) was
visiting the Max Planck Institute fur Kernphysik, Heidelberg
and thanks the Institute for hospitality during the period of
his visit there. He also acknowledges support from the
Alexander von Humboldt Foundation during the period of this
visit.

\section{Appendix  A}
In this section we define the notation for the components of
the vector superfield and for the chiral superfield used in
the text. For the vector superfield we have the expansion
\begin{eqnarray}
\widehat{\mathsf V}=C(x)+i\theta\chi (x)-i\bar {\theta}\overline {\chi}(x)
+\frac {i}{2}\theta^2\left[M(x)+iN(x)\right]-\frac{i}{2}\bar {\theta}^2
\left[M(x)-iN(x)\right]
-\theta\sigma^A\bar {\theta}{\cal V}_A(x)\nonumber\\
+i\theta^2\bar {\theta}\left[\overline
{\lambda}(x)+\frac{i}{2}\overline{\sigma}^A\partial_A\chi(x)\right]
-i\bar {\theta}^2\theta\left[\lambda(x)+\frac{i}{2}\sigma^A\partial_A
\overline{\chi}(x)\right]
+\frac{1}{2}\theta^2\bar
{\theta}^2\left[D(x)+\frac{1}{2}\partial_A\partial^AC(x)\right]
\end{eqnarray}
while for the chiral superfields we have the expansion
\begin{eqnarray}
\widehat{\Phi}_a=A_a(x)+\sqrt{2}\theta\psi_a+\theta^2F_a(x)+i\theta\sigma^A\bar
{\theta}\partial_AA_a(x)\nonumber\\
+\frac{i}{\sqrt{2}}\theta^2\bar {\theta}\overline
{\sigma}^A\partial_A\psi_a(x)+\frac{1}{4}\theta^2\bar
{\theta}^2\partial_A\partial^AA_a(x)\nonumber\\
\widehat{\Phi}_a^{\dagger}=A_a^{\dagger}(x)+\sqrt{2}\bar{\theta}\overline{\psi}
_a+\bar{\theta}^2F_a^{\dagger}(x)-i\theta\sigma^A\bar
{\theta}\partial_AA_a^{\dagger}(x)\nonumber\\
+\frac{i}{\sqrt{2}}\bar{\theta}^2 {\theta}
{\sigma}^A\partial_A\overline{\psi}_a(x)+\frac{1}{4}\theta^2\bar
{\theta}^2\partial_A\partial^AA_a^{\dagger}(x)
\end{eqnarray}

\section{Appendix B}
In this appendix we normalize the irreducible SU(5) tensors
contained in a $210$ vector ${\cal V}^{A} _{\mu\nu\rho\sigma}$,
a $210$ scalar $B_{\mu\nu\rho\sigma}$, and a $210$
spinor $\Lambda_{\mu\nu\rho\sigma}$. Latin
letters ($i, j, k, ...$) are used to denote the SU(5) indices.
 The normalized SU(5) gauge tensors
appearing in ${\cal V}^A _{\mu\nu\rho\sigma}$ are
\begin{eqnarray}
{\cal V}_A=4\sqrt{\frac{10}{3}}{\cal V}^{'}_A;~~~~{\cal
V}_A^i=8\sqrt{6}{\cal V}^{'i}_A;~~~~{\cal V}_{Ai}=8\sqrt{6}{\cal
V}^{'}_{Ai}\nonumber\\
{\cal V}^{ij}_A=\sqrt{2}{\cal V}^{'ij}_A;~~~~
 {\cal
V}_{Aij}=\sqrt{2}{\cal V}^{'}_{Aij};~~~~{\cal
V}^{j}_{Ai}=\sqrt{2}{\cal V}^{'j}_{Ai}\nonumber\\
{\cal V}^{ijk}_{Al}=\sqrt{\frac{2}{3}}{\cal
V}^{'ijk}_{Al};~~~~{\cal V}^{i}_{Ajkl}=\sqrt{\frac{2}{3}}{\cal
V}^{'i}_{Ajkl};~~~~{\cal V}^{ij}_{Akl}=\sqrt{\frac{2}{3}}{\cal
V}^{'ij}_{Akl}
\end{eqnarray}
so that
\begin{eqnarray}
-\frac{1}{4}{\cal V}^{AB} _{\mu\nu\rho\sigma}{\cal V}
_{AB\mu\nu\rho\sigma}=-\frac{1}{2}{\cal V}'_{AB}{\cal
V}^{'AB\dagger}-\frac{1}{2}{\cal V}^{'i}_{AB}{\cal
V}^{'ABi\dagger}-\frac{1}{2!}\frac{1}{2}{\cal V}^{'ij}_{AB}{\cal
V}^{'ABij\dagger}\nonumber\\
-\frac{1}{4}{\cal V}^{'i}_{ABj}{\cal
V}^{'ABj}_i-\frac{1}{3!}\frac{1}{2}{\cal V}^{'ijk}_{ABl}{\cal
V}^{'ABijk\dagger}_l-\frac{1}{2!}\frac{1}{2!}\frac{1}{4}{\cal
V}^{'ij}_{ABkl}{\cal V}^{'ABkl}_{ij}
\end{eqnarray}
The normalized SU(5) fields appearing in $B _{\mu\nu\rho\sigma}$
are
\begin{eqnarray}
B=4\sqrt{\frac{10}{3}}B^{'};~~~~B^i=8\sqrt{6}B^{'i};~~~~B_{i}=8\sqrt{6}B^{'}_{i}\nonumber\\
B^{ij}=\sqrt{2}B^{'ij};~~~~ B_{ij}=\sqrt{2}B^{'}_{ij};~~~~B^{j}_{i}=\sqrt{2}B^{'j}_{i}\nonumber\\
B^{ijk}_{l}=\sqrt{\frac{2}{3}}B^{'ijk}_{l};~~~~B_{jkl}=\sqrt{\frac{2}{3}}B^{'i}_{jkl};~~~~B^{ij}_{kl}=\sqrt{\frac{2}{3}}
B^{'ij}_{kl}
\end{eqnarray}
so that
\begin{eqnarray}
-\frac{1}{2}\partial^AB_{\mu\nu\rho\sigma}
\partial_AB_{\mu\nu\rho\sigma}=-\partial^AB^{'}\partial_AB^{'\dagger}-\partial^AB^{'i}\partial_AB
^{'i\dagger}-\frac{1}{2!}\partial^AB^{'ij}\partial_AB^{'ij\dagger}\nonumber\\
-\frac{1}{2}\partial^AB^{'i}_{j}\partial_AB^{'j}_i-\frac{1}{3!}\partial^AB^{'ijk}_{l}\partial_AB^{'ijk\dagger}_l
-\frac{1}{2!}\frac{1}{2!}\frac{1}{2}\partial^AB^{'ij}_{kl}\partial_AB^{'kl}_{ij}
\end{eqnarray}
The normalized SU(5) fields appearing in $\Lambda
_{\mu\nu\rho\sigma}$ are
\begin{eqnarray}
\Lambda=4\sqrt{\frac{5}{3}}\Lambda^{'};~~~~\Lambda_i=8\sqrt{6}\Lambda^{'i};~~~~\Lambda_{i}=8\sqrt{6}\Lambda^{'}_{i}\nonumber\\
\Lambda^{ij}=\sqrt{2}\Lambda^{'ij};~~~~\Lambda_{ij}=\sqrt{2}\Lambda^{'}_{ij};~~~~\Lambda^{j}_{i}=\sqrt{2}\Lambda^{'j}_{i}\nonumber\\
\Lambda^{ijk}_{l}=\sqrt{\frac{2}{3}}\Lambda^{'ijk}_{l};~~~~\Lambda_{jkl}=\sqrt{\frac{2}{3}}\Lambda^{'i}_{jkl}
;~~~~\Lambda^{ij}_{kl}=\sqrt{\frac{2}{3}} \Lambda^{'ij}_{kl}
\end{eqnarray}
so that
\begin{eqnarray}
-i\overline{\Lambda}_{\mu\nu\rho\sigma}\gamma^A
\partial_A\Lambda_{\mu\nu\rho\sigma}=-i\overline{\Lambda}^{'}\gamma^A\partial_A\Lambda^{'}
-i\overline{\Lambda}^{'}_{i}\gamma^A\partial_A\Lambda{'}_{i}-i\overline{\Lambda}^{'i}\gamma^A\partial_A\Lambda^{'i}\nonumber\\
-\frac{1}{2!}i\overline{\Lambda}^{'}_{ij}\gamma^A\partial_A\Lambda^{'}_{ij}
-\frac{1}{2!}i\overline{\Lambda}^{'ij}\gamma^A\partial_A\Lambda^{'ij}
-i\overline{\Lambda}^{'i}_j\gamma^A\partial_A\Lambda^{'i}_j\nonumber\\
-\frac{1}{3!}i\overline{\Lambda}^{'ijk}_l\gamma^A\partial_A\Lambda^{'ijk}_l
-\frac{1}{3!}i\overline{\Lambda}^{'l}_{ijk}\gamma^A\partial_A\Lambda^{'l}_{ijk}
-\frac{1}{2!}\frac{1}{2!}i\overline{\Lambda}^{'ij}_{kl}\gamma^A\partial_A\Lambda^{'ij}_{kl}
\end{eqnarray}

\section{Appendix C}
In this appendix we eliminate the auxiliary fields,
$D_{\mu\nu\rho\sigma}$ and ${\bf F}_{(\pm)}$ of section 2.
We find
\begin{eqnarray}
{\mathsf L}^{(210)}_{(3)auxiliary}+{\mathsf
L}^{(210)}_{(4)auxiliary}+{\mathsf
L}^{(210)}_{(5)auxiliary}~~~~~~~~~~~~~~~~~~~~~~~~~~~~~~~~~~~~~~~~~~~~~~~~~\nonumber\\
=-\frac{1}{4608} {\mathsf
g}^{^{(210)}2}h_{ab}^{^{(210+)}}h_{cd}^{^{(210+)}}<A_{(+)a}|\Gamma_{[\mu}\Gamma_{\nu}\Gamma_{\rho}
\Gamma_{\lambda]}|A_{(+)b}><A_{(+)c}|\Gamma_{[\mu}\Gamma_{\nu}\Gamma_{\rho}
\Gamma_{\lambda]}|A_{(+)d}>
\nonumber\\
-\frac{1}{4608} {\mathsf
g}^{^{(210)}2}h_{ab}^{^{(210-)}}h_{cd}^{^{(210-)}}<A_{(-)a}|\Gamma_{[\mu}\Gamma_{\nu}\Gamma_{\rho}
\Gamma_{\lambda]}|A_{(-)b}><A_{(-)c}|\Gamma_{[\mu}\Gamma_{\nu}\Gamma_{\rho}
\Gamma_{\lambda]}|A_{(-)d}>\nonumber\\
-\frac{1}{2304} {\mathsf
g}^{^{(210)}2}h_{ab}^{^{(210+)}}h_{cd}^{^{(210-)}}<A_{(+)a}|\Gamma_{[\mu}\Gamma_{\nu}\Gamma_{\rho}
\Gamma_{\lambda]}|A_{(+)b}><A_{(-)c}|\Gamma_{[\mu}\Gamma_{\nu}\Gamma_{\rho}
\Gamma_{\lambda]}|A_{(-)d}>
\end{eqnarray}
SU(5)expansion of these expressions gives
\begin{eqnarray}
-\frac{1}{4608} {\mathsf
g}^{^{(210)}2}h_{ab}^{^{(210+)}}h_{cd}^{^{(210+)}}<A_{(+)a}|\Gamma_{[\mu}\Gamma_{\nu}\Gamma_{\rho}
\Gamma_{\lambda]}|A_{(+)b}><A_{(+)c}|\Gamma_{[\mu}\Gamma_{\nu}\Gamma_{\rho}
\Gamma_{\lambda]}|A_{(+)d}>\nonumber\\
={\mathsf
g}^{^{(210)}2}\{-\frac{1}{6144}\left(\eta_{ab,cd}^{^{(210++)}}+8\eta_{ad,cb}^{^{(210++)}}\right){\bf
A}^{\dagger}_{(+)aij}{\bf A}_{(+)b}^{ij}{\bf
A}^{\dagger}_{(+)ckl}{\bf A}_{(+)d}^{kl}\nonumber\\
-\frac{1}{768}\left(\eta_{ab,cd}^{^{(210++)}}+8\eta_{ad,cb}^{^{(210++)}}\right){\bf
A}^{\dagger}_{(+)a}{\bf A}_{(+)b}{\bf A}^{i\dagger}_{(+)c}{\bf
A}_{(+)di}
\nonumber\\
+\frac{1}{512}\left(11\eta_{ab,cd}^{^{(210++)}}-2\eta_{ad,cb}^{^{(210++)}}\right){\bf
A}^{\dagger}_{(+)aij}{\bf A}_{(+)b}^{ij}{\bf
A}^{k\dagger}_{(+)c}{\bf A}_{(+)dk}\nonumber\\
+\frac{1}{192}\left(\eta_{ab,cd}^{^{(210++)}}+2\eta_{ad,cb}^{^{(210++)}}\right){\bf
A}^{\dagger}_{(+)aij}{\bf A}_{(+)b}^{jk}{\bf
A}^{i\dagger}_{(+)c}{\bf
A}_{(+)dk}\nonumber\\
+\frac{1}{1536}\left(25\eta_{ab,cd}^{^{(210++)}}+18\eta_{ad,cb}^{^{(210++)}}\right)
{\bf A}^{i\dagger}_{(+)a}{\bf A}_{(+)bi}{\bf
A}^{j\dagger}_{(+)c}{\bf A}_{(+)dj}\nonumber\\
+\frac{1}{1536}\left(\eta_{ab,cd}^{^{(210++)}}-6\eta_{ad,cb}^{^{(210++)}}\right){\bf
A}^{\dagger}_{(+)a}{\bf A}_{(+)b}{\bf A}^{\dagger}_{(+)cij}{\bf
A}_{(+)d}^{ij}
\nonumber\\
+\eta_{ab,cd}^{^{(210++)}}[\frac{1}{1536}\epsilon_{ijklm}{\bf
A}^{\dagger}_{(+)a}{\bf A}_{(+)b}^{ij}{\bf
A}^{k\dagger}_{(+)c}{\bf
A}_{(+)d}^{lm}+\frac{1}{1536}\epsilon^{ijklm}{\bf
A}^{\dagger}_{(+)aij}{\bf A}_{(+)bk}{\bf A}^{\dagger}_{(+)clm}{\bf
A}_{(+)d}\nonumber\\
+\frac{1}{768}{\bf A}^{\dagger}_{(+)aij}{\bf A}_{(+)b}^{jk}{\bf
A}^{\dagger}_{(+)ckl}{\bf A}_{(+)d}^{li}-\frac{5}{1536}{\bf
A}^{\dagger}_{(+)a}{\bf A}_{(+)b}{\bf A}^{\dagger}_{(+)c}{\bf
A}_{(+)d}]\}
\end{eqnarray}
\begin{eqnarray}
-\frac{1}{4608} {\mathsf
g}^{^{(210)}2}h_{ab}^{^{(210-)}}h_{cd}^{^{(210-)}}<A_{(-)a}|\Gamma_{[\mu}\Gamma_{\nu}\Gamma_{\rho}
\Gamma_{\lambda]}|A_{(-)b}><A_{(-)c}|\Gamma_{[\mu}\Gamma_{\nu}\Gamma_{\rho}
\Gamma_{\lambda]}|A_{(-)d}>\nonumber\\
={\mathsf
g}^{^{(210)}2}\{-\frac{1}{6144}\left(\eta_{ab,cd}^{^{(210--)}}+16\eta_{ad,cb}^{^{(210--)}}\right){\bf
A}^{ij\dagger}_{(-)a}{\bf A}_{(-)bij}{\bf
A}^{kl\dagger}_{(-)c}{\bf A}_{(-)d}^{kl}\nonumber\\
-\frac{1}{768}\left(11\eta_{ab,cd}^{^{(210--)}}+8\eta_{ad,cb}^{^{(210--)}}\right){\bf
A}^{\dagger}_{(-)a}{\bf A}_{(-)b}{\bf A}^{\dagger}_{(-)ci}{\bf
A}_{(-)d}^i
\nonumber\\
-\frac{1}{1536}\left(5\eta_{ab,cd}^{^{(210--)}}+6\eta_{ad,cb}^{^{(210--)}}\right){\bf
A}^{ij\dagger}_{(-)a}{\bf A}_{(-)bij}{\bf
A}^{\dagger}_{(-)ck}{\bf A}_{(-)d}^k\nonumber\\
+\frac{1}{384}\left(\eta_{ab,cd}^{^{(210--)}}+\eta_{ad,cb}^{^{(210--)}}\right){\bf
A}^{ij\dagger}_{(-)a}{\bf A}_{(-)bjk}{\bf A}^{\dagger}_{(-)ci}{\bf
A}_{(-)d}^k\nonumber\\
+\frac{1}{1536}\left(\eta_{ab,cd}^{^{(210--)}}-6\eta_{ad,cb}^{^{(210--)}}\right)
{\bf A}^{\dagger}_{(-)ai}{\bf A}_{(-)b}^i{\bf
A}^{\dagger}_{(-)cj}{\bf A}_{(-)d}^j\nonumber\\
-\frac{1}{1536}\left(29\eta_{ab,cd}^{^{(210-)}}+18\eta_{ad,cb}^{^{(210--)}}\right){\bf
A}^{\dagger}_{(-)a}{\bf A}_{(-)b}{\bf A}^{ij\dagger}_{(-)c}{\bf
A}_{(-)dij}
\nonumber\\
+\eta_{ab,cd}^{^{(210--)}}[\frac{5}{1536}\epsilon_{ijklm}{\bf
A}^{ij\dagger}_{(-)a}{\bf A}_{(-)b}{\bf A}^{kl\dagger}_{(-)c}{\bf
A}_{(-)d}^{m}+\frac{5}{1536}\epsilon^{ijklm}{\bf
A}^{\dagger}_{(-)ai}{\bf A}_{(-)bjk}{\bf A}^{\dagger}_{(-)c}{\bf
A}_{(-)dlm}\nonumber\\
-\frac{1}{256}{\bf A}^{ij\dagger}_{(-)a}{\bf A}_{(-)bjk}{\bf
A}^{kl\dagger}_{(-)c}{\bf A}_{(-)dli}+\frac{25}{1536}{\bf
A}^{\dagger}_{(-)a}{\bf A}_{(-)b}{\bf A}^{\dagger}_{(-)c}{\bf
A}_{(-)d}]\}
\end{eqnarray}
\begin{eqnarray}
-\frac{1}{2304} {\mathsf
g}^{^{(210)}2}h_{ab}^{^{(210+)}}h_{cd}^{^{(210-)}}<A_{(+)a}|\Gamma_{[\mu}\Gamma_{\nu}\Gamma_{\rho}
\Gamma_{\lambda]}|A_{(+)b}><A_{(-)c}|\Gamma_{[\mu}\Gamma_{\nu}\Gamma_{\rho}
\Gamma_{\lambda]}|A_{(-)d}>\nonumber\\
=\frac{{\mathsf g}^{^{(210)}2}\eta_{ab,cd}^{^{(210+-)}}}{192}\{
-10{\bf A}^{\dagger}_{(+)a}{\bf A}_{(+)b}{\bf
A}^{\dagger}_{(-)c}{\bf A}_{(-)d}+2{\bf A}^{\dagger}_{(+)a}{\bf
A}_{(+)b}{\bf A}^{ij\dagger}_{(-)c}{\bf A}_{(-)dij}\nonumber\\
-4{\bf A}^{\dagger}_{(+)a}{\bf A}_{(+)b}{\bf
A}^{\dagger}_{(-)ci}{\bf A}_{(-)d}^i-32{\bf
A}^{\dagger}_{(+)a}{\bf
A}_{(+)bi}{\bf A}^{\dagger}_{(-)c}{\bf A}_{(-)d}^i\nonumber\\
-4{\bf A}^{\dagger}_{(+)a}{\bf A}_{(+)b}^{ij}{\bf
A}^{\dagger}_{(-)c}{\bf A}_{(-)dij}-32{\bf
A}^{i\dagger}_{(+)a}{\bf
A}_{(+)b}{\bf A}^{\dagger}_{(-)ci}{\bf A}_{(-)d}\nonumber\\
+12{\bf A}^{i\dagger}_{(+)a}{\bf A}_{(+)b}^{jk}{\bf
A}^{\dagger}_{(-)ci}{\bf A}_{(-)djk}-8{\bf
A}^{i\dagger}_{(+)a}{\bf
A}_{(+)b}^{jk}{\bf A}^{\dagger}_{(-)cj}{\bf A}_{(-)dki}\nonumber\\
+18{\bf A}^{i\dagger}_{(+)a}{\bf A}_{(+)bi}{\bf
A}^{jk\dagger}_{(-)c}{\bf A}_{(-)djk}+4{\bf
A}^{i\dagger}_{(+)a}{\bf
A}_{(+)bi}{\bf A}^{\dagger}_{(-)cj}{\bf A}_{(-)d}^j\nonumber\\
-56{\bf A}^{i\dagger}_{(+)a}{\bf A}_{(+)bj}{\bf
A}^{jk\dagger}_{(-)c}{\bf A}_{(-)dki}-24{\bf
A}^{i\dagger}_{(+)a}{\bf
A}_{(+)bj}{\bf A}^{\dagger}_{(-)ci}{\bf A}_{(-)d}^j\nonumber\\
-164{\bf A}^{i\dagger}_{(+)a}{\bf A}_{(+)bi}{\bf
A}^{\dagger}_{(-)c}{\bf A}_{(-)d}-4{\bf A}^{\dagger}_{(+)aij}{\bf
A}_{(+)b}{\bf A}^{ij\dagger}_{(-)c}{\bf A}_{(-)d}\nonumber\\
-8{\bf A}^{\dagger}_{(+)aij}{\bf A}_{(+)bk}{\bf
A}^{jk\dagger}_{(-)c}{\bf A}_{(-)d}^i+8{\bf
A}^{\dagger}_{(+)aij}{\bf
A}_{(+)bk}{\bf A}^{ij\dagger}_{(-)c}{\bf A}_{(-)d}^k\nonumber\\
-{\bf A}^{\dagger}_{(+)aij}{\bf A}_{(+)b}^{ij}{\bf
A}^{kl\dagger}_{(-)c}{\bf A}_{(-)dkl}+8{\bf
A}^{\dagger}_{(+)aij}{\bf
A}_{(+)b}^{jk}{\bf A}^{il\dagger}_{(-)c}{\bf A}_{(-)dlk}\nonumber\\
+8{\bf A}^{\dagger}_{(+)aij}{\bf A}_{(+)b}^{jk}{\bf
A}^{\dagger}_{(-)ck}{\bf A}_{(-)d}^i-38{\bf
A}^{\dagger}_{(+)aij}{\bf
A}_{(+)b}^{ij}{\bf A}^{\dagger}_{(-)c}{\bf A}_{(-)d}\nonumber\\
+2{\bf A}^{\dagger}_{(+)aij}{\bf A}_{(+)b}^{ij}{\bf
A}^{\dagger}_{(-)ck}{\bf A}_{(-)d}^k-8{\bf
A}^{\dagger}_{(+)aij}{\bf
A}_{(+)b}^{kl}{\bf A}^{ij\dagger}_{(-)c}{\bf A}_{(-)dkl}\nonumber\\
+14\epsilon^{ijklm}{\bf A}^{\dagger}_{(+)aij}{\bf A}_{(+)bk}{\bf
A}^{\dagger}_{(-)c}{\bf A}_{(-)dlm}+14\epsilon_{ijklm}{\bf
A}^{i\dagger}_{(+)a}{\bf A}_{(+)b}^{jk}{\bf
A}^{lm\dagger}_{(-)c}{\bf A}_{(-)d}\nonumber\\
-2\epsilon^{ijklm}{\bf A}^{\dagger}_{(+)aij}{\bf A}_{(+)b}{\bf
A}^{\dagger}_{(-)ck}{\bf A}_{(-)dlm}-2\epsilon_{ijklm}{\bf
A}^{\dagger}_{(+)a}{\bf A}_{(+)b}^{ij}{\bf
A}^{kl\dagger}_{(-)c}{\bf A}_{(-)d}^m\}
\end{eqnarray}
where $\eta$'s are defined by
\begin{eqnarray}
\eta_{ab,cd}^{^{(210++)}}=h_{ab}^{^{(210+)}}h_{cd}^{^{(210+)}};~~~~~\eta_{ab,cd}^{^{(210--)}}=h_{ab}^{^{(210-)}}
h_{cd}^{^{(210-)}} ;~~~~~
\eta_{ab,cd}^{^{(210+-)}}=h_{ab}^{^{(210+)}}h_{cd}^{^{(210-)}}
\end{eqnarray}
We also find
\begin{eqnarray}
{\mathsf L}^{(210)}_{(1)auxiliary}+{\mathsf
L}^{(210)}_{(2)auxiliary}+{\mathsf
L}^{(210)}_{(6)auxiliary}~~~~~~~~~~~~~~~~~~~~~~~~~~~~~~~~~~~~~~~~~~~~~~~~~\nonumber\\
=-\left(\mu^*
[h^{^{(210-)}}]^{\bf{-1}}[h^{^{(210-)}}]^{\bf{T}}[h^{^{(210-)}}]^{\bf{-1}}\mu\right)_{ab}\left[{\bf
A}^{\dagger}_{(+)a}{\bf A}_{(+)b}+\frac{1}{4}{\bf
A}^{\dagger}_{(+)aij}{\bf A}_{(+)b}^{ij}+{\bf
A}^{i\dagger}_{(+)a}{\bf A}_{(+)bi}\right]\nonumber\\
-\left(\mu[h^{^{(210+)}}]^{\bf{-1T}}h^{^{(210+)}}[h^{^{(210+)}}]^{\bf{-1T}}\mu^*\right)_{ab}
\left[ {\bf A}^{\bf{T}}_{(-)a}{\bf A}_{(-)b}^*+\frac{1}{4}{\bf
A}^{\bf{T}}_{(-)aij}{\bf A}_{(-)b}^{ij*}+{\bf
A}^{i\bf{T}}_{(-)a}{\bf A}_{(-)bi}^*\right]
\end{eqnarray}
\section{Appendix D}
In this appendix we exhibit, for the benefit of the reader, a few
SU(5) expansions of the terms appearing in our final Lagrangian
Eq.(14). We begin by noting that any of the chiral fields $S$
($\equiv A, \Psi, F$) can be expanded in terms of its SU(5)
components as
\begin{equation}
|S_a>=|0>{\bf S}_a+\frac{1}{2}b_i^{\dagger}b_j^{\dagger}|0>{\bf
S}_a^{ij}+\frac{1}{24}\epsilon^{ijklm}b_j^{\dagger}b_k^{\dagger}b_l^{\dagger}b_m^{\dagger}|0>{\bf
S}_{ai}
\end{equation}
Together with the normalizations of appendix B and the basic
theorem given in Ref.\cite{ns1}, we can expand terms such as
\begin{eqnarray}
\frac{h_{ab}^{^{(210)}}}{48m\sqrt 2}\left(\overline
{\Psi}_{aL}\gamma^A\widetilde{\Gamma}_{\mathtt{XY}}\Lambda_{L\mathtt{XY}}\right)\partial_AA_b=
\frac{h_{ab}^{^{(210)}}}{48m\sqrt 2}\left(\overline
{\Psi}_{aL}\gamma^A\Gamma_{[\mu}\Gamma_{\nu}\Gamma_{\rho}
\Gamma_{\lambda]}\partial_AA_b\right)\Lambda_{L\mu\nu\rho\sigma}\nonumber\\
=\frac{h_{ab}^{^{(210)}}}{m}[\frac{1}{8\sqrt{30}}\left(10\overline
{\bf{\Psi}}_{aL}\gamma^A\partial_A{\bf A}_b-\overline
{\bf{\Psi}}_{aijL}\gamma^A\partial_A{\bf A}_b^{ij}+2\overline
{\bf{\Psi}}_{aL}^i\gamma^A\partial_A{\bf A}_{bi}\right)\Lambda^{'}_{L}\nonumber\\
+\frac{1}{2\sqrt 3}\left(\overline
{\bf{\Psi}}_{aL}^i\gamma^A\partial_A{\bf
A}_b\right)\Lambda^{'}_{iL}+\frac{1}{2\sqrt 3}\left(\overline
{\bf{\Psi}}_{aL}\gamma^A\partial_A{\bf
A}_{bi}\right)\Lambda^{'i}_{L}\nonumber\\
+\frac{1}{48}\left(-6\overline
{\bf{\Psi}}_{almL}^i\gamma^A\partial_A{\bf
A}_b+\epsilon_{ijklm}\overline
{\bf{\Psi}}_{aL}^i\gamma^A\partial_A{\bf
A}_b^{jk}\right)\Lambda^{'lm}_{L}\nonumber\\
 +\frac{1}{48}\left(-6\overline
{\bf{\Psi}}_{aL}\gamma^A\partial_A{\bf
A}_b^{lm}+\epsilon^{ijklm}\overline
{\bf{\Psi}}_{aijL}\gamma^A\partial_A{\bf
A}_{bk}\right)\Lambda^{'}_{lmL}\nonumber\\
 +\frac{1}{12}\left(-3\overline
{\bf{\Psi}}_{aL}^j\gamma^A\partial_A{\bf A}_b^{i}+\overline
{\bf{\Psi}}_{aikL}\gamma^A\partial_A{\bf
A}_{b}^{kj}\right)\Lambda^{'i}_{jL}\nonumber\\
+\frac{1}{12\sqrt 3}\left(\epsilon_{ijklm}\overline
{\bf{\Psi}}_{aL}^i\gamma^A\partial_A{\bf
A}_{b}^{jn}\right)\Lambda^{'klm}_{nL}-\frac{1}{12\sqrt 3
}\left(\epsilon^{ijklm}\overline
{\bf{\Psi}}_{ainL}\gamma^A\partial_A{\bf
A}_{bj}\right)\Lambda^{'n}_{klmL}\nonumber\\
-\frac{1}{8\sqrt 3}\left(\overline
{\bf{\Psi}}_{aijL}\gamma^A\partial_A{\bf
A}_{b}^{kl}\right)\Lambda^{'ij}_{klL}]
\end{eqnarray}
\begin{eqnarray}
B_{\mathtt{WY}}{\cal V}_{\mathtt{YZ}}^A{\cal
V}_{A\mathtt{ZW}}=B_{\mu\nu\rho\sigma}{\cal
V}_{\rho\sigma\lambda\tau}^A{\cal V}_{A\lambda\tau\mu\nu}\nonumber
=B_{\mu\nu\rho\sigma}{\mathit V}_{\mu\nu\rho\sigma}\nonumber\\
=\frac{1}{16}\{B_{c_ic_jc_kc_l}{\mathit V}_{{\bar c}_i{\bar
c}_j{\bar c}_k{\bar c}_l}+B_{{\bar c}_i{\bar c}_j{\bar c}_k{\bar
c}_l}{\mathit V}_{c_ic_jc_kc_l}+4B_{c_ic_jc_k{\bar c}_l}{\mathit
V}_{{\bar c}_i{\bar c}_j{\bar c}_k c_l}\nonumber\\
+4B_{{\bar c}_i{\bar c}_j{\bar c}_k c_l}{\mathit
V}_{c_ic_jc_k{\bar c}_l}+6B_{c_ic_j{\bar c}_k{\bar c}_l}{\mathit
V}_{{\bar c}_i{\bar c}_jc_k c_l}\}
\end{eqnarray}
 where${\mathit V}_{\mu\nu\rho\sigma}={\cal
V}_{\rho\sigma\lambda\tau}^A{\cal V}_{A\lambda\tau\mu\nu}$.
Thus we find
\begin{eqnarray}
B_{c_ic_jc_kc_l}{\mathit V}_{{\bar c}_i{\bar c}_j{\bar c}_k{\bar
c}_l}=\frac{1}{2}B_{c_ic_jc_kc_l}\left({\cal V}_{A{\bar c}_i{\bar
c}_jc_mc_n}{\cal V}_{{\bar c}_m{\bar c}_n{\bar c}_k{\bar c}_l}^A
+{\cal V}_{A{\bar c}_i{\bar c}_j{\bar c}_mc_n}{\cal V}_{c_m{\bar
c}_n{\bar c}_k{\bar c}_l}^A \right)\nonumber\\
=\frac{1}{144}B_i{\cal V}_A^j{\cal V}_j^{Ai}-\frac{1}{480}B_i{\cal
V}_A^i{\cal V}^{A}-\frac{1}{48}\epsilon^{ijklm}B_i{\cal
V}_{Ajkn}^p{\cal V}_{plm}^{An}\nonumber\\
+\frac{1}{36}\epsilon^{ijklm}B_i{\cal V}_{Ajkl}^n{\cal V}_{nm}^{A}
-\frac{1}{144}\epsilon^{ijklm}B_i{\cal V}_{Ajk}{\cal V}_{lm}^{A}
\end{eqnarray}
\begin{eqnarray}
B_{{\bar c}_i{\bar c}_j{\bar c}_k{\bar c}_l}{\mathit
V}_{c_ic_jc_kc_l}=\frac{1}{2}B_{{\bar c}_i{\bar c}_j{\bar
c}_k{\bar c}_l}\left({\cal V}_{A c_i c_jc_mc_n}{\cal V}_{{\bar
c}_m{\bar c}_n c_k c_l}^A +{\cal V}_{Ac_i c_j{\bar c}_mc_n}{\cal
V}_{c_m{\bar
c}_nc_kc_l}^A \right)\nonumber\\
=\frac{1}{144}B^i{\cal V}_{Aj}{\cal
V}_i^{Aj}-\frac{1}{480}B^i{\cal V}_{Ai}{\cal
V}^{A}-\frac{1}{48}\epsilon_{ijklm}B^i{\cal
V}^{jkn}_{Ap}{\cal V}^{Aplm}_{n}\nonumber\\
+\frac{1}{36}\epsilon_{ijklm}B^i{\cal V}^{jkl}_{An}{\cal V}^{Anm}
-\frac{1}{144}\epsilon_{ijklm}B^i{\cal V}^{ij}_A{\cal V}^{Akl}
\end{eqnarray}
\begin{eqnarray}
4B_{c_ic_jc_k{\bar c}_l}{\mathit V}_{{\bar c}_i{\bar c}_j{\bar
c}_k c_l}=B_{c_ic_jc_k{\bar c}_l}({\cal V}_{A {\bar c}_i{\bar
c}_jc_mc_n}{\cal V}_{{\bar c}_m{\bar c}_n {\bar c}_k c_l}^A +{\cal
V}_{A{\bar c}_i{\bar c}_j{\bar c}_m{\bar c_n}}{\cal V}_{c_m
c_n{\bar c}_kc_l}^A\nonumber\\
+2{\cal V}_{A{\bar c}_i{\bar c}_j{\bar c}_m c_n}{\cal V}_{c_m
{\bar c}_n{\bar c}_kc_l}^A)\nonumber\\
=B^{ijk}_l{\cal V}_{Aij}^{mn}{\cal V}_{mnk}^{Al}+2B^{ijk}_l{\cal
V}_{Akn}^{lm}{\cal V}_{mij}^{An}-\frac{2}{3}B^{ijk}_l{\cal
V}_{Ajk}^{lm}{\cal V}_{mi}^{A}\nonumber\\
-\frac{2}{3}B^{ijk}_l{\cal V}_{Ajkm}^{l}{\cal
V}_{i}^{Am}+\frac{2}{3}B^{ijk}_l{\cal V}_{Aijk}^{m}{\cal
V}_{m}^{Al}+\frac{2}{9}B^{ijk}_l{\cal V}_{Ajk}{\cal
V}_{i}^{Al}\nonumber\\
-\frac{2}{3}B^{ij}{\cal V}_{Ajlm}^{k}{\cal
V}_{ki}^{Alm}-\frac{2}{3}B^{ij}{\cal V}_{Aijl}^{k}{\cal
V}_{k}^{Al}+\frac{7}{9}B^{ij}{\cal V}_{Aij}^{kl}{\cal
V}_{kl}^{A}\nonumber\\
+\frac{28}{27}B^{ij}{\cal V}_{Ajk}{\cal
V}_{i}^{Ak}-\frac{1}{10}B^{ij}{\cal V}_{Aij}{\cal
V}^{A}-\frac{1}{24}\epsilon_{ijklm}B^{ijn}_p{\cal V}_{A}^{k}{\cal
V}_{n}^{Almp}\nonumber\\
+\frac{1}{36}\epsilon_{ijklm}B^{ijk}_n{\cal V}_{A}^{l}{\cal
V}^{Amn}-\frac{1}{36}\epsilon_{ijklm}B^{in}{\cal V}_{A}^{j}{\cal
V}^{Aklm}_n+\frac{1}{36}\epsilon_{ijklm}B^{ij}{\cal
V}_{A}^{k}{\cal V}^{Alm}
\end{eqnarray}
\begin{eqnarray}
4B_{{\bar c}_i{\bar c}_j{\bar c}_k c_l}{\mathit V}_{c_ic_jc_k{\bar
c}_l}=B_{{\bar c}_i{\bar c}_j{\bar c}_k c_l}({\cal V}_{A c_i
c_jc_mc_n}{\cal V}_{{\bar c}_m{\bar c}_n c_k {\bar c}_l}^A +{\cal
V}_{A c_ic_j{\bar c}_m{\bar c_n}}{\cal V}_{c_m
c_nc_k{\bar c}_l}^A\nonumber\\
+2{\cal V}_{Ac_ic_j{\bar c}_m c_n}{\cal V}_{c_m
{\bar c}_n c_k{\bar c}_l}^A)\nonumber\\
=B_{ijk}^l{\cal V}^{Aij}_{mn}{\cal V}^{mnk}_{Al}+2B_{ijk}^l{\cal
V}^{Akn}_{lm}{\cal V}^{mij}_{An}-\frac{2}{3}B_{ijk}^l{\cal
V}^{Ajk}_{lm}{\cal V}^{mi}_{A}\nonumber\\
-\frac{2}{3}B_{ijk}^l{\cal V}^{Ajkm}_{l}{\cal
V}^{i}_{Am}+\frac{2}{3}B_{ijk}^l{\cal V}^{Aijk}_{m}{\cal
V}^{m}_{Al}+\frac{2}{9}B_{ijk}^l{\cal V}^{Ajk}{\cal
V}^{i}_{Al}\nonumber\\
-\frac{2}{3}B_{ij}{\cal V}^{Ajlm}_{k}{\cal
V}^{ki}_{Alm}-\frac{2}{3}B_{ij}{\cal V}^{Aijl}_{k}{\cal
V}^{k}_{Al}+\frac{7}{9}B_{ij}{\cal V}^{Aij}_{kl}{\cal
V}^{kl}_{A}\nonumber\\
+\frac{28}{27}B_{ij}{\cal V}^{Ajk}{\cal
V}^{i}_{Ak}-\frac{1}{10}B_{ij}{\cal V}^{Aij}{\cal
V}_{A}-\frac{1}{24}\epsilon^{ijklm}B_{ijn}^p{\cal V}^{A}_{k}{\cal
V}^{n}_{Almp}\nonumber\\
+\frac{1}{36}\epsilon^{ijklm}B_{ijk}^n{\cal V}^{A}_{l}{\cal
V}_{Amn}-\frac{1}{36}\epsilon^{ijklm}B_{in}{\cal V}^{A}_{j}{\cal
V}_{Aklm}^n+\frac{1}{36}\epsilon^{ijklm}B_{ij}{\cal
V}^{A}_{k}{\cal V}_{Alm}
\end{eqnarray}
\begin{eqnarray}
6B_{c_ic_j{\bar c}_k{\bar c}_l}{\mathit V}_{{\bar c}_i{\bar
c}_jc_k c_l}=B_{c_ic_j{\bar c}_k{\bar c}_l}(\frac{3}{2}{\cal V}_{A
{\bar c}_i {\bar c}_jc_mc_n}{\cal V}_{{\bar c}_m{\bar c}_n c_k
c_l}^A +\frac{3}{2}{\cal V}_{A {\bar c}_i{\bar c}_j{\bar c}_m{\bar
c_n}}{\cal V}_{c_m
c_nc_k c_l}^A\nonumber\\
+3{\cal V}_{A{\bar c}_i{\bar c}_j{\bar c}_m c_n}{\cal V}_{c_m
{\bar c}_n c_kc_l}^A)\nonumber\\
=\frac{3}{2}B_{kl}^{ij}{\cal V}^{Amn}_{ij}{\cal
V}^{kl}_{Amn}+2B_{kl}^{ij}{\cal V}^{Aln}_{ij}{\cal
V}^{k}_{An}-\frac{3}{20}B_{kl}^{ij}{\cal V}^{Akl}_{ij}{\cal
V}_{A}\nonumber\\
+2B_{kl}^{ij}{\cal V}^{Akl}_{jm}{\cal
V}^{m}_{Ai}+\frac{4}{3}B_{kl}^{ij}{\cal V}^{Al}_{j}{\cal
V}^{k}_{Ai}+2B_{j}^{i}{\cal V}^{Akl}_{im}{\cal
V}_{Akl}^{mj}\nonumber\\
+\frac{4}{3}B_{j}^{i}{\cal V}^{Ajk}_{il}{\cal
V}^{l}_{Ak}-\frac{4}{9}B_{j}^{i}{\cal V}^{Aj}_{k}{\cal
V}^{k}_{Ai}-\frac{2}{15}B_{j}^{i}{\cal V}^{Aj}_{i}{\cal
V}_{A}\nonumber\\
-\frac{3}{20}B{\cal V}^{Aij}_{kl}{\cal
V}^{kl}_{Aij}-\frac{2}{15}B{\cal V}^{Ai}_{j}{\cal
V}^{j}_{Ai}-\frac{3}{200}B{\cal V}^{A}{\cal
V}_{A}\nonumber\\
+\frac{1}{48}B_{j}^{i}{\cal V}^{Aj}{\cal
V}_{Ai}-\frac{1}{480}B{\cal V}^{Ai}{\cal V}_{Ai}+3B^{ij}_{kl}{\cal
V}^{An}_{ijm}{\cal
V}_{An}^{mkl}\nonumber\\
+2B_{kl}^{ij}{\cal V}^{Al}_{ijm}{\cal
V}_{A}^{mk}+2B^{ij}_{kl}{\cal V}^{Aklm}_j{\cal
V}_{Ami}+\frac{1}{3}B^{ij}_{kl}{\cal V}^{A}_{ij}{\cal
V}_{A}^{kl}\nonumber\\
-4B_{m}^{i}{\cal V}^{Aj}_{ikl}{\cal
V}_{Aj}^{klm}-\frac{4}{3}B^{i}_{j}{\cal V}^{Aj}_{ikl}{\cal
V}_{A}^{kl}-\frac{4}{3}B^{i}_{j}{\cal V}^{Ajkl}_{i}{\cal
V}_{Akl}\nonumber\\
+2B_{k}^{i}{\cal V}^{A}_{ij}{\cal V}_{A}^{jk}-\frac{3}{10}B{\cal
V}^{Ai}_{jkl}{\cal V}_{Ai}^{jkl}-\frac{1}{6}B{\cal
V}^{A}_{ij}{\cal V}_{A}^{ij}
\end{eqnarray}

\section{Appendix E}
In this appendix we give the complete supersymmetric couplings
containing the singlet of SO(10).
\begin{equation}
{\mathsf L}={\mathsf L}_{V}^{^{(1~K.E.)}} +{\mathsf
L}_{V+\Phi}^{^{(1~Interaction)}}+{\mathsf L}_{\mathsf W}^{(1)}
\end{equation}
where
\begin{equation}
{\mathsf L}_{V}^{^{(1~K.E.)}}=\frac{1}{64}\left[\widehat{\cal
W}^{\tilde{\alpha}} \widehat{\cal
W}_{\tilde{\alpha}}|_{\theta^2}+\widehat{\overline{\cal
W}}_{\dot{\tilde{\alpha}}}\widehat{\overline{\cal
W}}^{\dot{\tilde{\alpha}}}|_{\bar {\theta}^2}\right]
\end{equation}
\begin{equation}
\widehat{\cal W}^{\tilde{\alpha}} =\overline {\mathsf D}^2\mathsf
D_{\tilde{\alpha}}\widehat{\mathsf V}
\end{equation}
Thus we have
\begin{eqnarray}
{\mathsf L}_{V}^{^{(1~K.E.)}}=-\frac{1}{4}{\cal V}_{AB}{\cal
V}^{AB}-\frac{i}{2}\overline {\Lambda}\gamma^A{\cal
D}_A\Lambda+{\mathsf L}_{(1)auxiliary}^{(1)}\nonumber\\
{\cal V}^{AB}=\partial^A{\cal V}^B_{\mu\nu}-\partial^B{\cal
V}^A_{\mu\nu}\nonumber\\
{\mathsf L}_{(1)auxiliary}^{(1)}=\frac{1}{2}D^2\nonumber\\
\Lambda=\left(\matrix{\lambda_{\tilde{\alpha}}\cr
\overline{\lambda}^{\dot{\tilde\alpha}}}\right)
\end{eqnarray}
Further,
\begin{eqnarray}
{\mathsf
L}_{V+\Phi}^{^{(1~Interaction)}}=h_{ab}^{^{(1+)}}<\widehat{\Phi}_{(+)a}|
e^{{\mathsf g}^{^{(1)}}q^{^{(+)}}\widehat{{\mathsf V}}
}|\widehat{\Phi}_{(+)b}>|_{\theta^2\bar
{\theta}^2}\nonumber\\
+h_{ab}^{^{(1-)}}<\widehat{\Phi}_{(-)a}|e^{{\mathsf
g}^{^{(1)}}q^{^{(-)}}\widehat{{\mathsf V}}
}|\widehat{\Phi}_{(-)b}>|_{\theta^2\bar {\theta}^2}
\end{eqnarray}
where $q^{^{(\pm)}}$ are the U(1) charges. Expanding $e^{{\mathsf
g}^{^{(1)}}q^{^{(\pm)}}}$ we have
\begin{eqnarray}
{\mathsf
L}_{V+\Phi}^{^{(1~Interaction)}}=h_{ab}^{^{(1+)}}[<\widehat{\Phi}_{(+)a}|
\widehat{\Phi}_{(+)b}>+{\mathsf
g}^{^{(1)}}q^{^{(+)}}<\widehat{\Phi}_{(+)a}|{\widehat{\mathsf
V}}|\widehat{\Phi}_{(+)b}>\nonumber\\
+\frac{1}{2}{\mathsf g}^{^{(1)}2}q^{^{(+)}2}
<\widehat{\Phi}_{(+)a}|{\widehat{\mathsf
V}}^2|\widehat{\Phi}_{(+)b}>]|_{\theta^2\bar {\theta}^2}
+h_{ab}^{^{(1-)}}[<\widehat{\Phi}_{(-)a}|
\widehat{\Phi}_{(-)b}>\nonumber\\
+{\mathsf
g}^{^{(1)}}q^{^{(-)}}<\widehat{\Phi}_{(-)a}|{\widehat{\mathsf
V}}|\widehat{\Phi}_{(-)b}> +\frac{1}{2}{\mathsf
g}^{^{(1)}2}q^{^{(-)}2}<\widehat{\Phi}_{(-)a}|{\widehat{\mathsf
V}}^2|\widehat{\Phi}_{(-)b}>]|_{\theta^2\bar {\theta}^2}
\end{eqnarray}
where the quantities entering Eq.(47) are determined by Eqs.(48)
-(54) below
\begin{eqnarray}
h_{ab}^{^{(1+)}}<\widehat{\Phi}_{(+)a}|
\widehat{\Phi}_{(+)b}>|_{\theta^2\bar
{\theta}^2}=h_{ab}^{^{(1+)}}[-\partial_A{\bf
A}^{\dagger}_{(+)a}\partial^A{\bf A}_{(+)b}
-\partial_A{\bf A}^{i\dagger}_{(+)a}\partial^A{\bf A}_{(+)bi}\nonumber\\
-\partial_A{\bf A}^{\dagger}_{(+)aij}\partial^A{\bf
A}_{(+)b}^{ij}-i\overline{\bf
{\Psi}}_{(+)aL}\gamma^A\partial_A{\bf
{\Psi}}_{(+)bL}\nonumber\\
-i\overline{\bf {\Psi}}_{(+)aL}^i\gamma^A\partial_A{\bf
{\Psi}}_{(+)biL}-i\overline{\bf
{\Psi}}_{(+)aijL}\gamma^A\partial_A{\bf
{\Psi}}_{(+)bL}^{ij}]
+{\mathsf L}_{(2)auxiliary}^{(1)}\nonumber\\
{\mathsf
L}_{(2)auxiliary}^{(1)}=h_{ab}^{^{(1+)}}<F_{(+)a}|F_{(+)b}>
\end{eqnarray}
\begin{eqnarray}
h_{ab}^{^{(1-)}}<\widehat{\Phi}_{(-)a}|
\widehat{\Phi}_{(-)b}>|_{\theta^2\bar
{\theta}^2}=h_{ab}^{^{(1-)}}[-\partial_A{\bf
A}^{\dagger}_{(-)a}\partial^A{\bf A}_{(-)b}
-\partial_A{\bf A}^{\dagger}_{(-)ai}\partial^A{\bf A}_{(-)b}^i\nonumber\\
-\partial_A{\bf A}^{ij\dagger}_{(-)a}\partial^A{\bf
A}_{(-)bij}-i\overline{\bf {\Psi}}_{(-)aL}\gamma^A\partial_A{\bf
{\Psi}}_{(-)bL}\nonumber\\
-i\overline{\bf {\Psi}}_{(-)aiL}\gamma^A\partial_A{\bf
{\Psi}}_{(-)bL}^i-i\overline{\bf
{\Psi}}_{(-)aL}^{ij}\gamma^A\partial_A{\bf {\Psi}}_{(-)bijL}]
+{\mathsf L}_{(3)auxiliary}^{(1)}\nonumber\\
{\mathsf
L}_{(3)auxiliary}^{(1)}=h_{ab}^{^{(1-)}}<F_{(-)a}|F_{(-)b}>
\end{eqnarray}

\begin{equation}
{\bf {\Psi}}_{(\pm)a}=\left(\matrix{{\bf
\psi}_{(\pm)a\tilde{\alpha}}\cr
\overline{{\bf\psi}}^{\dot{\tilde\alpha}}_{(\pm)a}}\right)
\end{equation}
\begin{eqnarray}
h_{ab}^{^{(1+)}}{\mathsf
g}^{^{(1)}}q^{^{(+)}}<\widehat{\Phi}_{(+)a}|{\widehat{\mathsf
V}}|\widehat{\Phi}_{(+)b}>|_{\theta^2\bar {\theta}^2}~~~~~\nonumber\\
=h_{ab}^{^{(1+)}}{\mathsf g}^{^{(1)}}q^{^{(+)}}\{[\frac{
1}{2}\left(i{\bf
A}^{\dagger}_{(+)a}\stackrel{\leftrightarrow}{\partial}_A {\bf
A}_{(+)b}-\overline{\bf {\Psi}}_{(+)aL}\gamma_A{\bf
{\Psi}}_{(+)bL}\right)\nonumber\\
+\frac{1}{4}\left(i{\bf
A}^{\dagger}_{(+)aij}\stackrel{\leftrightarrow}{\partial}_A{\bf
A}_{(+)b}^{ij}-\overline{\bf {\Psi}}_{(+)aijL}\gamma_A{\bf
{\Psi}}_{(+)bL}^{ij}\right)\nonumber\\
+\frac{1}{2}\left(i{\bf
A}^{i\dagger}_{(+)a}\stackrel{\leftrightarrow}{\partial}_A{\bf
A}_{(+)bi}-\overline{\bf {\Psi}}_{(+)aL}^i\gamma_A{\bf
{\Psi}}_{(+)biL}\right)]{\cal V}^{A}\nonumber\\
+\frac{i}{\sqrt {2}}\left[{\bf A}^{\dagger}_{(+)a}\overline{\bf
{\Psi}}_{(+)bR}+\frac{1}{2}{\bf A}^{\dagger}_{(+)aij}\overline{\bf
{\Psi}}_{(+)bR}^{ij}+{\bf A}^{i\dagger}_{(+)a}\overline{\bf
{\Psi}}_{(+)biR}\right]{\Lambda}_{L}\nonumber\\
-\frac{i}{\sqrt{2}}\left[\overline{\bf {\Psi}}_{(+)aL}{\bf
A}_{(+)b}+\frac{1}{12}\overline{\bf {\Psi}}_{(+)aijL}{\bf
A}^{ij}_{(+)b}+\overline{\bf {\Psi}}_{(+)aL}^i{\bf
A}_{(+)bi}\right]{\Lambda}_{R}\}
+{\mathsf L}_{(4)auxiliary}^{(1)}\nonumber\\
{\mathsf L}_{(4)auxiliary}^{(1)}=\frac{h_{ab}^{^{(1+)}}{\mathsf
g}^{^{(1)}}q^{^{(+)}}}{2}<A_{(+)a}|A_{(+)b}>D
\end{eqnarray}
\begin{eqnarray}
h_{ab}^{^{(1-)}}{\mathsf
g}^{^{(1)}}q^{^{(-)}}<\widehat{\Phi}_{(-)a}|{\widehat{\mathsf
V}}|\widehat{\Phi}_{(-)b}>|_{\theta^2\bar
{\theta}^2}~~~~~\nonumber\\
 =h_{ab}^{^{(1-)}}{\mathsf
g}^{^{(1)}}q^{^{(-)}}\{[\frac{1}{2}\left(i{\bf
A}^{\dagger}_{(-)a}\stackrel{\leftrightarrow}{\partial}_A {\bf
A}_{(-)b}-\overline{\bf {\Psi}}_{(-)aL}\gamma_A{\bf
{\Psi}}_{(-)bL}\right)\nonumber\\
+\frac{1}{4}\left({\bf
A}^{ij\dagger}_{(-)a}\stackrel{\leftrightarrow}{\partial}_A{\bf
A}_{(-)bij}+i\overline{\bf {\Psi}}_{(-)aL}^{ij}\gamma_A{\bf
{\Psi}}_{(-)bijL}\right)\nonumber\\
+\frac{1}{2}\left({\bf
A}^{\dagger}_{(-)ai}\stackrel{\leftrightarrow}{\partial}_A{\bf
A}_{(-)b}^{i}+i\overline{\bf {\Psi}}_{(-)aiL}\gamma_A{\bf
{\Psi}}_{(-)bL}^i\right)]{\cal V}^{A}\nonumber\\
+\frac{i}{\sqrt{2}}\left[{\bf A}^{\dagger}_{(-)a}\overline{\bf
{\Psi}}_{(-)bR}+\frac{1}{2}{\bf A}^{ij\dagger}_{(-)a}\overline{\bf
{\Psi}}_{(-)bijR}+{\bf A}^{\dagger}_{(-)ai}\overline{\bf
{\Psi}}_{(-)bR}^i\right]{\Lambda}_{L}\nonumber\\
-\frac{i}{\sqrt{2}}\left[\overline{\bf {\Psi}}_{(-)aL}{\bf
A}_{(-)b}+\frac{1}{2}\overline{\bf {\Psi}}_{(-)aL}^{ij}{\bf
A}_{(-)bij}+\overline{\bf {\Psi}}_{(-)aiL}{\bf
A}_{(-)b}^i\right]{\Lambda}_{R}\}
+{\mathsf L}_{(5)auxiliary}^{(1)}\nonumber\\
{\mathsf L}_{(5)auxiliary}^{(1)}=\frac{h_{ab}^{^{(1-)}}{\mathsf
g}^{^{(1)}}q^{^{(-)}}}{2}<A_{(-)a}|A_{(-)b}>D
\end{eqnarray}
\begin{eqnarray}
\frac{1}{2} h_{ab}^{^{(1+)}}{\mathsf
g}^{^{(1)}2}q^{^{(+)}2}<\widehat{\Phi}_{(+)a}|{\widehat{\mathsf
V}}^2|\widehat{\Phi}_{(+)b}>|_{\theta^2\bar
{\theta}^2}~~~~~\nonumber\\
=\frac{h_{ab}^{^{(1+)}}{\mathsf g}^{^{(1)}2}q^{^{(+)}2}}{4} [{\bf
A}^{\dagger}_{(+)a}{\bf A}_{(+)b} +\frac{1}{2}{\bf
A}^{\dagger}_{(+)aij}{\bf A}_{(+)b}^{ij}+{\bf
A}^{i\dagger}_{(+)a}{\bf A}_{(+)bi}]{\cal V}_A{\cal V}^{A}
\end{eqnarray}
\begin{eqnarray}
\frac{1}{2} h_{ab}^{^{(1-)}}{\mathsf
g}^{^{(1)}2}q^{^{(-)}2}<\widehat{\Phi}_{(-)a}|{\widehat{\mathsf
V}}^2|\widehat{\Phi}_{(-)b}>|_{\theta^2\bar
{\theta}^2}~~~~~~\nonumber\\
= \frac{h_{ab}^{^{(1-)}}{\mathsf g}^{^{(1)}2}q^{^{(-)}2}}{4} [{\bf
A}^{\dagger}_{(-)a}{\bf A}_{(-)b} +\frac{1}{2}{\bf
A}^{ij\dagger}_{(-)a}{\bf A}_{(-)bij}+{\bf
A}^{\dagger}_{(-)ai}{\bf A}_{(-)b}^i]{\cal V}_A{\cal V}^{A}
\end{eqnarray}
Finally, ${\mathsf L}_{\mathsf W}^{(1)}$ appearing in Eq.(42) is
given by
\begin{eqnarray}
{\mathsf L}_{\mathsf
W}^{(1)}=\mu_{ab}<\widehat{\Phi}_{(-)a}^*|B|\widehat{\Phi}_{(+)b}>|_{\theta^2}+
h.c.
\end{eqnarray}
Evaluation of Eq.(55) gives
\begin{eqnarray}
{\mathsf L}_{\mathsf W}^{(1)}=-i\mu_{ab}\left( \overline{\bf
{\Psi}}_{(-)aR}{\bf {\Psi}}_{(+)bL}+\overline{\bf
{\Psi}}_{(-)aR}^i{\bf {\Psi}}_{(+)biL}-\frac{1}{2}\overline{\bf
{\Psi}}_{(-)aijR}{\bf
{\Psi}}_{(+)bL}^{ij}\right)\nonumber\\
+i\mu_{ab}^{*}\left( \overline{\bf {\Psi}}_{(-)aL}{\bf
{\Psi}}_{(+)bR}+\overline{\bf {\Psi}}_{(-)aiL}{\bf
{\Psi}}_{(+)bR}^i-\frac{1}{2}\overline{\bf
{\Psi}}_{(-)aL}^{ij}{\bf {\Psi}}_{(+)bijR}\right)
+{\mathsf L}_{(6)auxiliary}^{(1)}\nonumber\\
{\mathsf L}_{(6)auxiliary}^{(1)}=i\mu_{ab}[{\bf F}_{(-)a}{\bf
A}_{(+)b}+{\bf A}_{(-)a}^{\bf{T}}{\bf F}_{(+)b}-\frac{1}{2}{\bf
F}_{(-)aij}{\bf A}_{(+)b}^{ij}\nonumber\\
-\frac{1}{2}{\bf A}_{(-)aij}^{\bf{T}}{\bf F}_{(+)b}^{ij} + {\bf
F}_{(-)a}^i{\bf A}_{(+)bi}+{\bf A}_{(-)a}^{i\bf{T}}{\bf
F}_{(+)bi}]
+h.c.
\end{eqnarray}
Elimination of the auxiliary fields ${\bf F}_{(\pm)}$ through
their field equations gives
\begin{eqnarray}
{\mathsf L}_{(2)auxiliary}^{(1)}+{\mathsf
L}_{(3)auxiliary}^{(1)}+{\mathsf
L}_{(6)auxiliary}^{(1)}~~~~~~~~~~~~~~~~~~~~~~~~~~~~~~~~~~~~~~~~~~~~~~~~~\nonumber\\
=-\left(\mu^*
[h^{^{(1-)}}]^{\bf{-1}}[h^{^{(1-)}}]^{\bf{T}}[h^{^{(1-)}}]^{\bf{-1}}\mu\right)_{ab}\left[{\bf
A}^{\dagger}_{(+)a}{\bf A}_{(+)b}+\frac{1}{4}{\bf
A}^{\dagger}_{(+)aij}{\bf A}_{(+)b}^{ij}+{\bf
A}^{i\dagger}_{(+)a}{\bf A}_{(+)bi}\right]\nonumber\\
-\left(\mu[h^{^{(1+)}}]^{\bf{-1T}}h^{^{(1+)}}[h^{^{(1+)}}]^{\bf{-1T}}\mu^*\right)_{ab}
\left[ {\bf A}^{\bf{T}}_{(-)a}{\bf A}_{(-)b}^*+\frac{1}{4}{\bf
A}^{\bf{T}}_{(-)aij}{\bf A}_{(-)b}^{ij*}+{\bf
A}^{i\bf{T}}_{(-)a}{\bf A}_{(-)bi}^*\right]
\end{eqnarray}
Similarly, after eliminating the field $D$ we get
\begin{eqnarray}
{\mathsf L}_{(1)auxiliary}^{(1)}+{\mathsf
L}_{(4)auxiliary}^{(1)}+{\mathsf
L}_{(5)auxiliary}^{(1)}~~~~~~~~~~~~~~~~~~~~~~~~~~~~~~~~~~~~~~~~~~~~~~~~~\nonumber\\
=-\frac{1}{8} {\mathsf
g}^{^{(1)}2}h_{ab}^{^{(1+)}}h_{cd}^{^{(1+)}}q^{^{(+)}2}<A_{(+)a}|A_{(+)b}><A_{(+)c}|
A_{(+)d}>\nonumber\\
-\frac{1}{8} {\mathsf
g}^{^{(1)}2}h_{ab}^{^{(1-)}}h_{cd}^{^{(1-)}}q^{^{(-)}2}<A_{(-)a}|A_{(-)b}><A_{(-)c}|A_{(-)d}>\nonumber\\
-\frac{1}{4} {\mathsf
g}^{^{(45)}2}h_{ab}^{^{(1+)}}h_{cd}^{^{(1-)}}q^{^{(+)}}q^{^{(-)}}<A_{(+)a}|A_{(+)b}><A_{(-)c}|A_{(-)d}>
\end{eqnarray}
\begin{eqnarray}
-\frac{1}{8} {\mathsf
g}^{^{(1)}2}h_{ab}^{^{(1+)}}h_{cd}^{^{(1+)}}q^{^{(+)}2}<A_{(+)a}|A_{(+)b}><A_{(+)c}|
A_{(+)d}>~~~~~~~~~~~~~~~~~~~\nonumber\\
=-\frac{1}{8} {\mathsf
g}^{^{(1)}2}h_{ab}^{^{(1+)}}h_{cd}^{^{(1+)}}q^{^{(+)}2}[{\bf
A}^{\dagger}_{(+)a}{\bf A}_{(+)b}{\bf A}^{\dagger}_{(+)c}{\bf
A}_{(+)d}+\frac{1}{4}{\bf A}^{\dagger}_{(+)aij}{\bf
A}_{(+)b}^{ij}{\bf A}^{\dagger}_{(+)ckl}{\bf
A}_{(+)d}^{kl}\nonumber\\
+{\bf A}^{i\dagger}_{(+)a}{\bf A}_{(+)bi}{\bf
A}^{j\dagger}_{(+)c}{\bf A}_{(+)dj}+{\bf A}^{\dagger}_{(+)a}{\bf
A}_{(+)b}{\bf A}^{\dagger}_{(+)cij}{\bf A}_{(+)d}^{ij}\nonumber\\
+2{\bf A}^{\dagger}_{(+)a}{\bf A}_{(+)b}{\bf
A}^{i\dagger}_{(+)c}{\bf A}_{(+)di}+{\bf A}^{i\dagger}_{(+)a}{\bf
A}_{(+)bi}{\bf A}^{\dagger}_{(+)cjk}{\bf A}_{(+)d}^{jk}]
\end{eqnarray}
\begin{eqnarray}
-\frac{1}{8} {\mathsf
g}^{^{(1)}2}h_{ab}^{^{(1-)}}h_{cd}^{^{(1-)}}q^{^{(-)}2}<A_{(-)a}|A_{(-)b}><A_{(-)c}|
A_{(-)d}>~~~~~~~~~~~~~~~~~~~\nonumber\\
=-\frac{1}{8} {\mathsf
g}^{^{(1)}2}h_{ab}^{^{(1-)}}h_{cd}^{^{(1-)}}q^{^{(-)}2}[{\bf
A}^{\dagger}_{(-)a}{\bf A}_{(-)b}{\bf A}^{\dagger}_{(-)c}{\bf
A}_{(-)d}+\frac{1}{4}{\bf A}^{ij\dagger}_{(-)a}{\bf
A}_{(-)bij}{\bf A}^{kl\dagger}_{(-)c}{\bf
A}_{(-)dkl}\nonumber\\
+{\bf A}^{\dagger}_{(-)ai}{\bf A}_{(-)b}^i{\bf
A}^{\dagger}_{(-)cj}{\bf A}_{(-)d}^j+{\bf A}^{\dagger}_{(-)a}{\bf
A}_{(-)b}{\bf A}^{ij\dagger}_{(-)c}{\bf A}_{(-)dij}\nonumber\\
+2{\bf A}^{\dagger}_{(-)a}{\bf A}_{(-)b}{\bf
A}^{\dagger}_{(-)ci}{\bf A}_{(-)d}^i+{\bf A}^{\dagger}_{(-)ai}{\bf
A}_{(-)b}^i{\bf A}^{jk\dagger}_{(-)c}{\bf A}_{(-)djk}]
\end{eqnarray}
\begin{eqnarray}
-\frac{1}{4} {\mathsf
g}^{^{(1)}2}h_{ab}^{^{(1+)}}h_{cd}^{^{(1-)}}q^{^{(+)}}q^{^{(-)}}<A_{(+)a}|A_{(+)b}><A_{(-)c}|
A_{(-)d}>~~~~~~~~~~~~~~~~~~~\nonumber\\
=-\frac{1}{4} {\mathsf
g}^{^{(1)}2}h_{ab}^{^{(1-)}}h_{cd}^{^{(1-)}}q^{^{(+)}}q^{^{(-)}}[{\bf
A}^{\dagger}_{(+)a}{\bf A}_{(+)b}{\bf A}^{\dagger}_{(-)c}{\bf
A}_{(-)d}+\frac{1}{2}{\bf A}^{\dagger}_{(+)a}{\bf A}_{(+)b}{\bf
A}^{ij\dagger}_{(-)c}{\bf A}_{(-)dij}\nonumber\\
+{\bf A}^{\dagger}_{(+)a}{\bf A}_{(+)b}{\bf
A}^{\dagger}_{(-)ci}{\bf A}_{(-)d}^i+\frac{1}{2}{\bf
A}^{\dagger}_{(+)aij}{\bf A}_{(+)b}^{ij}{\bf
A}^{\dagger}_{(-)c}{\bf A}_{(-)d}\nonumber\\
+\frac{1}{4}{\bf A}^{\dagger}_{(+)aij}{\bf A}_{(+)b}^{ij}{\bf
A}^{kl\dagger}_{(-)c}{\bf A}_{(-)dkl}+\frac{1}{2}{\bf
A}^{\dagger}_{(+)aij}{\bf A}_{(+)b}^{ij}{\bf
A}^{\dagger}_{(-)ck}{\bf A}_{(-)d}^k\nonumber\\
+{\bf A}^{i\dagger}_{(+)a}{\bf A}_{(+)bi}{\bf
A}^{\dagger}_{(-)c}{\bf A}_{(-)d}+\frac{1}{2}{\bf
A}^{i\dagger}_{(+)a}{\bf A}_{(+)bi}{\bf A}^{kl\dagger}_{(-)c}{\bf
A}_{(-)dkl} +{\bf A}^{i\dagger}_{(+)a}{\bf A}_{(+)bi}{\bf
A}^{\dagger}_{(-)cj}{\bf A}_{(-)d}^j]
\end{eqnarray}

\section{Appendix F}
In this appendix we give the complete supersymmetric vector couplings
for the 45-dimensional tensor of SO(10) in the Wess-Zumino gauge
\begin{equation}
{\mathsf L}^{45}={\mathsf L}_{V}^{^{(45~K.E.)}} +{\mathsf
L}_{V+\Phi}^{^{(45~Interaction)}}+{\mathsf L}_{\mathsf W}^{(45)}
\end{equation}
where
\begin{eqnarray}
{\mathsf L}_{V}^{^{(45~K.E.)}}=\frac{1}{64}\left[\widehat{\cal
W}^{\tilde{\alpha}} \widehat{\cal
W}_{\tilde{\alpha}}|_{\theta^2}+\widehat{\overline{\cal
W}}_{\dot{\tilde{\alpha}}}\widehat{\overline{\cal
W}}^{\dot{\tilde{\alpha}}}|_{\bar {\theta}^2}\right]\nonumber\\
 \widehat{\cal
W}_{\tilde{\alpha}}=\frac{1}{{\mathsf g}^{^{(45)}}}\overline
{\mathsf D}^2e^{-{\mathsf g}^{^{(45)}}\widehat{\mathsf
V}_{\mu\nu}M_{\mu\nu}}{\mathsf D}_{\tilde{\alpha}}e^{{\mathsf
g}^{^{(45)}}\widehat{\mathsf V}_{\rho\lambda}M_{\rho\lambda}}
\end{eqnarray}
where
$M_{\mu\nu}$ are the 45 generators in the vector (10-dimensional)
representation that satisfy the following Lie algebra
\begin{equation}
[M_{\alpha\beta},M_{\gamma\rho}]=-i\left(\delta_{\beta\gamma}M_{\alpha\rho}+\delta_{\alpha\rho}M_{\beta\gamma}
-\delta_{\alpha\gamma}M_{\beta\rho}-\delta_{\beta\rho}M_{\alpha\gamma}\right)
\end{equation}
and take on the form
\begin{equation}
\left(M_{\mu\nu}\right)_{\alpha\beta}=-i\left(\delta_{\mu\alpha}\delta_{\nu\beta}-\delta_{\mu\beta}\delta_{\nu\alpha}
\right)
\end{equation}
and $\widehat{\mathsf V}_{\mu\nu}$ in the Wess-Zumino gauge is
given by
\begin{eqnarray}
\widehat{\mathsf V}_{\mu\nu}=-\theta\sigma^A\bar {\theta}{\cal
V}_{A\mu\nu}  +i\theta^2\bar {\theta}\overline {\lambda}_{\mu\nu}
-i\bar {\theta}^2\theta\lambda_{\mu\nu} +\frac{1}{2}\theta^2\bar
{\theta}^2D_{\mu\nu}
\end{eqnarray}
 Finally we have
\begin{eqnarray}
{\mathsf L}_{V}^{^{(45~K.E.)}}=-\frac{1}{4}{\cal
V}_{AB\mu\nu}{\cal V}^{AB}_{\mu\nu}-\frac{i}{2}\overline
{\Lambda}_{\mu\nu}\gamma^A{\cal D}_A\Lambda_{\mu\nu}+{\mathsf
L}_{(1)auxiliary}^{(45)}\nonumber\\
{\cal V}_{\mu\nu}^{AB}=\partial^A{\cal
V}^B_{\mu\nu}-\partial^B{\cal V}^A_{\mu\nu}+{\mathsf
g}^{^{(45)}}\left({\cal V}^A_{\mu\alpha}{\cal
V}^B_{\alpha\nu}-{\cal V}^B_{\mu\alpha}{\cal
V}^A_{\alpha\nu}\right)\nonumber\\
{\cal D}^A=\partial^A+\frac{i{\mathsf
g}^{^{(45)}}}{2}M_{\mu\nu}{\cal V}_{\mu\nu}^A\nonumber\\
{\cal D}^A\Lambda_{\mu\nu}=\partial^A\Lambda_{\mu\nu}+{\mathsf
g}^{^{(45)}}\left({\cal V}^A_{\mu\alpha}\Lambda_{\alpha\nu}-{\cal
V}^A_{\nu\alpha}\Lambda_{\nu\alpha}\right)\nonumber\\
{\mathsf L}_{(1)auxiliary}^{(45)}=\frac{1}{2}D_{\mu\nu}D_{\mu\nu}\nonumber\\
\Lambda_{\mu\nu}=\left(\matrix{\lambda_{\tilde{\alpha}\mu\nu}\cr
\overline{\lambda}_{\mu\nu}^{\dot{\tilde\alpha}}}\right)
\end{eqnarray}
To normalize the SU(5) fields appearing in $-\frac{1}{4}{\cal
V}_{AB\mu\nu}{\cal V}^{AB}_{\mu\nu}$, we carry out a field
redefinition
\begin{eqnarray}
{\cal V}_A=2\sqrt{5}{\cal V}^{'}_A;~~~~~{\cal
V}^{j}_{Ai}=\sqrt{2}{\cal
V}^{'j}_{Ai};~~~~~
{\cal V}^{ij}_A=\sqrt{2}{\cal V}^{'ij}_A;~~~~
 {\cal V}_{Aij}=\sqrt{2}{\cal V}^{'}_{Aij}\nonumber\\
{\Lambda}=\sqrt{10}{\Lambda}^{'};~~~~~{\Lambda}^{j}_{i}=
\sqrt{2}{\Lambda}^{'j}_{i};~~~~~
{\Lambda}^{ij}=\sqrt{2}{\Lambda}^{'ij};~~~~
 {\Lambda}_{ij}=\sqrt{2}{\Lambda}^{'}_{ij}
\end{eqnarray}
so that
\begin{eqnarray}
-\frac{1}{4}{\cal V}^{AB} _{\mu\nu}{\cal V}
_{AB\mu\nu}=-\frac{1}{2}{\cal V}'_{AB}{\cal
V}^{'AB\dagger}-\frac{1}{2!}\frac{1}{2}{\cal V}^{'ij}_{AB}{\cal
V}^{'ABij\dagger} -\frac{1}{4}{\cal V}^{'i}_{ABj}{\cal V}^{'ABj}_i\nonumber\\
-\frac{i}{2}\overline {\Lambda}_{\mu\nu}\gamma^A{\cal
D}_A\Lambda_{\mu\nu}=-\frac{i}{2}\frac{1}{2!}\overline
{\Lambda}_{ij}^{'}\gamma^A{\cal
D}_A\Lambda_{ij}^{'}-\frac{i}{2}\frac{1}{2!}\overline
{\Lambda}^{'ij}\gamma^A{\cal
D}_A\Lambda^{'ij}\nonumber\\
-\frac{i}{2}\overline
{\Lambda}^{'i}_j\gamma^A{\cal
D}_A\Lambda^{'i}_j-\frac{i}{2}\overline {\Lambda}^{'}\gamma^A{\cal
D}_A\Lambda^{'}
\end{eqnarray}
Next we look at the second term in Eq.(62)
\begin{eqnarray}
{\mathsf
L}_{V+\Phi}^{^{(45~Interaction)}}=h_{ab}^{^{(45+)}}<\widehat{\Phi}_{(+)a}|
e^{\frac{1}{2!}{\mathsf g}^{^{(45)}}\widehat{{\mathsf
V}}_{\mu\nu}\Sigma_{\mu\nu}
}|\widehat{\Phi}_{(+)b}>|_{\theta^2\bar
{\theta}^2}\nonumber\\
+h_{ab}^{^{(45-)}}<\widehat{\Phi}_{(-)a}|e^{\frac{1}{2!}{\mathsf
g}^{^{(45)}}\widehat{{\mathsf V}}_{\mu\nu}\Sigma_{\mu\nu}
}|\widehat{\Phi}_{(-)b}>|_{\theta^2\bar {\theta}^2}
\end{eqnarray}
$\Sigma_{\mu\nu}$ being the 45 generators in the spinorial
representation. We find
\begin{eqnarray}
{\mathsf
L}_{V+\Phi}^{^{(45~Interaction)}}=h_{ab}^{^{(45+)}}[<\widehat{\Phi}_{(+)a}|
\widehat{\Phi}_{(+)b}>+\frac{1}{2}{\mathsf
g}^{^{(45)}}<\widehat{\Phi}_{(+)a}|{\widehat{\mathsf
V}}_{\mu\nu}\Sigma_{\mu\nu}|\widehat{\Phi}_{(+)b}>\nonumber\\
+\frac{1}{8}{\mathsf
g}^{^{(45)}2}<\widehat{\Phi}_{(+)a}|{\widehat{\mathsf
V}}_{\mu\nu}\Sigma_{\mu\nu}{\widehat{\mathsf
V}}_{\rho\lambda}\Sigma_{\rho\lambda}|\widehat{\Phi}_{(+)b}>]|_{\theta^2\bar
{\theta}^2} +h_{ab}^{^{(45-)}}[<\widehat{\Phi}_{(-)a}|
\widehat{\Phi}_{(-)b}>\nonumber\\
+\frac{1}{2}{\mathsf
g}^{^{(45)}}<\widehat{\Phi}_{(-)a}|{\widehat{\mathsf
V}}_{\mu\nu}\Sigma_{\mu\nu}|\widehat{\Phi}_{(-)b}>
+\frac{1}{8}{\mathsf
g}^{^{(45)}2}<\widehat{\Phi}_{(-)a}|{\widehat{\mathsf
V}}_{\mu\nu}\Sigma_{\mu\nu}{\widehat{\mathsf
V}}_{\rho\lambda}\Sigma_{\rho\lambda}|\widehat{\Phi}_{(-)b}>]|_{\theta^2\bar
{\theta}^2}
\end{eqnarray}
where
\begin{eqnarray}
h_{ab}^{^{(45+)}}<\widehat{\Phi}_{(+)a}|
\widehat{\Phi}_{(+)b}>|_{\theta^2\bar
{\theta}^2}=h_{ab}^{^{(45+)}}[-\partial_A{\bf
A}^{\dagger}_{(+)a}\partial^A{\bf A}_{(+)b}
-\partial_A{\bf A}^{i\dagger}_{(+)a}\partial^A{\bf A}_{(+)bi}\nonumber\\
-\partial_A{\bf A}^{\dagger}_{(+)aij}\partial^A{\bf
A}_{(+)b}^{ij}-i\overline{\bf
{\Psi}}_{(+)aL}\gamma^A\partial_A{\bf
{\Psi}}_{(+)bL}\nonumber\\
-i\overline{\bf {\Psi}}_{(+)aL}^i\gamma^A\partial_A{\bf
{\Psi}}_{(+)biL}-i\overline{\bf
{\Psi}}_{(+)aijL}\gamma^A\partial_A{\bf
{\Psi}}_{(+)bL}^{ij}]
+{\mathsf L}_{(2)auxiliary}^{(45)}\nonumber\\
{\mathsf
L}_{(2)auxiliary}^{(45)}=h_{ab}^{^{(45+)}}<F_{(+)a}|F_{(+)b}>
\end{eqnarray}
\begin{eqnarray}
h_{ab}^{^{(45-)}}<\widehat{\Phi}_{(-)a}|
\widehat{\Phi}_{(-)b}>|_{\theta^2\bar
{\theta}^2}=h_{ab}^{^{(45-)}}[-\partial_A{\bf
A}^{\dagger}_{(-)a}\partial^A{\bf A}_{(-)b}
-\partial_A{\bf A}^{\dagger}_{(-)ai}\partial^A{\bf A}_{(-)b}^i\nonumber\\
-\partial_A{\bf A}^{ij\dagger}_{(-)a}\partial^A{\bf
A}_{(-)bij}-i\overline{\bf {\Psi}}_{(-)aL}\gamma^A\partial_A{\bf
{\Psi}}_{(-)bL}\nonumber\\
-i\overline{\bf {\Psi}}_{(-)aiL}\gamma^A\partial_A{\bf
{\Psi}}_{(-)bL}^i-i\overline{\bf
{\Psi}}_{(-)aL}^{ij}\gamma^A\partial_A{\bf {\Psi}}_{(-)bijL}]
+{\mathsf L}_{(3)auxiliary}^{(45)}\nonumber\\
{\mathsf
L}_{(3)auxiliary}^{(45)}=h_{ab}^{^{(45-)}}<F_{(-)a}|F_{(-)b}>
\end{eqnarray}
\begin{equation}
{\bf {\Psi}}_{(\pm)a}=\left(\matrix{{\bf
\psi}_{(\pm)a\tilde{\alpha}}\cr
\overline{{\bf\psi}}^{\dot{\tilde\alpha}}_{(\pm)a}}\right)
\end{equation}
\begin{eqnarray}
\frac{1}{2}h_{ab}^{^{(45+)}}{\mathsf
g}^{^{(45)}}<\widehat{\Phi}_{(+)a}|{\widehat{\mathsf
V}}_{\mu\nu}\Sigma_{\mu\nu}|\widehat{\Phi}_{(+)b}>|_{\theta^2\bar
{\theta}^2}~~~~~\nonumber\\
=h_{ab}^{^{(45+)}}{\mathsf g}^{^{(45)}}\{[-\frac{\sqrt
5}{2}\left({\bf
A}^{\dagger}_{(+)a}\stackrel{\leftrightarrow}{\partial}_A {\bf
A}_{(+)b}+i\overline{\bf {\Psi}}_{(+)aL}\gamma_A{\bf
{\Psi}}_{(+)bL}\right)\nonumber\\
-\frac{1}{4\sqrt 5}\left({\bf
A}^{\dagger}_{(+)aij}\stackrel{\leftrightarrow}{\partial}_A{\bf
A}_{(+)b}^{ij}+i\overline{\bf {\Psi}}_{(+)aijL}\gamma_A{\bf
{\Psi}}_{(+)bL}^{ij}\right)\nonumber\\
+\frac{3}{2\sqrt 5}\left({\bf
A}^{i\dagger}_{(+)a}\stackrel{\leftrightarrow}{\partial}_A{\bf
A}_{(+)bi}+i\overline{\bf {\Psi}}_{(+)aL}^i\gamma_A{\bf
{\Psi}}_{(+)biL}\right)]{\cal V}^{'A}\nonumber\\
+[\frac{1}{2\sqrt 2}\left({\bf
A}^{\dagger}_{(+)alm}\stackrel{\leftrightarrow}{\partial}_A{\bf
A}_{(+)b}+i\overline{\bf {\Psi}}_{(+)almL}\gamma_A{\bf
{\Psi}}_{(+)bL}\right)\nonumber\\
+\frac{1}{4\sqrt 2}\epsilon_{ijklm}\left({\bf
A}^{i\dagger}_{(+)a}\stackrel{\leftrightarrow}{\partial}_A{\bf
A}_{(+)b}^{jk}+i\overline{\bf {\Psi}}_{(+)aL}^i\gamma_A{\bf
{\Psi}}_{(+)bL}^{jk}\right)]{\cal V}^{'Alm}\nonumber\\
+[-\frac{1}{2\sqrt 2}\left({\bf
A}^{\dagger}_{(+)a}\stackrel{\leftrightarrow}{\partial}_A{\bf
A}_{(+)b}^{lm}+i\overline{\bf {\Psi}}_{(+)aL}\gamma_A{\bf
{\Psi}}_{(+)bL}^{lm}\right)\nonumber\\
-\frac{1}{4\sqrt 2}\epsilon^{ijklm}\left({\bf
A}^{\dagger}_{(+)aij}\stackrel{\leftrightarrow}{\partial}_A{\bf
A}_{(+)bk}+i\overline{\bf {\Psi}}_{(+)aijL}\gamma_A{\bf
{\Psi}}_{(+)bkL}\right)]{\cal V}^{'A}_{lm}\nonumber\\
 +[-\frac{1}{\sqrt
2}\left({\bf
A}^{\dagger}_{(+)aik}\stackrel{\leftrightarrow}{\partial}_A{\bf
A}_{(+)b}^{kj}+i\overline{\bf {\Psi}}_{(+)aikL}\gamma_A{\bf
{\Psi}}_{(+)bL}^{kj}\right)\nonumber\\
-\frac{1}{\sqrt 2}\left({\bf
A}^{j\dagger}_{(+)a}\stackrel{\leftrightarrow}{\partial}_A{\bf
A}_{(+)bi}+i\overline{\bf {\Psi}}_{(+)aL}^j\gamma_A{\bf
{\Psi}}_{(+)biL}\right)]{\cal V}^{'iA}_j\nonumber\\
+\frac{\sqrt 5}{2}\left[-{\bf A}^{\dagger}_{(+)a}\overline{\bf
{\Psi}}_{(+)bR}-\frac{1}{10}{\bf
A}^{\dagger}_{(+)aij}\overline{\bf
{\Psi}}_{(+)bR}^{ij}+\frac{3}{5}{\bf
A}^{i\dagger}_{(+)a}\overline{\bf
{\Psi}}_{(+)biR}\right]{\Lambda}^{'}_{L}\nonumber\\
+\frac{1}{2}\left[{\bf A}^{\dagger}_{(+)alm}\overline{\bf
{\Psi}}_{(+)bR}+\frac{1}{2}\epsilon_{ijklm}{\bf
A}^{i\dagger}_{(+)a}\overline{\bf
{\Psi}}_{(+)bR}^{jk}\right]{\Lambda}^{'lm}_{L}\nonumber\\
-\frac{1}{2}\left[{\bf A}^{\dagger}_{(+)a}\overline{\bf
{\Psi}}_{(+)bR}^{lm}+\frac{1}{2}\epsilon^{ijklm}{\bf
A}^{\dagger}_{(+)aij}\overline{\bf
{\Psi}}_{(+)bkR}\right]{\Lambda}^{'}_{lmL}\nonumber\\
-\left[{\bf A}^{\dagger}_{(+)aik}\overline{\bf
{\Psi}}_{(+)bR}^{kj}+{\bf A}^{j\dagger}_{(+)a}\overline{\bf
{\Psi}}_{(+)biR}\right]{\Lambda}^{'i}_{jL}\nonumber\\
-\frac{\sqrt 5}{2}\left[-\overline{\bf {\Psi}}_{(+)aL}{\bf
A}_{(+)b}-\frac{1}{10}\overline{\bf {\Psi}}_{(+)aijL}{\bf
A}^{ij}_{(+)b}+\frac{3}{5}\overline{\bf {\Psi}}_{(+)aL}^i{\bf
A}_{(+)bi}\right]{\Lambda}^{'}_{R}\nonumber\\
-\frac{1}{2}\left[\overline{\bf {\Psi}}_{(+)almL}{\bf
A}_{(+)b}+\frac{1}{2}\epsilon_{ijklm}\overline{\bf
{\Psi}}_{(+)aL}^i{\bf A}^{jk}_{(+)b}\right]{\Lambda}^{'lm}_{R}\nonumber\\
+\frac{1}{2}\left[\overline{\bf {\Psi}}_{(+)aL}{\bf
A}_{(+)b}^{lm}+\frac{1}{2}\epsilon^{ijklm}\overline{\bf
{\Psi}}_{(+)aijL}{\bf A}_{(+)bk}\right]{\Lambda}^{'}_{lmR}\nonumber\\
+\left[\overline{\bf {\Psi}}_{(+)aikL}{\bf
A}_{(+)b}^{kj}+\overline{\bf {\Psi}}_{(+)aL}^j{\bf
A}_{(+)bi}\right]{\Lambda}^{'i}_{jR}\}
+{\mathsf L}_{(4)auxiliary}^{(45)}\nonumber\\
{\mathsf L}_{(4)auxiliary}^{(45)}=\frac{h_{ab}^{^{(45+)}}{\mathsf
g}^{^{(45)}}}{4}<A_{(+)a}|\Sigma_{\mu\nu}|A_{(+)b}>D_{\mu\nu}
\end{eqnarray}
where
\begin{equation}
{\bf
A}^{i\dagger}_{(+)a}\stackrel{\leftrightarrow}{\partial}_A{\bf
A}_{(+)bi}\stackrel{def}={\bf A}^{i\dagger}_{(+)a}{\partial}_A{\bf
A}_{(+)bi}-\left({\partial}_A{\bf A}^{i\dagger}_{(+)a}\right) {\bf
A}_{(+)bi}
\end{equation}
Similarly for $\overline{16}_{-}16_{-}-$ couplings we have
\begin{eqnarray}
\frac{1}{2}h_{ab}^{^{(45-)}}{\mathsf
g}^{^{(45)}}<\widehat{\Phi}_{(-)a}|{\widehat{\mathsf
V}}_{\mu\nu}\Sigma_{\mu\nu}|\widehat{\Phi}_{(-)b}>|_{\theta^2\bar
{\theta}^2}~~~~~\nonumber\\
 =h_{ab}^{^{(45-)}}{\mathsf
g}^{^{(45)}}\{[\frac{\sqrt 5}{2}\left({\bf
A}^{\dagger}_{(-)a}\stackrel{\leftrightarrow}{\partial}_A {\bf
A}_{(-)b}+i\overline{\bf {\Psi}}_{(-)aL}\gamma_A{\bf
{\Psi}}_{(-)bL}\right)\nonumber\\
+\frac{1}{4\sqrt 5}\left({\bf
A}^{ij\dagger}_{(-)a}\stackrel{\leftrightarrow}{\partial}_A{\bf
A}_{(-)bij}+i\overline{\bf {\Psi}}_{(-)aL}^{ij}\gamma_A{\bf
{\Psi}}_{(-)bijL}\right)\nonumber\\
-\frac{3}{2\sqrt 5}\left({\bf
A}^{\dagger}_{(-)ai}\stackrel{\leftrightarrow}{\partial}_A{\bf
A}_{(-)b}^{i}+i\overline{\bf {\Psi}}_{(-)aiL}\gamma_A{\bf
{\Psi}}_{(-)bL}^i\right)]{\cal V}^{'A}\nonumber\\
+[\frac{1}{2\sqrt 2}\left({\bf
A}^{\dagger}_{(-)a}\stackrel{\leftrightarrow}{\partial}_A{\bf
A}_{(-)blm}+i\overline{\bf {\Psi}}_{(-)aL}\gamma_A{\bf
{\Psi}}_{(-)blmL}\right)\nonumber\\
+\frac{1}{4\sqrt 2}\epsilon_{ijklm}\left({\bf
A}^{ij\dagger}_{(-)a}\stackrel{\leftrightarrow}{\partial}_A{\bf
A}_{(-)b}^{k}+i\overline{\bf {\Psi}}_{(-)aL}^{ij}\gamma_A{\bf
{\Psi}}_{(-)bL}^{k}\right)]{\cal V}^{'Alm}\nonumber\\
+[-\frac{1}{2\sqrt 2}\left({\bf
A}^{lm\dagger}_{(-)a}\stackrel{\leftrightarrow}{\partial}_A{\bf
A}_{(-)b}+i\overline{\bf {\Psi}}_{(-)aL}^{lm}\gamma_A{\bf
{\Psi}}_{(-)bL}\right)\nonumber\\
-\frac{1}{4\sqrt 2}\epsilon^{ijklm}\left({\bf
A}^{\dagger}_{(-)ai}\stackrel{\leftrightarrow}{\partial}_A{\bf
A}_{(-)bjk}+i\overline{\bf {\Psi}}_{(-)aiL}\gamma_A{\bf
{\Psi}}_{(-)bjkL}\right)]{\cal V}^{'A}_{lm}\nonumber\\
 +[\frac{1}{\sqrt
2}\left({\bf
A}^{jk\dagger}_{(-)a}\stackrel{\leftrightarrow}{\partial}_A{\bf
A}_{(-)bki}+i\overline{\bf {\Psi}}_{(-)aL}^{jk}\gamma_A{\bf
{\Psi}}_{(-)bkiL}\right)\nonumber\\
+\frac{1}{\sqrt 2}\left({\bf
A}^{\dagger}_{(-)ai}\stackrel{\leftrightarrow}{\partial}_A{\bf
A}_{(-)b}^{j}+i\overline{\bf {\Psi}}_{(-)aiL}\gamma_A{\bf
{\Psi}}_{(-)bL}^j\right)]{\cal V}^{'iA}_j\nonumber\\
-\frac{\sqrt 5}{2}\left[-{\bf A}^{\dagger}_{(-)a}\overline{\bf
{\Psi}}_{(-)bR}-\frac{1}{10}{\bf
A}^{ij\dagger}_{(-)a}\overline{\bf
{\Psi}}_{(-)bijR}+\frac{3}{5}{\bf
A}^{\dagger}_{(-)ai}\overline{\bf
{\Psi}}_{(-)bR}^i\right]{\Lambda}^{'}_{L}\nonumber\\
+\frac{1}{2}\left[{\bf A}^{\dagger}_{(-)a}\overline{\bf
{\Psi}}_{(-)blmR}+\frac{1}{2}\epsilon_{ijklm}{\bf
A}^{ij\dagger}_{(-)a}\overline{\bf
{\Psi}}_{(-)bR}^{k}\right]{\Lambda}^{'lm}_{L}\nonumber\\
-\frac{1}{2}\left[{\bf A}^{lm\dagger}_{(-)a}\overline{\bf
{\Psi}}_{(-)bR}+\frac{1}{2}\epsilon^{ijklm}{\bf
A}^{\dagger}_{(-)ai}\overline{\bf
{\Psi}}_{(-)bjkR}\right]{\Lambda}^{'}_{lmL}\nonumber\\
+\left[{\bf A}^{jk\dagger}_{(-)a}\overline{\bf
{\Psi}}_{(-)bkiR}+{\bf A}^{\dagger}_{(-)ai}\overline{\bf
{\Psi}}_{(-)bR}^j\right]{\Lambda}^{'i}_{jL}\nonumber\\
+\frac{\sqrt 5}{2}\left[-\overline{\bf {\Psi}}_{(-)aL}{\bf
A}_{(-)b}-\frac{1}{10}\overline{\bf {\Psi}}_{(-)aL}^{ij}{\bf
A}_{(-)bij}+\frac{3}{5}\overline{\bf {\Psi}}_{(-)aiL}{\bf
A}_{(-)b}^i\right]{\Lambda}^{'}_{R}\nonumber\\
-\frac{1}{2}\left[\overline{\bf {\Psi}}_{(-)aL}{\bf
A}_{(-)blm}+\frac{1}{2}\epsilon_{ijklm}\overline{\bf
{\Psi}}_{(-)aL}^{ij}{\bf A}^{k}_{(-)b}\right]{\Lambda}^{'lm}_{R}\nonumber\\
+\frac{1}{2}\left[\overline{\bf {\Psi}}_{(-)aL}^{lm}{\bf
A}_{(-)b}+\frac{1}{2}\epsilon^{ijklm}\overline{\bf
{\Psi}}_{(-)aiL}{\bf A}_{(-)bjk}\right]{\Lambda}^{'}_{lmR}\nonumber\\
-\left[\overline{\bf {\Psi}}_{(-)aL}^{jk}{\bf
A}_{(-)bki}+\overline{\bf {\Psi}}_{(-)aiL}{\bf
A}_{(-)b}^j\right]{\Lambda}^{'i}_{jR}\}
+{\mathsf L}_{(5)auxiliary}^{(45)}\nonumber\\
{\mathsf L}_{(5)auxiliary}^{(45)}=\frac{h_{ab}^{^{(45-)}}{\mathsf
g}^{^{(45)}}}{4}<A_{(-)a}|\Sigma_{\mu\nu}|A_{(-)b}>D_{\mu\nu}
\end{eqnarray}
Next we evaluate couplings to  $\overline{16}_+16_+$ of  matter
 which are quadratic in the vector
multiplet  fields.   We have
\begin{eqnarray}
\frac{1}{8} h_{ab}^{^{(45+)}}{\mathsf
g}^{^{(45)}2}<\widehat{\Phi}_{(+)a}|{\widehat{\mathsf
V}}_{\mu\nu}\Sigma_{\mu\nu}{\widehat{\mathsf
V}}_{\rho\lambda}\Sigma_{\rho\lambda}|\widehat{\Phi}_{(+)b}>|_{\theta^2\bar
{\theta}^2}\nonumber\\
= h_{ab}^{^{(45+)}}{\mathsf g}^{^{(45)}2}
 \{[-\frac{5}{4}[{\bf
A}^{\dagger}_{(+)a}{\bf A}_{(+)b} +\frac{1}{50}{\bf
A}^{\dagger}_{(+)aij}{\bf A}_{(+)b}^{ij}+\frac{9}{25}{\bf
A}^{i\dagger}_{(+)a}{\bf A}_{(+)bi}]{\cal V}^{'}_A{\cal V}^{'A}\nonumber\\
+\frac{1}{2}[-{\bf A}^{m\dagger}_{(+)a}{\bf
A}_{(+)bm}\delta_i^l\delta_j^k+\left({\bf
A}^{\dagger}_{(+)aim}{\bf A}_{(+)b}^{ml}-{\bf
A}^{l\dagger}_{(+)a}{\bf A}_{(+)bi}\right)\delta_j^k+{\bf
A}^{\dagger}_{(+)aij}{\bf A}_{(+)b}^{kl}]{\cal V}^{'i}_{Ak}{\cal
V}^{'Aj}_{l}\nonumber\\
+\frac{1}{4}[\left(2{\bf A}^{\dagger}_{(+)a}{\bf
A}_{(+)b}^{km}+\epsilon^{ijpkm}{\bf A}^{\dagger}_{(+)aij}{\bf
A}_{(+)bp}\right)\delta_n^l-2\epsilon^{ijklm}{\bf
A}^{\dagger}_{(+)ain}{\bf A}_{(+)bj}]{\cal V}^{'}_{Akl}{\cal
V}^{'An}_{m}\nonumber\\
+\frac{1}{4}[\left(-2{\bf A}^{\dagger}_{(+)akm}{\bf
A}_{(+)b}-\epsilon_{ijpkm}{\bf A}^{i\dagger}_{(+)a}{\bf
A}_{(+)b}^{jp}\right)\delta_l^n+2\epsilon_{ijklm}{\bf
A}^{i\dagger}_{(+)a}{\bf A}_{(+)b}^{nj}]{\cal V}^{'kl}_{A}{\cal
V}^{'Am}_{n}\nonumber\\
+\frac{1}{4\sqrt {10}}[-6{\bf A}^{\dagger}_{(+)a}{\bf
A}_{(+)b}^{lm}+\epsilon^{ijklm}{\bf A}^{\dagger}_{(+)aij}{\bf
A}_{(+)bk}]{\cal V}^{'}_{Alm}{\cal
V}^{'A}\nonumber\\
+\frac{1}{4\sqrt {10}}[6{\bf A}^{\dagger}_{(+)alm}{\bf
A}_{(+)b}-\epsilon_{ijklm}{\bf A}^{i\dagger}_{(+)a}{\bf
A}_{(+)b}^{jk}]{\cal V}^{'lm}_{A}{\cal
V}^{'A}\nonumber\\
+\frac{1}{\sqrt{10}}[3{\bf A}^{j\dagger}_{(+)a}{\bf
A}_{(+)bi}+5{\bf A}^{\dagger}_{(+)aik}{\bf A}_{(+)b}^{kj}]{\cal
V}^{'i}_{Aj}{\cal
V}^{'A}\nonumber\\
-\frac{1}{8}[\epsilon^{ijklm}{\bf A}^{\dagger}_{(+)a}{\bf
A}_{(+)bi}]{\cal V}^{'}_{Ajk}{\cal V}^{'A}_{lm}
-\frac{1}{8}[\epsilon_{ijklm}{\bf A}^{i\dagger}_{(+)a}{\bf
A}_{(+)b}]{\cal V}^{'jk}_{A}{\cal V}^{'Alm}\nonumber\\
+\frac{1}{8}[\left( 2{\bf A}^{\dagger}_{(+)a} {\bf A}_{(+)b}+{\bf
A}^{\dagger}_{(+)amn} {\bf
A}_{(+)b}^{mn}\right)\delta_i^k\delta_j^l +2{\bf
A}^{\dagger}_{(+)aij} {\bf A}_{(+)b}^{kl}\nonumber\\
+2\left({\bf A}^{\dagger}_{(+)aim} {\bf A}_{(+)b}^{ml}-{\bf
A}^{l\dagger}_{(+)a} {\bf A}_{(+)bi}\right)\delta_j^k ]{\cal
V}^{'ij}_{A}{\cal V}^{'A}_{kl}\}
\end{eqnarray}
Similarly couplings to  $\overline{16}_-16_-$ of  matter
 which are quadratic in the vector
multiplet  fields are given by
\begin{eqnarray}
\frac{1}{8} h_{ab}^{^{(45-)}}{\mathsf
g}^{^{(45)}2}<\widehat{\Phi}_{(-)a}|{\widehat{\mathsf
V}}_{\mu\nu}\Sigma_{\mu\nu}{\widehat{\mathsf
V}}_{\rho\lambda}\Sigma_{\rho\lambda}|\widehat{\Phi}_{(-)b}>|_{\theta^2\bar
{\theta}^2}\nonumber\\
= h_{ab}^{^{(45-)}}{\mathsf g}^{^{(45)}2}
 \{[-\frac{5}{4}[{\bf
A}^{\dagger}_{(-)a}{\bf A}_{(-)b} +\frac{1}{50}{\bf
A}^{ij\dagger}_{(-)a}{\bf A}_{(-)bij}+\frac{9}{25}{\bf
A}^{\dagger}_{(-)ai}{\bf A}_{(-)b}^i]{\cal V}^{'}_A{\cal V}^{'A}\nonumber\\
+\frac{1}{2}[\left({\bf A}^{im\dagger}_{(-)a}{\bf A}_{(-)bml}-{\bf
A}^{\dagger}_{(-)al}{\bf A}_{(-)b}^i\right)\delta_k^j-{\bf
A}^{ij\dagger}_{(-)a}{\bf A}_{(-)bkl}]{\cal V}^{'k}_{Ai}{\cal
V}^{'Al}_{j}\nonumber\\
+\frac{1}{4}[\left(2{\bf A}^{km\dagger}_{(-)a}{\bf
A}_{(-)b}+\epsilon^{ijpkm}{\bf A}^{\dagger}_{(-)ai}{\bf
A}_{(-)bjp}\right)\delta_n^l+\epsilon^{ijklm}{\bf
A}^{\dagger}_{(-)an}{\bf A}_{(-)bij}]{\cal V}^{'}_{Akl}{\cal
V}^{'An}_{m}\nonumber\\
-\frac{1}{4}[\left(2{\bf A}^{\dagger}_{(-)a}{\bf
A}_{(-)bkm}+\epsilon_{ijpkm}{\bf A}^{ij\dagger}_{(-)a}{\bf
A}_{(-)b}^{p}\right)\delta_l^n+\epsilon_{ijklm}{\bf
A}^{ij\dagger}_{(-)a}{\bf A}_{(-)b}^{n}]{\cal V}^{'kl}_{A}{\cal
V}^{'Am}_{n}\nonumber\\
+\frac{1}{4\sqrt {10}}[6{\bf A}^{lm\dagger}_{(-)a}{\bf
A}_{(-)b}-\epsilon^{ijklm}{\bf A}^{\dagger}_{(-)ai}{\bf
A}_{(-)bjk}]{\cal V}^{'}_{Alm}{\cal
V}^{'A}\nonumber\\
+\frac{1}{4\sqrt {10}}[-6{\bf A}^{\dagger}_{(-)a}{\bf
A}_{(-)blm}+\epsilon_{ijklm}{\bf A}^{ij\dagger}_{(-)a}{\bf
A}_{(-)b}^{k}]{\cal V}^{'lm}_{A}{\cal
V}^{'A}\nonumber\\
+\frac{1}{\sqrt{10}}[3{\bf A}^{\dagger}_{(-)ai}{\bf
A}_{(-)b^j}+{\bf A}^{jk\dagger}_{(-)a}{\bf A}_{(-)bki}]{\cal
V}^{'i}_{Aj}{\cal
V}^{'A}\nonumber\\
-\frac{1}{8}[\epsilon^{ijklm}{\bf A}^{\dagger}_{(-)ai}{\bf
A}_{(-)b}]{\cal V}^{'}_{Ajk}{\cal V}^{'A}_{lm}
-\frac{1}{8}[\epsilon_{ijklm}{\bf A}^{\dagger}_{(-)a}{\bf
A}_{(-)b}^i]{\cal V}^{'jk}_{A}{\cal V}^{'Alm}\nonumber\\
+\frac{1}{4}[\left({\bf A}^{\dagger}_{(-)a} {\bf A}_{(-)b}+{\bf
A}^{mn\dagger}_{(-)a}{\bf A}_{(-)bmn}+{\bf A}^{\dagger}_{(-)am}
{\bf A}_{(-)b}^m\right)\delta_k^i\delta_l^j+{\bf
A}^{ij\dagger}_{(-)a} {\bf A}_{(-)bkl}\nonumber\\
+\left(-3{\bf A}^{im\dagger}_{(-)a} {\bf A}_{(-)bml}+{\bf
A}^{\dagger}_{(-)al} {\bf A}_{(-)b}^i\right)\delta_k^j] {\cal
V}^{'}_{Aij}{\cal V}^{'Akl}\}
\end{eqnarray}
\begin{eqnarray}
{\mathsf L}_{\mathsf W}^{(45)}={\mathsf
W}(\widehat{\Phi}_{(+)},\widehat{\Phi}_{(-)})|_{\theta^2}+ h.c.
\end{eqnarray}
where
\begin{eqnarray}
{\mathsf
W}(\widehat{\Phi}_{(+)},\widehat{\Phi}_{(-)})=
\mu_{ab}<\widehat{\Phi}_{(-)a}^*|B|\widehat{\Phi}_{(+)b}>\nonumber\\
{\mathsf W}({\bf A}_{(+)},{\bf A}_{(-)}) =i\mu_{ab}\left({\bf
A}_{(-)a}^{\bf{T}}{\bf A}_{(+)b}-\frac{1}{2}{\bf
A}_{(-)aij}^{\bf{T}}{\bf A}_{(+)b}^{ij}+{\bf
A}_{(-)a}^{i\bf{T}}{\bf A}_{(+)bi}\right)
\end{eqnarray}
and where $\mu_{ab}$ is taken to be a symmetric tensor.
Thus we have
\begin{eqnarray}
{\mathsf L}_{\mathsf W}^{(45)}=-i\mu_{ab}\left( \overline{\bf
{\Psi}}_{(-)aR}{\bf {\Psi}}_{(+)bL}+\overline{\bf
{\Psi}}_{(-)aR}^i{\bf {\Psi}}_{(+)biL}-\frac{1}{2}\overline{\bf
{\Psi}}_{(-)aijR}{\bf
{\Psi}}_{(+)bL}^{ij}\right)\nonumber\\
+i\mu_{ab}^{*}\left( \overline{\bf {\Psi}}_{(-)aL}{\bf
{\Psi}}_{(+)bR}+\overline{\bf {\Psi}}_{(-)aiL}{\bf
{\Psi}}_{(+)bR}^i-\frac{1}{2}\overline{\bf
{\Psi}}_{(-)aL}^{ij}{\bf {\Psi}}_{(+)bijR}\right)
+{\mathsf L}_{(6)auxiliary}^{(45)}\nonumber\\
{\mathsf L}_{(6)auxiliary}^{(45)}=i\mu_{ab}[{\bf F}_{(-)a}{\bf
A}_{(+)b}+{\bf A}_{(-)a}^{\bf{T}}{\bf F}_{(+)b}-\frac{1}{2}{\bf
F}_{(-)aij}{\bf A}_{(+)b}^{ij}\nonumber\\
-\frac{1}{2}{\bf A}_{(-)aij}^{\bf{T}}{\bf F}_{(+)b}^{ij} + {\bf
F}_{(-)a}^i{\bf A}_{(+)bi}+{\bf A}_{(-)a}^{i\bf{T}}{\bf
F}_{(+)bi}]
+h.c.
\end{eqnarray}
Eliminating the fields, ${\bf F}_{(\pm)}$ through their field equations
we get
\begin{eqnarray}
{\mathsf L}_{(2)auxiliary}^{(45)}+{\mathsf
L}_{(3)auxiliary}^{(45)}+{\mathsf
L}_{(6)auxiliary}^{(45)}~~~~~~~~~~~~~~~~~~~~~~~~~~~~~~~~~~~~~~~~~~~~~~~~~\nonumber\\
=-\left(\mu^*
[h^{^{(45-)}}]^{\bf{-1}}[h^{^{(45-)}}]^{\bf{T}}[h^{^{(45-)}}]^{\bf{-1}}\mu\right)_{ab}\left[{\bf
A}^{\dagger}_{(+)a}{\bf A}_{(+)b}+\frac{1}{4}{\bf
A}^{\dagger}_{(+)aij}{\bf A}_{(+)b}^{ij}+{\bf
A}^{i\dagger}_{(+)a}{\bf A}_{(+)bi}\right]\nonumber\\
-\left(\mu[h^{^{(45+)}}]^{\bf{-1T}}h^{^{(45+)}}[h^{^{(45+)}}]^{\bf{-1T}}\mu^*\right)_{ab}
\left[ {\bf A}^{\bf{T}}_{(-)a}{\bf A}_{(-)b}^*+\frac{1}{4}{\bf
A}^{\bf{T}}_{(-)aij}{\bf A}_{(-)b}^{ij*}+{\bf
A}^{i\bf{T}}_{(-)a}{\bf A}_{(-)bi}^*\right]
\end{eqnarray}
Similarly, eliminating the auxiliary field $D_{\mu\nu}$ we get
\begin{eqnarray}
{\mathsf L}_{(1)auxiliary}^{(45)}+{\mathsf
L}_{(4)auxiliary}^{(45)}+{\mathsf
L}_{(5)auxiliary}^{(45)}~~~~~~~~~~~~~~~~~~~~~~~~~~~~~~~~~~~~~~~~~~~~~~~~~\nonumber\\
=-\frac{1}{32} {\mathsf
g}^{^{(45)}2}h_{ab}^{^{(45+)}}h_{cd}^{^{(45+)}}<A_{(+)a}|\Sigma_{\mu\nu}|A_{(+)b}><A_{(+)c}|\Sigma_{\mu\nu}|A_{(+)d}>
\nonumber\\
-\frac{1}{32} {\mathsf
g}^{^{(45)}2}h_{ab}^{^{(45-)}}h_{cd}^{^{(45-)}}<A_{(-)a}|\Sigma_{\mu\nu}|A_{(-)b}><A_{(-)c}|\Sigma_{\mu\nu}|A_{(-)d}>\nonumber\\
-\frac{1}{16} {\mathsf
g}^{^{(45)}2}h_{ab}^{^{(45+)}}h_{cd}^{^{(45-)}}<A_{(+)a}|\Sigma_{\mu\nu}|A_{(+)b}><A_{(-)c}|\Sigma_{\mu\nu}|A_{(-)d}>
\end{eqnarray}
The terms above when expanded in terms of SU(5) fields give
\begin{eqnarray}
-\frac{1}{32} {\mathsf
g}^{^{(45)}2}h_{ab}^{^{(45+)}}h_{cd}^{^{(45+)}}<A_{(+)a}|\Sigma_{\mu\nu}|A_{(+)b}><A_{(+)c}|\Sigma_{\mu\nu}|A_{(+)d}>
~~~~~~~~~~~~~~~~~~~~~~~~~~~~~~~~~\nonumber\\
={\mathsf
g}^{^{(45)}2}\{-\frac{1}{16}\left(\eta_{ab,cd}^{^{(45++)}}+4\eta_{ad,cb}^{^{(45++)}}\right)({\bf
A}^{\dagger}_{(+)a}{\bf A}_{(+)b}{\bf A}^{\dagger}_{(+)cij}{\bf
A}_{(+)d}^{ij}+{\bf A}^{i\dagger}_{(+)a}{\bf A}_{(+)bi}{\bf
A}^{j\dagger}_{(+)c}{\bf A}_{(+)dj}\nonumber\\
+{\bf A}^{i\dagger}_{(+)a}{\bf A}_{(+)bi}{\bf
A}^{\dagger}_{(+)cjk}{\bf A}_{(+)d}^{jk})
-\frac{1}{2}\left(\eta_{ab,cd}^{^{(45++)}}+\eta_{ad,cb}^{^{(45++)}}\right){\bf
A}^{i\dagger}_{(+)a}{\bf A}_{(+)bj}{\bf A}^{\dagger}_{(+)cik}{\bf
A}_{(+)d}^{kj}\nonumber\\
+\eta_{ab,cd}^{^{(45++)}}[-\frac{1}{8}\epsilon_{ijklm}{\bf
A}^{\dagger}_{(+)a}{\bf A}_{(+)b}^{ij}{\bf
A}^{k\dagger}_{(+)c}{\bf
A}_{(+)d}^{lm}-\frac{1}{8}\epsilon^{ijklm}{\bf
A}^{\dagger}_{(+)aij}{\bf A}_{(+)bk}{\bf A}^{\dagger}_{(+)clm}{\bf
A}_{(+)d}\nonumber\\
-\frac{1}{4}{\bf A}^{\dagger}_{(+)aik}{\bf A}_{(+)b}^{kj}{\bf
A}^{\dagger}_{(+)cjl}{\bf A}_{(+)d}^{li}+\frac{3}{64}{\bf
A}^{\dagger}_{(+)aij}{\bf A}_{(+)b}^{ij}{\bf
A}^{\dagger}_{(+)ckl}{\bf A}_{(+)d}^{kl}\nonumber\\
+\frac{3}{8}{\bf A}^{\dagger}_{(+)a}{\bf A}_{(+)b}{\bf
A}^{i\dagger}_{(+)c}{\bf A}_{(+)di}-\frac{5}{16}{\bf
A}^{\dagger}_{(+)a}{\bf A}_{(+)b}{\bf A}^{\dagger}_{(+)c}{\bf
A}_{(+)d}]\}
\end{eqnarray}
\begin{eqnarray}
-\frac{1}{32} {\mathsf
g}^{^{(45)}2}h_{ab}^{^{(45-)}}h_{cd}^{^{(45-)}}<A_{(-)a}|\Sigma_{\mu\nu}|A_{(-)b}><A_{(-)c}|\Sigma_{\mu\nu}|A_{(-)d}>
~~~~~~~~~~~~~~~~~~~~~~~~~~~~~~~~~\nonumber\\
={\mathsf
g}^{^{(45)}2}\{-\frac{1}{16}\left(\eta_{ab,cd}^{^{(45--)}}+4\eta_{ad,cb}^{^{(45--)}}\right)({\bf
A}^{\dagger}_{(-)ai}{\bf A}_{(-)b}^{i}{\bf
A}^{jk\dagger}_{(-)c}{\bf A}_{(-)djk} +{\bf
A}^{\dagger}_{(-)ai}{\bf A}_{(-)b}^{i}{\bf
A}^{\dagger}_{(-)cj}{\bf A}_{(-)d}^{j})\nonumber\\
-\frac{1}{16}\left(-11\eta_{ab,cd}^{^{(45--)}}+4\eta_{ad,cb}^{^{(45--)}}\right)
{\bf A}^{\dagger}_{(-)a}{\bf A}_{(-)b}{\bf
A}^{ij\dagger}_{(-)c}{\bf A}_{(-)dij}\nonumber\\
-\frac{1}{2}\left(\eta_{ab,cd}^{^{(45--)}}+\eta_{ad,cb}^{^{(45--)}}\right){\bf
A}^{\dagger}_{(-)ai}{\bf A}_{(-)b}^{j}{\bf
A}^{ik\dagger}_{(-)c}{\bf
A}_{(-)dkj}\nonumber\\
+\eta_{ab,cd}^{^{(45--)}}[-\frac{1}{8}\epsilon_{ijklm}{\bf
A}^{ij\dagger}_{(-)a}{\bf A}_{(-)b}{\bf A}^{kl\dagger}_{(-)c}{\bf
A}_{(-)d}^{m}-\frac{1}{8}\epsilon^{ijklm}{\bf
A}^{\dagger}_{(-)ai}{\bf A}_{(-)bjk}{\bf A}^{\dagger}_{(-)c}{\bf
A}_{(-)dlm}\nonumber\\
-\frac{1}{4}{\bf A}^{ik\dagger}_{(-)a}{\bf A}_{(-)bkj}{\bf
A}^{jl\dagger}_{(-)c}{\bf A}_{(-)dli}+\frac{3}{64}{\bf
A}^{ij\dagger}_{(-)a}{\bf A}_{(-)bij}{\bf
A}^{kl\dagger}_{(-)c}{\bf A}_{(-)dkl}\nonumber\\
+\frac{7}{8}{\bf A}^{\dagger}_{(-)a}{\bf A}_{(-)b}{\bf
A}^{\dagger}_{(-)ci}{\bf A}_{(-)d}^i+\frac{15}{16}{\bf
A}^{\dagger}_{(-)a}{\bf A}_{(-)b}{\bf A}^{\dagger}_{(-)c}{\bf
A}_{(-)d}]\}
\end{eqnarray}
\begin{eqnarray}
-\frac{1}{16} {\mathsf
g}^{^{(45)}2}h_{ab}^{^{(45+)}}h_{cd}^{^{(45-)}}<A_{(+)a}|\Sigma_{\mu\nu}|A_{(+)b}><A_{(-)c}|\Sigma_{\mu\nu}|A_{(-)d}>
~~~~~~~~~~~~~~~~~~~~~~~~~~~~~~~~~\nonumber\\
={\mathsf g}^{^{(45)}2}\eta_{ab,cd}^{^{(45+-)}}\{\frac{5}{8}{\bf
A}^{\dagger}_{(+)a}{\bf A}_{(+)b}{\bf A}^{\dagger}_{(-)c}{\bf
A}_{(-)d}-\frac{3}{32}{\bf A}^{\dagger}_{(+)aij}{\bf
A}_{(+)b}^{ij}{\bf A}^{kl\dagger}_{(-)c}{\bf
A}_{(-)dkl}\nonumber\\
+\frac{1}{8}{\bf A}^{i\dagger}_{(+)a}{\bf A}_{(+)bi}{\bf
A}^{\dagger}_{(-)cj}{\bf A}_{(-)d}^j+\frac{1}{2}{\bf
A}^{j\dagger}_{(+)a}{\bf A}_{(+)bi}{\bf A}^{\dagger}_{(-)cj}{\bf
A}_{(-)d}^i\nonumber\\
+\frac{1}{2}{\bf A}^{\dagger}_{(+)aik}{\bf A}_{(+)b}^{kj}{\bf
A}^{il\dagger}_{(-)c}{\bf A}_{(-)dlj}-\frac{3}{8}{\bf
A}^{\dagger}_{(+)a}{\bf A}_{(+)b}{\bf A}^{\dagger}_{(-)ci}{\bf
A}_{(-)d}^i\nonumber\\
+\frac{13}{8}{\bf A}^{i\dagger}_{(+)a}{\bf A}_{(+)bi}{\bf
A}^{\dagger}_{(-)c}{\bf A}_{(-)d}+\frac{1}{16}{\bf
A}^{\dagger}_{(+)a}{\bf A}_{(+)b}{\bf A}^{ij\dagger}_{(-)c}{\bf
A}_{(-)dij}\nonumber\\
+\frac{9}{16}{\bf A}^{\dagger}_{(+)aij}{\bf A}_{(+)b}^{ij}{\bf
A}^{\dagger}_{(-)c}{\bf A}_{(-)d}-\frac{15}{16}{\bf
A}^{i\dagger}_{(+)a}{\bf A}_{(+)bi}{\bf A}^{jk\dagger}_{(-)c}{\bf
A}_{(-)djk}\nonumber\\
-\frac{1}{16}{\bf A}^{\dagger}_{(+)aij}{\bf A}_{(+)b}^{ij}{\bf
A}^{\dagger}_{(-)kc}{\bf A}_{(-)d}^k+\frac{1}{2}{\bf
A}^{j\dagger}_{(+)a}{\bf A}_{(+)bi}{\bf A}^{ik\dagger}_{(-)c}{\bf
A}_{(-)dkj}\nonumber\\
+\frac{1}{2}{\bf A}^{\dagger}_{(+)aik}{\bf A}_{(+)b}^{kj}{\bf
A}^{\dagger}_{(-)jc}{\bf A}_{(-)d}^i-\frac{1}{4}{\bf
A}^{\dagger}_{(+)aij}{\bf A}_{(+)b}{\bf A}^{ij\dagger}_{(-)c}{\bf
A}_{(-)d}\nonumber\\
-\frac{1}{4}{\bf A}^{\dagger}_{(+)a}{\bf A}_{(+)b}^{ij}{\bf
A}^{\dagger}_{(-)c}{\bf A}_{(-)dij}-\frac{1}{4}{\bf
A}^{i\dagger}_{(+)a}{\bf A}_{(+)b}^{jk}{\bf
A}^{\dagger}_{(-)ci}{\bf
A}_{(-)djk}\nonumber\\
-\frac{1}{4}{\bf A}^{\dagger}_{(+)aij}{\bf A}_{(+)bk}{\bf
A}^{ij\dagger}_{(-)c}{\bf A}_{(-)d}^k-\frac{1}{2}{\bf
A}^{i\dagger}_{(+)a}{\bf A}_{(+)b}^{jk}{\bf
A}^{\dagger}_{(-)cj}{\bf
A}_{(-)dki}\nonumber\\
-\frac{1}{2}{\bf A}^{\dagger}_{(+)aij}{\bf A}_{(+)bk}{\bf
A}^{ki\dagger}_{(-)c}{\bf
A}_{(-)d}^j-\frac{1}{8}\epsilon^{ijklm}{\bf
A}^{\dagger}_{(+)aij}{\bf A}_{(+)b}{\bf A}^{\dagger}_{(-)ck}{\bf
A}_{(-)dlm}\nonumber\\
-\frac{1}{8}\epsilon_{ijklm}{\bf A}^{\dagger}_{(+)a}{\bf
A}_{(+)b}^{ij}{\bf A}^{kl\dagger}_{(-)c}{\bf
A}_{(-)d}^m-\frac{1}{8}\epsilon_{ijklm}{\bf
A}^{i\dagger}_{(+)a}{\bf A}_{(+)b}^{jk}{\bf
A}^{lm\dagger}_{(-)c}{\bf
A}_{(-)d}\nonumber\\
-\frac{1}{8}\epsilon^{ijklm}{\bf A}^{\dagger}_{(+)aij}{\bf
A}_{(+)bk}{\bf A}^{\dagger}_{(-)c}{\bf A}_{(-)dlm}\}
\end{eqnarray}
where $\eta$'s are defined by
\begin{eqnarray}
\eta_{ab,cd}^{^{(45++)}}=h_{ab}^{^{(45+)}}h_{cd}^{^{(45+)}};~~~~~\eta_{ab,cd}^{^{(45--)}}=h_{ab}^{^{(45-)}}h_{cd}^{^{(45-)}}
;~~~~~
\eta_{ab,cd}^{^{(45+-)}}=h_{ab}^{^{(45+)}}h_{cd}^{^{(45-)}}
\end{eqnarray}

\section{Appendix G}
In this Appendix we discuss the coupling of the U(1) vector with matter
without imposition of the constraint of the Wess-Zumino gauge.
We consider the following Lagrangian  which couples the
vector multiplet with a scalar multiplet $\widehat{\Phi}$.
\begin{eqnarray}
{\mathsf L}^{(U(1))}={\mathsf L}_{V}^{^{(U(1)~K.E.)}}+{\mathsf
L}_{V}^{^{(U(1)~Mass)}}+ {\mathsf
L}_V^{^{(U(1)~Self-Interaction)}}
+{\mathsf L}_{V+\Phi}^{^{(U(1)~Interaction)}}+{\mathsf L}_{\Phi}^{(U(1))}\nonumber\\
{\mathsf L}_{V}^{^{(U(1)~K.E.)}}=\frac{1}{4}\left[\widehat{\cal
W}^{\tilde{\alpha}} \widehat{\cal
W}_{\tilde{\alpha}}|_{\theta^2}+\widehat{\overline{\cal
W}}_{\dot{\tilde{\alpha}}}\widehat{\overline{\cal
W}}^{\dot{\tilde{\alpha}}}|_{\bar {\theta}^2}\right]\nonumber\\
{\mathsf L}_{V}^{^{(U(1)~Mass)}}=m^2\widehat{\mathsf
V}^2|_{\theta^2\bar
{\theta}^2}\nonumber\\
{\mathsf
L}_V^{^{(U(1)~Self-Interaction)}}={\alpha}_1\widehat{\mathsf
V}^3|_{\theta^2\bar{\theta}^2}+{\alpha}_2\widehat{\mathsf
V}^4|_{\theta^2\bar{\theta}^2}\nonumber\\
{\mathsf
L}_{V+\Phi}^{^{(U(1)~Interaction)}}=h\widehat{\Phi}^{\dagger}_a\widehat{\mathsf
V}\widehat{\Phi}_a|_{\theta^2\bar {\theta}^2}\nonumber\\
{\mathsf
L}_{\Phi}^{(U(1))}=\widehat{\Phi}^{\dagger}_a\widehat{\Phi}_a|_{\theta^2\bar
{\theta}^2}+\left[{\mathsf
W}(\widehat{\Phi})|_{\theta^2}+h.c.\right]
\end{eqnarray}
where

\begin{eqnarray}
\widehat{\cal W}_{\tilde{\alpha}}=-\frac{1}{4}\overline {\mathsf
D}^2\mathsf
D_{\tilde{\alpha}}\widehat{\mathsf V};~~~~
\widehat{\overline{\cal
W}}_{\dot{\tilde{\alpha}}}=-\frac{1}{4}{\mathsf D}^2\overline
{\mathsf D}_{\dot{\tilde{\alpha}}}\widehat{\mathsf V}
\end{eqnarray}

Finally, the superpotential ${\mathsf W}(\widehat{\Phi})$ of the
theory is
\begin{equation}
{\mathsf W}(\widehat{\Phi})={\cal F}_a\widehat{\Phi}_a+\frac{1}{2}
{\cal M}_{ab}\widehat{\Phi}_a\widehat{\Phi}_b+\frac{1}{3}
{\cal G}_{abc}\widehat{\Phi}_a\widehat{\Phi}_b\widehat{\Phi}_c
\end{equation}
The couplings ${\cal M}_{ab}$ and ${\cal G}_{abc}$ are taken to be
completely symmetric tensors.
 Expansion in component form gives
\begin{equation}
{\mathsf L}_{V}^{^{(U(1)~K.E.)}}=-\frac{1}{4}{\cal V}_{AB}{\cal
V}^{AB}+\frac{1}{2}D^2-i\lambda\sigma^A\partial_A\overline
{\lambda}
\end{equation}
\begin{eqnarray}
{\mathsf
L}_{V}^{^{(U(1)~Mass)}}=m^2CD+\frac{1}{2}m^2\left(M^2+N^2\right)-m^2\left(\lambda\chi+
\overline{\lambda}\overline{\chi}\right)-\frac{1}{2}m^2\partial^AC\partial_AC
\nonumber\\
-im^2\chi\sigma^A\partial_A\overline{\chi}-\frac{1}{2}m^2{\cal V}_{A}{\cal V}^A
\end{eqnarray}
\begin{eqnarray}
{\mathsf L}_{V+\Phi}^{^{(U(1)~Interaction)}}=-\frac{h}{2\sqrt
2}\left(\chi\sigma^A\overline{\psi}_a\right)\partial_AA_a+\frac{h}{\sqrt
2}
\left(\chi\sigma^A\partial\overline{\psi}_a\right)A_a-\frac{ih}{\sqrt
2}\left(\overline{\lambda}\overline{\psi}_a\right)A_a\nonumber\\
-\frac{h}{2\sqrt
2}\left(\psi_a\sigma^A\overline{\chi}\right)\partial_AA^{\dagger}_a-\frac{h}{\sqrt
2}
\left(\psi_a\sigma^A\partial\overline{\chi}\right)A^{\dagger}_a+\frac{ih}{\sqrt
2}\left(\lambda\psi_a\right)A^{\dagger}_a\nonumber\\
+\frac{ih}{2}{\cal
V}_A\left[\left(\partial^AA_a\right)A^{\dagger}_a-
\left(\partial^AA^{\dagger}_a\right)A_a\right]+\frac{h}{2}{\cal
V}_A
\left(\psi_a\sigma^A\overline{\psi}_a\right)\nonumber\\
-hC\left(\partial_AA^{\dagger}_a\right)\left(\partial_AA_a\right)
-ihC\left(\psi_a\sigma^A\partial_A\overline{\psi}_a\right)-
\frac{h}{4}\partial^A\left(A_aA^{\dagger}_a\right)\partial_AC\nonumber\\
+hCF_aF^{\dagger}_a-\frac{ih}{\sqrt
2}\left(\chi\psi_a\right)F^{\dagger}_a +\frac{ih}{\sqrt
2}\left(\overline{\chi}\overline{\psi}_a\right)F_a
+\frac{h}{2}DA_aA^{\dagger}_a\nonumber\\
+\frac{ih}{2}\left(M+iN\right)A_aF^{\dagger}_a-\frac{ih}{2}\left(M-iN\right)A^{\dagger}_a
F_a
\end{eqnarray}
\begin{eqnarray}
{\mathsf
L}_V^{^{(U(1)~Self-Interaction)}}=-3\left(\frac{\alpha_1}{2}C+\alpha_2C^2
\right[{\cal V}_A{\cal V}^A+2\left(\lambda\chi+
\overline{\lambda}\overline{\chi}\right)+2i\chi\sigma^A\partial_A\overline{\chi}\nonumber\\
-\left(M^2+N^2\right)]
+3\left(\frac{\alpha_1}{4}+\alpha_2C\right)[i\left(M+iN\right)
\left(\overline{\chi}\overline{\chi}\right)-
i\left(M-iN\right)\left(\chi\chi\right)\nonumber\\
-2\left(\chi\sigma^A\overline{\chi}\right) {\cal V}_A]
+\left(\frac{3\alpha_1}{4}C+\alpha_2C^2\right)\left[2CD-\partial_AC\partial^AC
\right] + \frac{3\alpha_2}{2}
\left(\chi\chi\right)\left(\overline{\chi}\overline{\chi}\right)
\end{eqnarray}
\begin{eqnarray}
{\mathsf
L}_{\Phi}^{(U(1))}=-\partial_AA^{\dagger}_a\partial^AA_a-i\overline{\psi}_a
\overline{\sigma}^A\partial_A\psi_a+F^{\dagger}_aF_a\nonumber\\
-\left(\frac{1}{2}{\cal M}_{ab}+{\cal
G}_{abc}A_c\right)\psi_a\psi_b -\left(\frac{1}{2}{\cal
M}_{ab}^*+{\cal G}_{abc}^*A_c^{\dagger}
\right)\overline{\psi}_a\overline{\psi}_b\nonumber\\
+\left( {\cal F}_a+{\cal M}_{ab}A_b+{\cal G}_{abc}A_cA_c\right)F_a
+\left({\cal F}_a^*+{\cal M}_{ab}^*A_b^{\dagger}+{\cal
G}_{abc}^*A_c^{\dagger}A_c^{\dagger} \right)F_a^{\dagger}
\end{eqnarray}
Evaluation of Eq.(89) using Eqs.(90-96) gives in the
four-component notation
\begin{eqnarray}
{\mathsf L}^{(U(1))}= -\frac{1}{4}{\cal V}_{AB}{\cal
V}^{AB}-\frac{1}{2}m^2{\cal V}_{A}{\cal
V}^{A}-\frac{1}{2}\partial^AB\partial_AB -i\overline
{\Lambda}\gamma^A\partial_A\Lambda-m\overline
{\Lambda}\Lambda\nonumber\\
-\partial^AA^{\dagger}_a\partial_AA_a
 -\frac{h}{m}B\partial^AA^{\dagger}_a\partial_AA_a
\nonumber\\
-\frac{h}{4m}\partial^A\left(A_aA^{\dagger}_a\right)
\partial_AB+\frac{ih}{2}\left(A^{\dagger}_a\partial^AA_a-A_a\partial^A
A_a^{\dagger}\right)
{\cal V}_A\nonumber\\
+\frac{h}{2m\sqrt{2}}\left[\left(\overline
{\Psi}_{aL}\gamma^A\Lambda_{L}\right)\partial_AA_a+\left(\overline
{\Lambda}_{L}\gamma^A\Psi_{aL}\right)\partial_AA_a^{\dagger}\right]\nonumber\\
+\frac{h}{m\sqrt{2}}\left[\left(\overline
{\Psi}_{aL}\gamma^A\partial_A\Lambda_{L}\right)A_a-\left(\overline
{\Lambda}_{L}\gamma^A\partial_A\Psi_{aL}\right)A_a^{\dagger}\right]\nonumber\\
-\frac{ih}{\sqrt{2}}\left[\left(\overline
{\Psi}_{aL}\Lambda_{R}\right)A_a-\left(\overline
{\Lambda}_{R}\Psi_{aL}\right)A_a^{\dagger}\right]\nonumber\\
-\left[\left(\frac{1}{2}{\cal M}_{ab}+{\cal
G}_{abc}A_d\right)\overline{\Psi}_{aR}\Psi_{bL}
+\left(\frac{1}{2}{\cal M}_{ab}^*+{\cal G}_{abc}^*A_d^{\dagger}
\right)\overline{\Psi}_{aL}\Psi_{bR}\right]\nonumber\\
-i\left(1+\frac{h}{m}B\right)\overline {\Psi}_{aL}\gamma^A
{\cal D}_A\Psi_{aL}\nonumber\\
-\frac{1}{m^3}\left(\frac{3\alpha_1}{4}B+\frac{\alpha_2}{m}B^2\right)
\partial_AB\partial^AB-
\frac{3}{m}\left(\frac{\alpha_1}{2}B+\frac{\alpha_2}{m}B^2\right)
{\cal V}_A{\cal V}^A\nonumber\\
+\frac{3}{m^3}\left(\alpha_1B+\frac{2\alpha_2}{m}B^2\right)
\left(i\overline {\Lambda}_L\gamma^A\partial_A\Lambda_L
-m\overline{\Lambda}\Lambda\right)\nonumber\\
+\frac{3}{m^2}\left(\frac{\alpha_1}{2}+\frac{2\alpha_2}{m}B\right)
\left(\overline{\Lambda}_L\gamma^A\Lambda_L\right){\cal V}_A+
\frac{3\alpha_2}{2m^4}
\left(\overline{\Lambda}^c_R\Lambda_L\right)
\left(\overline{\Lambda}_L\Lambda_R^c\right) +{\mathsf
L}_{auxiliary}^{(U(1))}
\end{eqnarray}
where we have defined
\begin{eqnarray}
\Lambda=\left(\matrix{m\chi_{\tilde{\alpha}}\cr
\overline{\lambda}^{\dot{\tilde\alpha}}}\right),~~~
\Psi_a=\left(\matrix{\psi_{a\tilde{\alpha}}\cr
\overline{\psi}^{\dot{\tilde\alpha}}_a}\right),~~~ B=mC,~~~
 \Lambda^c={\cal C}\overline{\Lambda}^T,~~~ {\cal
C}=\left(\matrix{i\sigma^2&0\cr
0&i\overline{\sigma}^2}\right),\nonumber\\
 \overline{\Lambda}=\Lambda^{\dagger}\gamma^0,~~~\Lambda_{R,L}=\frac{1\pm
 \gamma_5}{2}\Lambda~~~,
 {\cal D}_A=\partial_A-\frac{ig}{2}\left(1+\frac{g}{m}B\right)^{-1}{\cal
V}_A
\end{eqnarray}
and
\begin{eqnarray}
{\mathsf
L}_{auxiliary}^{(U(1))}=\left(mB+\frac{3\alpha_1}{2m^2}B^2
+\frac{2\alpha_2}{m^3}B^3+\frac{h}{2}A^{\dagger}_aA_a\right)D
+\frac{1}{2}D^2\nonumber\\
+\left(\frac{1}{2}m^2+\frac{3\alpha_1}{2m}B+\frac{3\alpha_2}{m^2}B^2\right)
\left(M^2+N^2\right)\nonumber\\
+i\left[\frac{h}{2}A_aF^{\dagger}_a+\frac{3}{m^2}\left(\frac{\alpha_1}{4}+
\frac{\alpha_2}{m}B\right)\left(\overline{\Lambda}_L\Lambda_R^c\right)
\right]\left(M+iN\right)\nonumber\\
-i\left[\frac{h}{2}A^{\dagger}_aF_a+\frac{3}{m^2}\left(\frac{\alpha_1}{4}+
\frac{\alpha_2}{m}B\right)\left(\overline{\Lambda}^c_R\Lambda_L\right)
\right]\left(M-iN\right)\nonumber\\
+\left[{\cal F}_a+{\cal M}_{ab}A_b+{\cal G}_{abc}A_bA_c
 +\frac{ih}{m\sqrt 2}\left(\overline{\Lambda}_{L}\Psi_{aR}\right)\right]F_a
\nonumber\\
+\left[{\cal F}_a^*+{\cal M}_{ab}^*A_b^{\dagger}+{\cal
G}_{abc}^*A_c^{\dagger}A_d^{\dagger}-\frac{ih}{m\sqrt 2}
\left(\overline{\Psi}_{aR}\Lambda_L\right)\right]F_a^{\dagger}
+\left(\frac{h}{m}B+1\right)F^{\dagger}_aF_a
\end{eqnarray}
We next eliminate the auxiliary fields $M$, $N$, and $D$ through
their field equations  to get
\begin{eqnarray}
{\mathsf
L}_{auxiliary}^{(U(1))}=-\frac{1}{2}m^2B^2-\frac{3\alpha_1}{2m}B^3
-\left(\frac{9{\alpha_1}^2}{8m^4}+\frac{2\alpha_2}{m^2}\right)B^4
-\frac{3\alpha_1\alpha_2}{m^5}B^5-\frac{2{\alpha_2}^2}{m^6}B^6\nonumber\\
-\frac{h^2}{8}\left(A^{\dagger}_aA_a\right)\left(A^{\dagger}_bA_b\right)
-\frac{mh}{2}B\left(A^{\dagger}_aA_a\right)
-\frac{3h\alpha_1}{4m^2}B^2\left(A^{\dagger}_aA_a\right)
-\frac{h\alpha_2}{m^3}B^3\left(A^{\dagger}_aA_a\right)\nonumber\\
+ \left[{\cal F}_a^*+{\cal M}_{ab}^*A_b^{\dagger}+{\cal
G}_{abc}^*A_b^{\dagger}A_c^{\dagger}-\frac{ih}{m\sqrt 2}
\left(\overline{\Psi}_{aR}\Lambda_L\right) -\frac{hf_1(B)}{f_2(B)}
\left(\overline{\Lambda}^c_R\Lambda_L\right)\right]F_a^{\dagger}\nonumber\\
-\frac{2f_1^2(B)}{f_2(B)}
\left(\overline{\Lambda}^c_R\Lambda_L\right)
\left(\overline{\Lambda}_L\Lambda_R^c\right)
\end{eqnarray}
where
\begin{eqnarray}
f_1(B)=\frac{3}{m^2}\left(\frac{\alpha_1}{4}+\frac{\alpha_2}{m}B\right),~~~
f_2(B)=m^2+\frac{3\alpha_1}{m}B+\frac{6\alpha_2}{m^2}B^2
\end{eqnarray}
and the auxiliary field $F$ satisfies the field equation
\begin{eqnarray}
F^{\dagger}_b\left[\delta_{ab}\left(1+\frac{h}{m}B\right)-\frac{h^2}{2f_2(B)}
A^{\dagger}_aA_b\right]= -\frac{ih}{m\sqrt
2}\overline{\Lambda}_{L}\Psi_{aR}+
\frac{hf_1(B)}{f_2(B)}\left(\overline{\Lambda}_L\Lambda_R^c\right)
A^{\dagger}_a\nonumber\\
-{\cal F}_a-{\cal M}_{ab}A_b-{\cal G}_{abc}A_bA_c
\end{eqnarray}
Inverting this last equation we obtain
\begin{eqnarray}
F^{\dagger}_a= \left(1+\frac{h}{m}B\right)^{-1}\left[\delta_{ab}+
\frac{h^2A_a^{\dagger}A_b}{2f_2^2(B)\left(1+\frac{h}{m}B\right)-h^2A_c^{\dagger}A_c}
\right]\nonumber\\
\times \left[-\frac{ih}{m\sqrt 2}\overline{\Lambda}_{L}\Psi_{bR}+
\frac{hf_1(B)}{f_2(B)}\left(\overline{\Lambda}_L\Lambda_R^c\right)
A^{\dagger}_b -{\cal F}_b-{\cal M}_{bd}A_d-{\cal G}_{bde}A_dA_e
\right]
\end{eqnarray}
For the case when self-interactions of the vector multiplet
are absent (i.e., $\alpha_1=\alpha_2=0$), we get
\begin{eqnarray}
{\mathsf L}^{(U(1))}= -\frac{1}{4}{\cal V}_{AB}{\cal
V}^{AB}-\frac{1}{2}m^2{\cal V}_{A}{\cal
V}^{A}-\frac{1}{2}\partial^AB\partial_AB -i\overline
{\Lambda}\gamma^A\partial_A\Lambda-m\overline
{\Lambda}\Lambda\nonumber\\
-\partial^AA^{\dagger}_a\partial_AA_a
 -\frac{h}{m}B\partial^AA^{\dagger}_a\partial_AA_a
\nonumber\\
-\frac{h}{4m}\partial^A\left(A_aA^{\dagger}_a\right)
\partial_AB+\frac{ih}{2}\left(A^{\dagger}_a\partial^AA_a-A_a\partial^A
A_a^{\dagger}\right)
{\cal V}_A\nonumber\\
+\frac{h}{2m\sqrt{2}}\left[\left(\overline
{\Psi}_{aL}\gamma^A\Lambda_{L}\right)\partial_AA_a+\left(\overline
{\Lambda}_{L}\gamma^A\Psi_{aL}\right)\partial_AA_a^{\dagger}\right]\nonumber\\
+\frac{h}{m\sqrt{2}}\left[\left(\overline
{\Psi}_{aL}\gamma^A\partial_A\Lambda_{L}\right)A_a-\left(\overline
{\Lambda}_{L}\gamma^A\partial_A\Psi_{aL}\right)A_a^{\dagger}\right]\nonumber\\
-\frac{ih}{\sqrt{2}}\left[\left(\overline
{\Psi}_{aL}\Lambda_{R}\right)A_a-\left(\overline
{\Lambda}_{R}\Psi_{aL}\right)A_a^{\dagger}\right]\nonumber\\
-\left[\left(\frac{1}{2}{\cal M}_{ab}+{\cal
G}_{abc}A_c\right)\overline{\Psi}_{aR}\Psi_{bL}
+\left(\frac{1}{2}{\cal M}_{ab}^*+{\cal G}_{abc}^*A_c^{\dagger}
\right)\overline{\Psi}_{aL}\Psi_{bR}\right]\nonumber\\
-i\left(1+\frac{h}{m}B\right)\overline {\Psi}_{aL}\gamma^A
{\cal D}_A\Psi_{aL}\nonumber\\
 -\frac{1}{2}m^2B^2-\frac{h^2}{8}\left(A^{\dagger}_aA_a\right)
\left(A^{\dagger}_bA_b\right)-\frac{h}{2}mB\left(A^{\dagger}_aA_a\right)
\nonumber\\
-\left[\frac{\delta_{ab}}{1+\frac{h}{m}B}+
\frac{h^2A_a^{\dagger}A_b}{2m^2-h^2\left(1+\frac{h}{m}B\right)A_c^{\dagger}A_c}
\right] \left[{\cal F}_a+{\cal M}_{ad}A_d+{\cal G}_{ade}A_dA_e
+\frac{ih}{m\sqrt 2}\overline{\Lambda}_{L}\Psi_{aR}
\right]\nonumber\\
~~~~~~~~~~~~~~~~~~\times \left[{\cal F}_b^*+{\cal
M}_{bf}^*A_f^{\dagger}+{\cal
G}_{bfg}^*A_f^{\dagger}A_g^{\dagger}-\frac{ih}{m\sqrt 2}
\left(\overline{\Psi}_{bR}\Lambda_L\right)\right]\nonumber\\
\end{eqnarray}
As is evident the $U(1)$ invariant effective Lagrangian above is
highly nonlinear with infinite order nonlinearities.

\end{document}